\def\slash#1{\setbox0=\hbox{$#1$}#1\hskip-\wd0\dimen0=5pt\advance
       \dimen0 by-\ht0\advance\dimen0 by\dp0\lower0.5\dimen0\hbox
	 to\wd0{\hss\sl/\/\hss}}
\documentstyle[fleqn,epsf]{article}
\begin{document}

\def\gisd{{G_{D^{**} D \pi}}}
\def\gisb{{G_{B^{**} B \pi}}}
\def\lq{\left [}
\def\rq{\right ]}
\def\LL{{\cal L}}
\def\VV{{\cal V}}
\def\AA{{\cal A}}
\def\gi{{g_{P^* P \pi}}}
\def\gis{{g_{P^* P^* \pi}}} 
\def\lq{\left [}
\def\rq{\right ]}
\def\qq{<{\overline u}u>}
\def\dmu{\partial_{\mu}}
\def\dmus{\partial^{\mu}}
\def\gid{{g_{D^* D \pi}}}
\def\gib{{g_{B^* B \pi}}}
\def\gids{{g_{D^* D^* \pi}}}
\def\gibs{{g_{B^* B^* \pi}}}
\def\mbs{{M_{B^*}}}
\def\fbs{{f_{B^*}}}
\def\MM{{\cal M}}
\def\BB{{\cal B}}

\newcommand{\bra}[1]{\left\langle #1 \right|}
\newcommand{\ket}[1]{\left| #1 \right\rangle}
\newcommand{\spur}[1]{\not\! #1 \,}
\newcommand{\be}{\begin{equation}}
\newcommand{\ee}{\end{equation}}
\newcommand{\bea}{\begin{eqnarray}}
\newcommand{\eea}{\end{eqnarray}}
\newcommand{\nn}{\nonumber}
\newcommand{\dd}{\displaystyle}

\bigbreak
\begin{center}
  \begin{large}
  \begin{bf}
{PHENOMENOLOGY OF HEAVY MESON CHIRAL LAGRANGIANS} \\ 
\end{bf}
  \vspace{1.2cm}
R. Casalbuoni \\ 
{\small Dipartimento di Fisica, Universit\`a di Firenze, and INFN, 
largo E.Fermi, 2 I-50125 Firenze, Italy}
\\
A. Deandrea, N. Di Bartolomeo and R. Gatto \\
{\small D\'epartement de Physique Th\'eorique, 
24, quai Ernest-Ansermet, CH-1211 Gen\`eve 4, Switzerland}
\\
F. Feruglio \\
{\small Dipartimento di Fisica, Universit\`a di Padova, and INFN, 
via Marzolo, 8 I-35131 Padova, Italy}
\\
G. Nardulli \\
{\small Dipartimento di Fisica, Universit\`a di Bari, and INFN,
via Amendola, 173 I-70126 Bari, Italy}
\end{large}

\vspace{1.5cm}

\begin{abstract}
The approximate symmetries of Quantum ChromoDynamics in the infinite heavy 
quark ($Q=c,b$) mass limit ($m_Q \to \infty$) and in the chiral 
limit for the light quarks
($m_q \to 0,\;q=\,u,\,d,\,s$) can be used together to build up an effective
chiral lagrangian for heavy and light mesons describing strong interactions
among effective meson fields as well as their couplings to electromagnetic
and weak currents, including the relevant symmetry breaking terms.
The effective theory includes heavy ($Q \bar q$) mesons of both 
negative and positive parity, light pseudoscalars, as well as light vector 
mesons. We summarize the estimates for the parameters entering the
effective lagrangian and discuss in particular 
some phenomenologically important couplings, such as $g_{B^* B \pi}$.
The hyperfine splitting of heavy mesons is discussed in detail. 
The effective lagrangian allows for the possibility to describe consistently 
weak couplings of heavy ($B,\, D$) to light ($\pi,\, \rho, \,
K^*,\,$ etc.) mesons. The method has however its own limitations, 
due to the requirement that the light meson momenta should be small,
and we discuss how such limitations can be circumvented through reasonable 
ansatz on the form factors.
Flavour conserving (e. g. $B^* \to B\, \gamma$) and
flavour changing (e. g. $B \to K^* \, \gamma$) radiative decays
provide another field of applications of effective lagrangians; they
are discussed together with their phenomenological implications.
Finally we analyze effective lagrangians describing heavy charmonium- like
($\bar Q Q$) mesons and their strong and electromagnetic interactions.
The role of approximate heavy quark symmetries for this case and the
phenomenological tests of these models are also discussed.
\end{abstract}
\vspace{1cm}
UGVA-DPT 1996/05-928\\
BARI-TH/96-237\\
hep-ph/9605342
\end{center}
\newpage
\tableofcontents
\newpage

\section{Introduction}

There is a general agreement at the present time that quantum chromodynamics
(QCD) is the correct theory of strong interactions. Although QCD
is simple and elegant in its formulation, the derivation of its
physical predictions presents however arduous difficulties
because of long distance QCD effects that are essentially non perturbative. 
Related to them is, for example, the
most prominent expected implication of QCD, color confinement.

Inevitably, QCD effects enter any calculation of
processes involving hadrons, such as electroweak transitions
between hadronic states. Predictions for such transitions and
their comparison with data are essential to complete the program
of determining the parameters of the standard electroweak model.
The main source of uncertainty for such predictions is our inability
to calculate the relevant non perturbative QCD effects. 

The theoretical progress in the field has gone through various
directions, including lattice simulations and the use of 
sum rules, but one framework has emerged as basic to advance our
understanding, namely the one employing approximate symmetries, broken
explicitly or spontaneously, or both ways.

The empirical pattern of quark masses, 
that are widely different, is the essential
logical guide to the formulation of the  symmetries that
have been introduced. Historically, different roads were followed,
some symmetries being already known and investigated even before the
notion of quark was established.
The first important development was isotopic spin, vastly used already
in the physics of nuclei, suggested by the approximate equality
of proton and neutron mass. In the quark language it is the closeness
of the masses of the up and down quark that induces isotopic spin
symmetry. The strange quark being much heavier, the extension of the
$SU(2)$ isospin symmetry to $SU(3)$, to include the strange quark, then
necessarily implied dealing with stronger symmetry breaking effects.
Later on it was realized that there is a typical energy
scale of hadronic phenomena, such that it is the relative magnitude of
the symmetry breaking mass parameters, as compared to such a scale,
which suggests the degree of accuracy of the symmetry predictions.

 From this point of view the magnitude of the $SU(2)$ breaking
was generally expected to be related to the ratio of the up and
down quark mass difference to the hadronic scale (plus the effects
of electromagnetism, which breaks isospin as well). Both masses are now known
to be very small in comparison to the scale, which suggests a
larger symmetry, $SU(2)\times{SU(2)}$, the light quark chiral symmetry,
exactly valid in QCD in the limit when both the up and down quark have zero 
mass.

Spontaneous symmetry breaking takes place and breaks the chiral symmetry into 
isospin, thus explaining the better experimental viability of isospin in 
strong
phenomena as compared to chiral symmetry. Historically, the progress went
the other way around, with chiral symmetry proposed before
the quark mass values were roughly known. Basic to this progress was the
interpretation of the pion as the Goldstone boson of the spontaneous
symmetry breaking.

One can attempt to treat the strange quark as a massless quark in
some first approximation, ready to deal subsequently with
substantial deviations from the symmetry. The approximate chiral
symmetry is then extended to chiral $SU(3)\times{SU(3)}$.

Within such a frame, current algebra provides for a number of useful
results. The other useful approach is a systematical lagrangian
expansion, known as chiral perturbation theory. In this approach
the symmetry is used to provide for a {\it catalogue raisonn\'e} of
the terms appearing in the chiral expansion. In this way one can
determine which phenomenological inputs are needed to fix at
a given chiral order the full expansion and decide how to determine
them from experiment \cite{ltw}.

At the opposite side with respect to the hadronic scale of QCD, are
the heavier quark masses, i.e. those of  the beauty $b$
and charm $c$ quark. In the limit of
infinite masses ($m_c , m_b \to \infty$), three phenomena appear.

The first one consists in the
fact that the resulting effective lagrangian
exhibits a $SU(2)$ heavy flavour symmetry; 
this symmetry applies to quantities that remain finite in the limit 
$m_Q \to \infty$ and arises because, in such extreme limit, the exact value
of the heavy quark mass plays no role in its interaction with the
light sector.  For finite quark masses the heavy flavour symmetry
is broken, and the breaking can be
relevant especially in the charm sector, since the $c$ quark is
substantially lighter than the $b$ quark.

The second phenomenon is a heavy quark
velocity superselection rule, which is due
to the fact that the strong interactions of the heavy quark,
in the $m_Q \to \infty$ limit,
do not change its velocity $v_Q$ that always remain equal to 
the heavy meson velocity (only weak and electromagnetic 
interactions can change $v_Q$).
As a consequence of the velocity superselection rule, 
the effective lagrangian
describing strong interactions of the heavy quarks
should be written as a sum of terms that are diagonal in the
velocity dependent heavy quark field operators.

The last phenomenon appearing in the limit of infinite quark mass is the 
decoupling of the gluon from the quark spin; in other words the effective 
lagrangian is invariant under heavy quark spin transformations and has, 
therefore, a further $SU(2)$ spin symmetry. In conclusion, the
complete symmetry of the effective lagrangian
is a $SU(2 N_f)$ of flavor ($N_f$ is the number of heavy flavours) 
and spin for each
value of the heavy quark velocity. The resulting
effective theory is nowadays known as Heavy Quark Effective Theory
(HQET) (see \cite{hq1,hq2,hq3,hq4,hq5,hqrev,neubert}). In the physical world 
the
symmetry is broken explicitly because of the finite heavy quark masses.
Symmetry breaking terms are expected to be particularly
important for charm quark and they can be  systematically added to the 
lagrangian of HQET and parameterize, order by order, 
the deviations from the heavy mass limit.

One of the first and most important applications of the heavy quark symmetries
has been the study of the semileptonic decay of $B \to D^* l \bar{\nu}$. In
the infinite quark mass limit this process is described by one form factor 
whose
normalization is fixed at the kinematical point where the two heavy
quarks have the same velocity. The velocity of the heavy quark in this limit
is, as we have already stressed, the velocity of the meson.

To illustrate in more detail the usefulness of the heavy quark symmetry
one can consider the analogy between the determination of the $V_{us}$,
element of the CKM matrix from the semileptonic decay  $K\to
\pi e \bar{\nu}$ and the possible determination of the element $V_{cb}$
from the semileptonic decay of $B$ into $D^*$. For the $K$ to $\pi$
decay, a non renormalization theorem says that corrections to the
$SU(3)$ normalization of the form factor at the symmetry point, i.e. zero
momentum transfer, vanish at first order in the difference between the
strange quark mass and the nonstrange quark mass. For the heavy
transition the symmetry is the heavy quark symmetry, which is
valid for very large quark masses; in this limit some relevant form factors 
are
renormalized only at second order in the symmetry breaking parameter
(the inverse quark mass) at the relevant (zero velocity) symmetry point
\cite{luke90}.

The last example shows the usefulness of the heavy quark symmetry not
only to provide us with exact relations valid in the {\it terra firma} of
the exact limit, but also as a platform for studying corrections away
from the limit.

In some kinematical regions, which, at the same time,
are not very far from the heavy quark limit and from the chiral limit
for the light particles, one can try to use simultaneously both
the heavy quark and the chiral approach in the two distinct sectors and, as we 
have already discussed, the most economical way to do this consists in using 
phenomenological lagrangians. In other words, chiral $SU(3)\times SU(3)$ 
symmetry can be used together with the spin-flavour heavy quark symmetry of 
HQET and the velocity superselection rule to build up an effective lagrangian 
whose basic fields are heavy and light meson operators. 
This approach has been proposed in a number of papers 
\cite{wise,yan,burdman92,noilr,casa92,casa93,schechter,falkluke,kilian,ebert} 
and the purpose of this report is to review this method and its 
applications to the interactions among heavy and light mesons. 

The first results we describe are in the field of the strong 
interactions and concern the properties of the 
effective fields describing the heavy mesons as well 
as their couplings to the pseudoscalar octet of the Goldstone bosons. 
We also discuss the introduction  
in the lagrangian of the light vector resonances 
$\rho, K^*$ etc and the inclusion of positive parity
heavy meson states. Applications of these ideas
to heavy baryons containing one or more heavy quarks have been
also studied \cite{barwise,bargeorgi,barman,barhussain,HUSSAIN}, but they will 
not be reviewed here because the experimental situation 
concerning heavy baryons is still poor and there are 
therefore too few constraints
on the parameters of the resulting effective theory. 
  
The chiral lagrangian approach has the advantage of allowing for a 
perturbative theory including not only tree level contributions, but also loop 
calculations. Such calculations, at  present, can only give an order of 
magnitude estimate of the effects, because we do not have yet 
sufficient experimental data to fix the arbitrary coefficients
in the counterterms of the effective lagrangian. Nevertheless
they offer a clue to the size of the loop effects and can be
extremely useful in reconciling  data with 
theoretical expectations based on  tree level calculations.
For pedagogical purposes we shall present two explicit and 
detailed examples of these calculations;
the first one is the evaluation of the loop effects in the 
hyperfine $M_B^* - M_B$ mass splitting. The second example we
shall show is given by the chiral loop effects to the 
ratio ${f_{D_s}}/{f_D}$, where $f_D$ and $f_{D_s}$
are the $D$ and ${D_s}$ meson leptonic decay constants. Other examples
of computed chiral loop effects are given by the corrections
to the strong coupling constants $g_{D^* D \pi}$, to semileptonic
form factors and to $B$ and $D$ meson radiative and rare decays: they
will be also reviewed and, whenever possible, the results
of the lagrangian approach will be compared to other 
existing theoretical methods.

The main pitfall of the effective lagrangian approach is the abundance of 
coupling constants and parameters appearing in the lagrangian. Even if one 
works at the lowest order in the light meson derivatives and in the $1/m_Q$ 
expansion, one has to fix several couplings from  data. A typical
example is the already mentioned $D^* D \pi$ coupling
constant, whose experimental determination is still missing.
In absence of experimental inputs, one may rely on 
theoretical information coming, for example, from QCD sum
rules \cite{qcdsr} (for a review of this subject see 
\cite{M.Shifman}), or potential models \cite{defazio2} (for a review see 
\cite{rivnc}) or,
when available, on the results obtained by Lattice QCD
(for a review of the $D$ and $B$ meson phenomenology
on the lattice see \cite{MARTINELLI}). An alternative is provided by the use of
information coming not only from strong interactions, but also 
from weak and electromagnetic interactions 
among mesons. Actually the application of the chiral 
lagrangian to these processes offers the possibility not 
only to exploit experimental data to constrain the 
effective lagrangian, but also to relate different processes 
using the symmetries. This is the second main issue to be discussed in the 
present report. We shall see that two methods can be used to perform the task: 
the first one uses the chiral and heavy flavour symmetries to 
relate different weak and electromagnetic transitions by 
establishing scaling relations among them. The second 
method makes use of the chiral lagrangian 
to compute the different amplitudes. 
In both cases, however, some additional hypothesis
on the $q^2$-behaviour of the form factors must be made and we shall
discuss the different scenarios as well as their comparison with the data.

The third and final topic discussed in 
this paper is the application of the ideas of HQET to 
mesons made up by two heavy quarks  
(heavy quarkonium). The effective quark theory resulting from the 
$ m_Q \to \infty$ limit satisfies, as in the previous case, 
the velocity superselection rule and the spin symmetry, 
but not the heavy flavour symmetry. As a matter of fact, the non 
relativistic kinetic energy term of the effective QCD lagrangian, 
which is flavour dependent, 
cannot be neglected since it acts as an infrared regulator. 
Therefore the chiral effective lagrangian for light mesons and
heavy quarkonium-like mesons does not possess the $SU(2)$ heavy flavour 
symmetry; nevertheless, because of the spin and chiral symmetries,
it allows for a number of relations among different strong and electromagnetic 
decay amplitudes of heavy quarkonia states: they 
will be discussed and compared with the data whenever they are available.

In our opinion the chiral lagrangian approach to the interactions of the 
heavy mesons is a predictive method to relate a large amount of processes and 
decay rates of these states. We hope to convince
the reader by this work that the chiral lagrangian method for heavy
hadrons is a promising way to describe this most fascinating physics.

We conclude this introduction with a brief summary 
of the subsequent sections. In section \ref{chap-rob} we review the 
symmetries of the approach, we construct the 
effective chiral lagrangian for heavy mesons and we discuss 
the inclusion in the effective lagrangian
of the light vector mesons and the positive parity heavy 
meson resonances. In section \ref{sec:strongi} we discuss some problems 
related to the strong interactions effective lagrangian: the 
strong coupling constant $g_{B^* B \pi}$ and its possible 
determinations; the one loop calculation of the $B^* -B $ 
hyperfine splitting  and the strong
decays of positive parity states. In section \ref{sec:BD}, after a 
brief review of the $B \to D, D^* $ semileptonic transition, we 
discuss the effective weak $V-A$ current and 
the chiral corrections to the ratio ${f_{D_s}}/{f_D}$. Semileptonic heavy 
mesons decays into a final state containing one light meson are discussed in 
section \ref{chap-semilep}, where we also consider 
the constraints put on the $q^2$-behaviour of the form factors
by different theoretical approaches and by some weak non leptonic decay 
rates, most notably $B \to J/\psi K^*$.
In section \ref{chap-bep} we consider radiative heavy meson 
decays and we discuss the predictions arising from the 
chiral lagrangian approach.
Sections \ref{chap-al} and \ref{chap-al1} are devoted to 
heavy mesons containing two heavy quarks: we write down an effective
lagrangian describing their interactions and we use it to relate 
different decay processes of these states. In particular we also discuss 
processes characterized by the breaking of the symmetries of the 
effective theory: spin and chiral symmetry. Finally 
three appendices conclude the work: the first contains a list of Feynman rules 
used to compute the amplitudes; in the second, some integrals encountered in 
the loop calculations are listed; the last appendix contains the formalism for 
higher angular momentum quarkonium states.

\section{Heavy quark and chiral symmetry}
\label{chap-rob}

\subsection{Heavy Quark Effective Theory}
\label{sec:2.1}

The HQET describes processes where a heavy quark
interacts via soft gluons with  the light degrees of freedom. 
The heavy scale in this case is clearly $m_Q$, the heavy quark mass, and the 
other
physical scale for the processes of interest here is $\Lambda_{QCD}$.
The identification of the heavy degrees of freedom to be removed 
requires some care: we do
not want to integrate out completely the heavy quarks, being interested in
decays of heavy hadrons and therefore in matrix elements with heavy quarks on
the external legs. 
As we will see,  the so called small component of the heavy quark spinor 
field, 
describing fluctuations around the mass shell, has to be eliminated.

We indicate by $v_\mu$ the velocity of the hadron containing the heavy quark
$Q$. This is almost on shell and its momentum $p_Q$ can be written,
 introducing a
residual momentum $k$ of the order of $\Lambda_{QCD}$, as
\be
p_Q= m_Q v + k~~.
\label{1}
\ee
We now extract  the dominant part $m_Q v$ of the heavy quark momentum 
defining a new field $Q_v$
\be
Q_v(x) = \exp(i m_Q v x) Q(x)=h_v(x) + H_v (x)~~.
\label{2}
\ee
The field $h_v$ is the large component field,
 satisfying the constraint $\slash v h_v =  h_v$: if the quark $Q$
is exactly on shell, it is the only term present in  (\ref{2}).  
$H_v$, the small component field, is of the order $1/m_Q$ and
 satisfies $\slash v H_v = - H_v$: it is
integrated out when deriving the HQET effective lagrangian. 

 The non local effective lagrangian is derived by integrating out the heavy 
fields
in the QCD generating functional, as done in \cite{mannel}.
At tree level one has simply to solve the equation of motion for $H_v$ and  
substitute the result in the QCD lagrangian. The equation of motion is
\be
(2m_Q+ivD)H_v=\frac{1-\slash v} 2 i\slash D h_v
\ee
where,
\be
D_\mu=\partial_\mu+ig_s A_\mu^a T_a
\ee
with $T_a$ the generators of $SU(3)_c$ and $g_s=\sqrt{4\pi\alpha_s}$ the 
strong 
coupling constant.
We get
\be
{\cal L}_{eff} = \bar{h}_v (iv D) h_v + 
\bar{h}_v i \slash D \frac{(1- \slash v)}{2} \left( \frac{1}{2 m_Q + i v D} 
\right) i \slash D h_v~~.
\label{2bis}
\ee
where a sum over velocities is understood.
By using the following identity:
\be
\frac{1+\slash v} 2\gamma_\mu\frac{1-\slash v}2\gamma_\nu\frac{1+\slash v} 2=
\frac{1+\slash v}2 (g_{\mu\nu}-v_\mu v_\nu-i\sigma_{\mu\nu})\frac{1+\slash 
v}2~~,
\ee
we can write
\be
{\cal L}_{eff} = \bar{h}_v (iv D) h_v +   
\bar h_vi D_\mu(g^{\mu\nu}-v^\mu v^\nu-i\sigma^{\mu\nu})\frac{1}{2m_Q+ivD}
iD_\nu h_v~~.
\ee
The expansion of this lagrangian in $1/m_Q$ gives an infinite series of local
terms. 
The leading one is
\be
{\cal L} = \bar{h}_v (iv D) h_v
\label{3}
\ee

\noindent
which, being mass independent, 
clearly exhibits  the heavy-flavour symmetry.
Moreover, since there are no Dirac matrices in (\ref{3}), the heavy quark spin 
is not affected by the interaction of the quarks with gluons 
and therefore the lagrangian has a $SU(2)$-spin symmetry.

These symmetries are lost if we keep the next terms in the $1/m_Q$ expansion
\be
{\cal L}=\frac{1}{2 m_Q} \bar{h}_v (i D)^2 h_v + \frac{g_s}{4 m_Q}
\bar{h}_v \sigma_{\alpha\beta}G^{\alpha\beta} h_v + {\cal O}(1/m_Q^2)~~,
\label{4}
\ee
where we have used the equation of motion, $ivD h_v={\cal O}(1/m_Q)$, to get 
rid of the
term $ \bar h_v (vD)^2 h_v$.
The first term is the kinetic energy arising from the off-shell motion of the
heavy quark, the second one describes the chromomagnetic interaction of the
heavy quark spin with the gluon field. 

The last step in building up the effective lagrangian is the inclusion of QCD
radiative corrections. In (\ref{3}) and (\ref{4})  the Wilson
coefficients are taken at the matching scale $m_Q$, i.e. the scale at which 
the heavy
degrees of freedom are integrated out. The evolution down to a scale $\mu <
m_Q$ introduces logarithmic corrections. Details and references can be found 
in
\cite{hqrev,neubert}.  We shall summarize here only some results. 

The inclusion of quantum loop corrections due to hard gluon exchanges 
modifies the coefficients in the lagrangian (\ref{3}) and (\ref{4}),
giving:
\be
{\cal L}  =  
\bar{h}_v (iv D) h_v +
\frac{a_1}{2 m_Q} \bar{h}_v (i D)^2 h_v + \frac{g_s a_2}{4 m_Q}
\bar{h}_v \sigma_{\alpha\beta}G^{\alpha\beta} h_v~~. 
\label{4a}
\ee
The tree level matching gives $a_1(m_Q)=a_2(m_Q)=1$: in the leading logarithm
approximation one finds at the scale $\mu < m_Q$
\be
a_1 (\mu )=1~~, ~~~~~~~~~
a_2 (\mu)= \left[ \frac{\alpha_s (m_Q)}{\alpha_s (\mu)} 
\right]^{-9/(33- 2 N_f)}~~,
\ee
where $N_f$ is the number of active quark flavours in the range between $\mu$
and $m_Q$. Notice that $a_1 (\mu )=1$: this is a consequence of the so-called
reparametrization invariance \cite{ripinv}, which relates the term in $a_1$ to
the leading one. Such an invariance arises from the fact that the 
decomposition
(\ref{1}) of the heavy quark momentum is not unique. The transformation
\bea
v & \to & v + \frac{q}{m_Q} \nn \\
k & \to & k - q \; ,
\label{rep}
\eea
where $v \cdot q =0$ to satisfy the constraint $v^2=1$, is another possible
decomposition and it has to give rise to the same physical observables:  
only the heavy quark momentum is a well defined quantity.
The consequences of this invariance have been studied in ref. \cite{ripinv}.
The main results are as follows. First of all the velocity and the derivative
$iD$ should appear only in the combination
\be
v_\mu+i\frac{D_\mu}{M}
\label{velocity}
\ee
where $M$ is the mass of the field under consideration (in this case $m_Q$). 
Second one has to
modify the fields in the velocity representation, that is
\be
\phi_v(x)=\exp(iMvx)\phi(x)~~.
\ee
The scalar fields do not require any change, but a vector fields 
${P^*_v}^\mu$, 
at the order $1/M$ should appear in the combination
\be
{P'_v}^\mu={P_v^*}^\mu-v^\mu\frac {iD\cdot P_v^*} M~~.
\label{repinv}
\ee
This is because the field $P_v^*$ should satisfy the constraint $v_\mu 
{P_v^*}^\mu=0$ also after reparametrization. The field $P'_v$, as well as the 
scalar
field, have a very simple transformation law under the 
reparametrization (\ref{rep}). They pick
up a phase factor
\be
{P_v'}^\mu\to\exp{(iqx)}{P'_v}^\mu~~.
\ee
Invariant terms under reparametrization are then easily constructed. In
particular  one finds  $a_1 (\mu) =1$ from these constraints.

We want now to implement the symmetries discussed before in the spectrum of
physical states, in particular the pseudoscalar  $D$ and $B$ meson states and 
the
corresponding vector resonances $D^*$ and $B^*$. 
The wave function of a heavy meson has to be independent of flavour and spin
of the heavy quark: therefore it can be characterized by the total angular
momentum $s_\ell$ of the light degrees of freedom. To each value of $s_\ell$ 
corresponds
a degenerate doublet of states with spin $J=s_\ell \pm 1/2$.
The mesons $P$ and $P^*$ form the spin-symmetry
doublet corresponding to $s_\ell= 1/2$.

The negative parity 
spin doublet $(P,P^*)$ can be represented by a $4 \times 4$ Dirac-type
 matrix $H$,
with one spinor index for the heavy quark and the other for the light degrees
of freedom. Such wave functions transform under a Lorentz transformation
$\Lambda$ as
\be
H \to D(\Lambda) H D (\Lambda )^{-1}
\label{4bis}
\ee
where $D( \Lambda)$ is the usual $4 \times 4$ representation of the Lorentz
group. Under a heavy quark spin transformation $S$ belonging to $SU(2)$ one 
has:
\be
H \to S H~~,
\label{4tris}
\ee
where $S$ satisfies $[\slash v, S]=0$  to preserve the constraint $\slash v H
=H$. 

A matrix representation of current use is:
\bea
H &=& \frac{(1+\slash v)}{2}[P_{\mu}^*\gamma^\mu - P \gamma_5 ]\\
{\bar H} &=& \gamma_0 H^{\dagger} \gamma_0~~.
\label{5}
\eea
Here $v$ is the heavy meson velocity,
$v^\mu P^*_{a\mu}=0$ and $M_H=M_P=M_{P^*}$ (we shall use also the notation 
$M_H=M$).
Moreover $\slash v H=-H \slash v =H$, ${\bar H} \slash v=
-\slash v {\bar H}={\bar H}$.

$P^{*\mu}$ and $P$ are annihilation operators normalized as follows:
\bea
\langle 0|P| Q{\bar q} (0^-)\rangle & =&\sqrt{M_H}\\
\langle 0|{P^*}^\mu| Q{\bar q} (1^-)\rangle & = & \epsilon^{\mu}\sqrt{M_H}~~.
\label{6}
\eea
The general formalism for higher spin states is given in \cite{falk}.
Here we will consider only the extension to $P$-waves of the system $Q\bar q$.
The heavy quark effective theory predicts two distinct multiplets, one
containing a $0^+$ and a $1^+$ degenerate state, and the other one a $1^+$ and
a $2^+$ state. In matrix notations, analogous to the ones used for the 
negative
parity states, they are described by
\be
S=\frac 1 2 (1+\slash v)[D_1^\mu\gamma_\mu\gamma_5-D_0]
\ee
and
\be
T^\mu=\frac 1 2 (1+\slash v)\left[D_2^{\mu\nu}\gamma_\nu-\sqrt{\frac 3 
2}\tilde
D_{1\nu}\gamma_5\left(g^{\mu\nu}-\frac 1 3 \gamma^\nu(\gamma^\mu-v^\mu)\right)
\right]
\label{r:tmu}
\ee
with the following conditions:
\bea
\slash v S= S\slash v=S\nn\\
\slash v T^\mu=-T^\mu\slash v=T^\mu\nn\\
\slash v\bar S=\bar S\slash v=\bar S\nn\\
-\slash v{\bar T}^\mu={\bar T}^\mu\slash v={\bar T}^\mu~~.
\eea
These two multiplets have $s_\ell=1/2$ and $s_\ell=3/2$ respectively, where
$s_\ell$,  the angular momentum of the light degrees of freedom, is
conserved together with the spin $s_Q$ in the infinite quark mass limit 
because $\vec J={\vec s}_\ell+{\vec s}_Q$.

\subsection{Chiral symmetry}
\label{subsec:cs}

 From the point of view of HQET it is natural to divide quarks into two classes
by comparing their lagrangian mass with $\Lambda_{QCD}$. The $u$ and $d$ 
quarks
belong definitely to the light quark class, $m_u,~m_d\ll\Lambda_{QCD}$. The
situation for the strange quark is not so clear, but it is usually considered
to belong to the light quark class, though non negligible mass corrections are
expected. If we take the limit $m_u,~m_d,~m_s\to 0$, the QCD lagrangian for
these three quarks possesses a $SU(3)_L\otimes SU(3)_R\otimes U(1)_V$
symmetry which is spontaneously broken down to $SU(3)_V\otimes U(1)_V$.
The lightest pseudoscalar particles of the octect $\pi$, $K$, $\bar K$, $\eta$ 
are 
then
identified with the Goldstone bosons corresponding to the broken generators. 
Of
course, due to the explicit symmetry breaking given by the quark mass term, 
the
mesons acquire a mass.

As it is well known, the interactions among Goldstone bosons can be described
by the chiral perturbation theory \cite{GL}, that is a low momentum expansion
in momenta and meson masses. Chiral perturbation theory describes the
Goldstone bosons in terms of a $3\times 3$ matrix $\Sigma(x)\in SU(3)$
transforming under $SU(3)_L\otimes SU(3)_R$ as
 \be
\Sigma\to g_L\Sigma {g_R}^\dagger~~.
\label{1.0}
\ee 
The meson octect is introduced via the exponential representation
\be
\Sigma = \exp\left( \frac{2 i {\cal M}}{f} \right) \; \; \; \; \; \; \;
 f=132 \; MeV
\label{1.1}
\ee
where ${\cal M}$ is a $3\times 3$ hermitian, traceless matrix:
\be
{\cal M} = \left( \begin{array}{ccc} 
\sqrt{\frac{1}{2}}\pi^0 + \sqrt{\frac{1}{6}}\eta & \pi^+ & K^+ \nn\\
\pi^- & -\sqrt{\frac{1}{2}}\pi^0+\sqrt{\frac{1}{6}}\eta & K^0 \\ 
K^- & {\bar K}^0 &-\sqrt{\frac{2}{3}}\eta
\end{array} \right) ~~.
\label{1.2}
\ee
To the lowest order in the momenta and in the massless quark limit, the most
general invariant lagrangian is given by
\be
\LL = \frac{f^2}{8}Tr\left[\partial^\mu\Sigma\partial_\mu
\Sigma^\dagger\right]
\label{1.3}
\ee
where the constant $f^2/8$ has been chosen  such as  to get a canonical
kinetic term for the mesonic fields appearing inside the matrix ${\cal M}$.

Higher order terms in the momentum expansion are suppressed by powers of
$p/\Lambda_\chi$, where $p$ is the typical momentum scale of the process and
$\Lambda_\chi$ is the chiral symmetry breaking scale, which is evaluated to be
of the order of 1 GeV. As we have already noticed, chiral symmetry is not an
exact symmetry of QCD, being  explicitly  broken by the quark mass term
\be
\sum_{a=u,d,s} {\bar q}_a {\hat m}_{ab} q_b
\label{mass0}
\ee
where $\hat m$ is the light mass matrix:
\be
{\hat m}=
\left (\begin{array}{ccc} 
m_u & 0 & 0 \nn \\
0 & m_d & 0 \nn \\
0 & 0 & m_s 
\end{array}\right )~~.
\label{mass1}
\ee
The expression (\ref{mass0}) transforms as the representation $(\bar 3_L,3_R)$
$\oplus$ $(3_L,\bar 3_R)$. We can take into account this breaking, at the 
first
order in the quark masses, by adding to the chiral lagrangian a term
transforming exactly in the same way. This contribution can be written in the
form
\be
\lambda_0 Tr \left( {\hat m} \Sigma + \Sigma^{\dagger} {\hat m} \right)~~.
\label{mass2}
\ee
The Goldstone bosons receive a contribution to their square mass from this 
term.
This is the reason for treating formally the quark masses 
as second order terms in the
momentum expansion. Then, the tree diagrams generated by (\ref{1.3}) and 
(\ref{mass2}) reproduce the same results of the soft pion theorems.
Corrections to the leading terms come from higher derivative or mass terms and
from loop diagrams. It is also important to stress that chiral perturbation
theory is renormalizable at any fixed order in the momentum expansion.

The interactions of the Goldstone fields with  matter fields such as baryons,
heavy mesons or light vector mesons ($\rho$, $\omega$), can be described by 
using
the theory of non linear representations as discussed in the classical paper 
by
Callan, Coleman, Wess and Zumino (CCWZ) \cite{CCWZ}. The key ingredient in 
this theory
is the coset field $\xi(x)$, which is defined on the coset space
$SU(3)_L\otimes SU(3)_R/SU(3)_V$. In this context, the $\xi(x)$ field is
simply related to $\Sigma(x)$ by the relation
\be
\Sigma(x)=\xi^2(x)~~.
\label{1.4}
\ee
The transformation properties of $\xi(x)$ under chiral transformations (that 
is
transformations of $SU(3)_L\otimes SU(3)_R$) are
\be
\xi(x)\to g_L\xi(x) U^\dagger(x)=U(x)\xi(x) g_R^\dagger~~.
\label{1.5}
\ee
The matrix $U(x)$ belongs to the $SU(3)_V$ unbroken subgroup and it is defined
by the previous equation. As a consequence, $U(x)$ is generally a complicated
non-linear function of the coset field $\xi(x)$ itself, and, as such, 
space-time
dependent. The matter fields have definite transformation properties under the
unbroken $SU(3)_V$ group. For instance, a heavy meson made up by a heavy quark
$Q$ and a light antiquark $\bar q_a$ ($a=u,d,s$), transforms, under a chiral
transformation, according to the representation $\bar 3$ of $SU(3)_V$, that is
$(H_a\approx Q\bar q_a)$
\be
H_a\to H_b U_{ba}^\dagger(x)
\label{1.10}
\ee
where $U$ is the same matrix appearing in eq. (\ref{1.5}). In view of the
locality properties of the transformation $U(x)$, one needs covariant
derivatives or gauge fields, in order to be able to construct invariant
derivative couplings. This is provided by the vector current
\be
\VV_\mu=\frac{1}{2}\left(\xi^\dagger\partial_\mu \xi+ \xi
\partial_\mu \xi^\dagger\right)
\label{1.6}
\ee
transforming under the chiral transformation of eq. (\ref{1.5}) as
\be
\VV_\mu\to U\VV_\mu U^\dagger+U\partial_\mu U^\dagger~~.
\label{1.8}
\ee
It is also possible to introduce an axial current, transforming
as the adjoint representation of $SU(3)_V$
\be
\AA_\mu=\frac{1}{2}\left(\xi^\dagger\partial_\mu \xi - \xi
\partial_\mu \xi^\dagger\right)
\label{1.7}
\ee
with
\be
\AA_\mu\to U\AA_\mu U^\dagger~~.
\label{1.9}
\ee

\subsection{ A chiral lagrangian for heavy mesons}
\label{sec:chiral}

The effective lagrangian for the strong interactions of heavy mesons
with light pseudoscalars must satisfy Lorentz and C, P, T invariance.
Furthermore, at the leading order in the $1/M$ expansion 
($M$ is the heavy meson mass), and in the massless
quark limit, we shall require flavour and spin symmetry in the heavy meson
sector, and chiral $SU(3)_L\otimes SU(3)_R$ invariance in the light one. The
most general lagrangian is then \cite{wise,yan,burdman92}:
\bea
{\cal L} &=& i < H_b v^\mu D_{\mu ba} {\bar H}_a > +
i g <H_b \gamma_\mu \gamma_5 \AA^\mu_{ba} {\bar H}_a> +\nn \\
&+ &  \frac{f^2}{8}\partial^\mu\Sigma_{ab}\partial_\mu
\Sigma_{ba}^\dagger
\label{wise}
\eea   
where $D_\mu = \partial_\mu + \VV_{\mu}$ and
$< \ldots >$ means trace over the $4 \times 4$ matrices. 
In (\ref{wise}) a sum over heavy meson velocities is understood.
The first term in the lagrangian contains the kinetic term for the heavy 
mesons
giving the $P$ and $P^*$ propagators,
\be
\frac i{2v\cdot k}
\ee
and
\be
- \frac {i(g^{\mu\nu}-v^\mu v^\nu)}{2v\cdot k}
\ee
respectively. The interactions among
heavy and light mesons are obtained by expanding the field
$\xi(x)=\exp(i{\cal M}(x)/f)$ and taking the traces. In the first term there 
are 
interactions among the heavy mesons and an even number of 
 pions coming from the expansion of
the vector current $\VV_\mu$. The interactions with an odd number of pions 
originate
from the second term. As an example, the first term in the expansion of the 
axial current gives 
\be
\AA_\mu\approx \frac i f\partial_\mu {\cal M}+\dots
\ee
The last term in (\ref{wise}) is the non-linear lagrangian discussed in the
previous section, describing the light meson self-interactions. Corrections to
this lagrangian originate from higher 
terms in the $1/M$ expansion and from chiral symmetry
breaking. Let us start with the last issue. We proceed as in the previous
section by considering, at the first order, breaking terms transforming as 
$(\bar 3_L,3_R)\oplus(3_L,\bar 3_R)$ under the chiral group. The most general
expression is
\bea
\LL_{\chi B} &=&\lambda_0 ( {\hat m}_{ab}\Sigma_{ba} + 
{\hat m}_{ab}\Sigma_{ba}^\dagger)
+\lambda_1<{\bar H}_a H_b(\xi {\hat m}\xi+\xi^\dagger {\hat 
m}\xi^\dagger)_{ba}>
\nn\\
&+&\lambda_1^\prime <{\bar H}_a H_a({\hat m}\Sigma+{\hat m}
\Sigma^\dagger)_{bb}>\nn \\
& + &  i \lambda_3 
<H_b \gamma_\mu \gamma_5 \AA^\mu_{bc} (\xi {\hat m}\xi+\xi^\dagger
{\hat m}\xi^\dagger)_{ca} {\bar H}_a> +\nn \\
& + &  i \lambda^\prime_3 
<H_a {\bar H}_a\gamma_\mu \gamma_5 \AA^\mu_{cd} (\xi {\hat m}\xi+\xi^\dagger
{\hat m}\xi^\dagger)_{dc}>~~. 
\label{chircorr}
\eea
Here we have neglected terms contributing to processes with more than one 
pion.
Notice that the coefficients $\lambda_3$ and $\lambda_3^\prime$ should be of
order $1/\Lambda_\chi$ because they multiply operators of dimension five.
In principle  there are other dimension five operators (see \cite{Boyd-Grin}),
which however contribute only to the order $1/\Lambda_\chi\times 1/M$
(neglecting again
interaction terms with more than one pion field). The $\lambda_1$ and
$\lambda_1^\prime$ terms give rise to a shift in the heavy meson propagators.
For instance, in the case of the strange heavy mesons, they produce the 
shift
$v\cdot k\to v\cdot k-\delta$ with $\delta= M_{D_s}-M_D=M_{B_s}-M_B$.

Let us now discuss the $1/M$ corrections (see refs. \cite{Boyd-Grin,cheng}). 
First, one has to take into
account the constraints coming from the reparametrization invariance that tie
together different orders in the expansion. In the present formalism one can
define $H$ fields transforming  by the simple phase factor 
$\exp(iqx)$ under the transformation (\ref{rep})
\be
H^\prime=H+\frac i{2M}D_\mu[\gamma^\mu,H]~~.
\ee
In fact, it is easily seen that, neglecting terms proportional to the form of 
the
free wave
equation (contributing to the next order in the expansion) is equivalent
to use the equation (\ref{velocity}) for the four-velocity and the equation
(\ref{repinv}) for the vector field in the definition of $H$. This 
substitution
modifies the zeroth order lagrangian in the way described in \cite{Boyd-Grin}.
We shall not report here the expressions because all the extra terms involve 
at
least two derivatives and they contribute only at the order
$1/M^2$. Finally we have  ${\cal O}(1/M)$ terms which are invariant under
four-velocity reparametrization, and therefore they appear with arbitrary
coefficients. In this discussion an important role is played by the
time-reversal invariance.  In our case we have to require invariance under the
following transformations
\bea
H_v(x)\to TH_{v_P}(-x_P)T^{-1}\nn\\
{\cal M }(x)\to - {\cal M }(-x_P)\nn\\
\AA^\mu(x)\to\AA_\mu(-x_P)
\eea
where $x_P$ and $v_P$ are the parity reflections of $x$ and $v$, that is,
$x_P^\mu=x_\mu$ and $v_P^\mu=v_\mu$. Also
\be
T\gamma^\mu T^{-1}=\gamma_\mu^*~~.
\ee
Taking into account this constraint and neglecting higher derivative terms
(which contribute to the order $1/M^2$), one finds \cite{Boyd-Grin}
\bea
{\cal L}_{1/M} &=& 
\frac{\lambda_2}{M}<{\bar H}_a\sigma_{\mu\nu} H_a\sigma^{\mu\nu}>\nn\\
& + & i \frac{ g_1}{M} 
<{H}_b \gamma_\mu \gamma_5 \AA^\mu_{ba}  {\bar H}_a>\nn\\ 
& + & i \frac{ g_2}{M} 
< \gamma_\mu \gamma_5 \AA^\mu_{ba}  H_b{\bar H}_a>~~.
\label{1/m}
\eea
By writing $\lambda_2=-M\Delta/2=-M(M_{P^*}-M_P)/2$, one sees that the effect
of the corresponding operator is to shift the $P$ and $P^*$ propagators to
\be
\frac i{2(v\cdot k+\frac 3 4\Delta)}
\ee
and
\be 
-\frac {i(g^{\mu\nu}-v^\mu v^\nu)}{2(v\cdot k-\frac 1 4\Delta)}
\ee
respectively. The couplings $g_1$ and $g_2$ renormalize the coupling $g$
appearing in equation (\ref{wise}) in different way for the $P^*P^*$and 
$P^*P$
couplings. More precisely one finds
\bea
g\to g_{P^*P^*}=g+\frac 1 M(g_1+g_2)\nn\\
g\to g_{P^*P}=g+\frac 1 M(g_1-g_2)~~.
\label{g:1/m}
\eea

\subsection{Light vector resonances}
\label{subsec:lvr}

We want now to introduce in the previous effective lagrangian,
equation (\ref{wise}), the light vector resonances, $\rho$, $K^*$, etc. We  
shall
make the hypothesis that they can be treated as light degrees of freedom. 
Therefore they could be introduced as matter fields by using the CCWZ 
formalism \cite{CCWZ}.
However we prefer here to make use of the hidden gauge symmetry approach
\cite{bando}, as done in \cite{noilr} (see also \cite{schechter,ko93}).
 The two methods are completely
equivalent, but the second one is  easier to deal with. The main idea lies in
the observation that any non-linear $\sigma$-model based on the quotient space
$G/H$, where $G$ is the symmetry group and $H$ the unbroken subgroup, is
equivalent to a linear model with enlarged symmetry $G\times H_{local}$, where
$H_{local}$ is a local symmetry group isomorphic to the unbroken group $H$.
In the linear model the fields have values in the group $G$, rather than in
$G/H$ as in the non-linear formulation. The extra degrees  of freedom can be
gauged away by taking advantage of the local invariance related to 
$H_{local}$. In
the unitary gauge one recovers the CCWZ formulation. However, the explicit
appearance, in the formalism, of a local invariance group gives room for the
introduction of gauge fields with values in the Lie algebra of $H$, which will
be interpreted as the light vector mesons. Again, one can show that in the
unitary gauge these fields correspond to  vector matter fields of the CCWZ
formulation \cite{bando}. Originally \cite{bando1}  it was proposed that the
$\rho$ meson was the dynamical gauge boson of the hidden local symmetry
$H_{local}=SU(2)_V$ in the $SU(2)_L\otimes SU(2)_R/SU(2)_V$ nonlinear
chiral lagrangian. The extension to $SU(3)$ is straightforward \cite{bando2}
and incorporates the $\rho$, $K^*$, $\bar K^*$ and $\phi-\omega$ mesons.
 
 Let us briefly describe the procedure.
 It
consists in using two new $SU(3)$ matrix-valued fields $L$ and $R$ to
build up $\Sigma$
\be
\Sigma=LR^\dagger~~.
\label{h1}
\ee
The chiral lagrangian in (\ref{1.3}) is then invariant under the
group $SU(3)_L\otimes SU(3)_R\otimes SU(3)_H$
\be
L\to g_L L h^\dagger(x),~~~~~~~R\to g_R R h^\dagger(x)
\label{h2}
\ee
where $h\in SU(3)_H$ is a local gauge transformation. The local symmetry
associated to the group $SU(3)_H$ is called 
$hidden$  because the
field $\Sigma$ belongs to the singlet representation. It should be
noticed that this description is equivalent to the previous one by
the gauge fixing $L=R^\dagger=\xi$, which can be reached through a gauge
transformation of $SU(3)_H$.
With the fields $L$ and $R$ we can construct two
currents
\be
\VV_\mu=\frac{1}{2}\left(L^\dagger\partial_\mu L+R^\dagger
\partial_\mu R\right)
\label{h3}
\ee
\be
\AA_\mu=\frac{1}{2}\left(L^\dagger\partial_\mu L-R^\dagger
\partial_\mu R\right)
\label{h4}
\ee
which are singlets under $SU(3)_L\otimes SU(3)_R$ and transform as
\be
\VV_\mu\to h\VV_\mu h^\dagger+h\partial_\mu h^\dagger
\label{h5}
\ee
\be
\AA_\mu\to h\AA_\mu h^\dagger
\label{h6}
\ee
under the local group $SU(3)_H$.  In the unitary gauge $L=R^\dagger=\xi$ 
they reduce to the $\VV$ and $\AA$ previously introduced in (\ref{1.6}) and
(\ref{1.7}).

In this notation, the transformation (\ref{1.10}) for $H_a$ reads
\be
H_a\to H_b h_{ba}^\dagger(x)
\label{h7}
\ee
and the covariant derivative is defined as
\be
D_\mu{\bar H}=(\partial_\mu+\VV_\mu){\bar H}~~.
\label{h8}
\ee

The octet of vector resonances ($\rho$, etc.) is introduced as the gauge 
multiplet associated to the group $SU(3)_H$. We put
\be
\rho_\mu=i\frac{g_V}{\sqrt{2}}\hat\rho_\mu
\label{h9}
\ee
where $\hat\rho$ is a hermitian $3\times 3$ matrices analogous to the one 
defined in equation (\ref{1.2}). This field transforms under the full symmetry
group as $\VV_\mu$
\be
\rho_\mu\to h\rho_\mu h^\dagger+h\partial_\mu h^\dagger~~.
\label{h10}
\ee
The vector particles
acquire a common mass through the breaking of
$SU(3)_L\otimes SU(3)_R\otimes SU(3)_H$ to $SU(3)_V$. In fact, 8 out of the
16 Goldstone bosons coming from the breaking are the light pseudoscalar
mesons, whereas the other 8 are absorbed by the $\rho$ field. 

We can now build a lagrangian describing the interactions of heavy mesons with 
low momentum  vector resonances, respecting chiral and heavy quark symmetries.
The new terms we have to add to (\ref{wise}) are : 
\bea
\LL_{\rho} &=& 
\frac{1}{2g_V^2}<F_{\mu\nu}(\rho)F^{\mu\nu}(\rho)> 
- \frac{f^2}{2} \left[ <(\AA_\mu)^2>+a <(\VV_\mu-
\rho_\mu)^2> \right] \nn \\
&+ &  
i\beta <H_bv^\mu\left(\VV_\mu-\rho_\mu\right)_{ba}{\bar H}_a>
+ i \lambda <H_b \sigma^{\mu\nu} F_{\mu\nu}(\rho)_{ba} {\bar H}_a>~~,
\label{1.11}
\eea   
where $F_{\mu\nu}(\rho)=\partial_\mu\rho_\nu-\partial_\nu\rho_\mu+
[\rho_\mu,\rho_\nu]$. 

In the first line in (\ref{1.11}) there is  the kinetic term for the
light vector resonances.
The second term gives back 
the non linear $\sigma$-model lagrangian (\ref{1.3}),
 as it can be seen by using the identity 
\be
\langle \AA_\mu\AA^\mu\rangle=-\frac 1 4\langle\partial_\mu \Sigma\partial^\mu
\Sigma^\dagger\rangle
\ee                     
 plus 
interactions among pions and $\rho$-like particles. The value of the
parameters $a$ and $g_V$ can be fixed 
by considering the electromagnetic couplings \cite{bando}. 
In this way one can see that the first KSRF relation \cite{KSRF}
is automatically satisfied
\be 
g_\rho=g_{\rho\pi\pi} f^2~~,
\ee
with $g_\rho$ is the $\rho-\gamma$ mixing parameter, and 
$g_{\rho\pi\pi}=ag_V/2$. 
Furthermore, from the second KSRF relation,
\be
m_\rho^2=g_{\rho\pi\pi}^2 f^2
\ee
and extracting the $\rho$ mass from (\ref{1.11}),
\be
m_\rho^2=\frac 1 2 a g_V^2 f^2~~,
\ee
we see that $a=2$, and that
\be
g_V=\frac{m_\rho} f\approx 5.8~~.
\label{rob:gv}
\ee

The terms proportional to $\beta$ and $\lambda$ give the couplings
of the light vector mesons with the heavy states, like $P P \rho$, $P P^* 
\rho$
etc. 

As usual in (\ref{1.11}) we have considered the lowest derivative
terms.
Explicit symmetry breaking terms can be introduced as in (\ref{chircorr})
and (\ref{1/m}).

As shown in \cite{gvect}, the hidden symmetry approach has an interesting 
limit
in which an additional symmetry appear, the so-called vector symmetry. 
In this limit the vector meson octet is massless and the chiral symmetry is
realized in an unbroken way: the longitudinal components of the vector mesons
are the chiral partners of the pions. A chiral lagrangian for heavy mesons
incorporating both heavy quark and vector symmetries has been written down in
\cite{manvec}: having an additional symmetry, there is a reduction of the
number of effective coupling constants, 
and in the exact symmetry limit only one
unknown coupling constant appears. 
However large symmetry breaking effects are expected and corrections to the
vector limit can be sizeable.

\subsection{The chiral lagrangian for the positive parity states}
\label{sec:2.5}

In the sequel we shall use also the chiral lagrangian for the positive 
parity states introduced in section \ref{sec:2.1}.
This lagrangian, containing the fields $S_a$ and $T^\mu_a$ as well as their
interactions with the Goldstone bosons and the fields $H_a$, has been 
derived in refs. \cite{falk,kilian}:
\be
\LL_3=\LL_{kin}+\LL_{1\pi}+\LL_s+\LL_d
\label{ll3}
\ee
\bea
\LL_{kin} & = & i<S_b (v \cdot D)_{ba} {\bar S}_a>+ i<T^\mu_b 
(v \cdot D)_{ba} {\bar T}_{\mu a}> \nn\\
&-&\delta m_s <S_a {\bar S}_a>-\delta m_T <T^\mu_a {\bar T}_{\mu a}>
\label{lkin}
\eea
\be
\LL_{1\pi}=ik<T^\mu_b \gamma_\lambda\gamma_5 \AA_{ba}^\lambda 
{\bar T}_{\mu a}>+
i\tilde k <S_b \gamma_\mu\gamma_5 \AA_{ba}^\mu {\bar S}_a>
\ee
\be
\LL_s=i h<S_b \gamma_\mu 
\gamma_5 \AA_{ba}^\mu {\bar H}_a>+
i \tilde h <T_b^\mu \AA_{\mu ba}\gamma_5 {\bar S}_a>+ h.c.
\label{eq:lls}
\ee
\bea
\LL_d &= & i \frac{h_1}{\Lambda_\chi}<T^\mu_b \gamma_\lambda \gamma_5 (
D_\mu \AA^{\lambda})_{ba} {\bar H}_a>\nn\\
&+& i \frac{h_2}{\Lambda_\chi}<T^\mu_b \gamma_\lambda \gamma_5 (
D^\lambda \AA_{\mu})_{ba} {\bar H}_a>~~.
\label{eq:lld}
\eea
In (\ref{lkin}) $\delta m_s=M_{D_0}-M_P=M_{D_1}-M_{P}$, 
$\delta m_T=M_{D_2}-M_P=M_{{\tilde D}_1}-M_{P}$ . 
A mixing term between the $S$ and $T_\mu$ field is absent at the leading 
order.
Indeed,  saturating the $\mu$ index of $T_\mu$ with $v_\mu$ or $\gamma_\mu$
gives a vanishing result, and derivative terms are forbidden by the 
reparametrization invariance \cite{falk,kilian}.
We can also 
introduce the couplings of the vector meson light resonances to the 
positive and negative parity states as follows
\be
\LL_4=\LL_{S\rho}+\LL_{T\rho}+\LL'
\ee
\be
\LL_{S\rho} = i \beta_1<S_b v^\mu (\VV_\mu-\rho_\mu)_{ba} {\bar S}_a> 
+i \bar\lambda_1 <S_b \sigma^{\mu\nu}F_{\mu\nu}(\rho)_{ba}{\bar S}_a>
\ee
\be
\LL_{T\rho} = i \beta_2 <T^{\lambda}_b v^\mu (\VV_\mu -\rho_\mu )_{ba}
{\bar T}_{a \lambda}> +i\bar\lambda_2 <T^{\lambda}_b \sigma^{\mu\nu} 
F_{\mu\nu}(\rho)_{ba}{\bar T}_{a \lambda}>
\ee
\bea
\LL' & = & i \zeta < {\bar S}_a H_b \gamma_\mu (\VV^\mu-\rho^\mu)_{ba}>
+i \mu <{\bar S}_a H_b\sigma^{\lambda\nu} F_{\lambda\nu}(\rho )_{ba}
> \nn \\
&+&i \zeta_1 < {\bar H}_a T^{\mu}_b \gamma_\mu 
(\VV^\mu -\rho^\mu)_{ba}> +i \mu_1 <{\bar H}_a T^{\mu}_b \gamma^\nu
 F_{\mu\nu}(\rho )_{ba}>~~.
\label{rob:zm}
\eea
We shall see in the sequel that some information on the coupling
constants $g$, $\mu$, $\lambda$, and $\zeta$ can be obtained by the analysis
of the semileptonic decays
\be
H  \to   P l {\bar \nu}_l,~~~~~~
H  \to   P^* l {\bar\nu}_l
\ee 
and from the radiative decay 
\be
P^*\to P\gamma~~.
\ee
As discussed in the next sections $g$ and $\lambda$ have been also evaluated
by potential models and QCD sum rules.

\section{Strong interactions}
\label{sec:strongi}

In the limit of exact chiral, heavy flavour, and spin symmetries, the 
low-energy interaction
among two heavy mesons and light pseudoscalars is governed by the
lagrangian (\ref{wise}).
The coupling constant $g$, 
describing the coupling of the heavy mesons to 
the pseudoscalar Goldstone bosons, 
is one of the fundamental parameters of the effective
lagrangian. As we shall see, via chiral loops, it enters into
a variety of corrections to both the chiral and the spin symmetry
limit of many quantities of interest.

For the time being we limit the discussion to the strong interaction
among the lowest lying, negative parity states, $P^a$ ($0^-$)
and ${P^a}^*$ ($1^-$), contained in the multiplet $H^a$. Later on we shall
discuss strong interactions involving excited states.

The terms containing one light pseudoscalar are readily obtained 
from the lagrangian (\ref{wise}). They read:
\bea
{\cal L} &=& - \dd\frac{g}{f} {\rm tr}({\overline H}_a H_b \gamma_\mu
\gamma_5) \partial^\mu {\cal M}_{ba}\nn\\
&=&\left[-\dd\frac{2g}{f} {P^*}_\mu \partial^\mu {\cal M} 
{P^\dagger} +{\rm  h.c.}\right]\nn\\
&+&\frac{2 g i}{f} \epsilon_{\alpha\beta\mu\nu} 
{{P^*}}^\beta \partial^\mu {\cal M} {{{P^*}^\dagger}}^\alpha v^\nu \; .
\label{ppm}
\eea
The interaction term $P P \pi$ is forbidden by parity;
the direct $P^*\to P \pi$ transition is not allowed in the $B$ system
because of lack of phase space. On the other hand, this transition
occurs for $D$ mesons. From eq. (\ref{ppm}) one obtains the 
partial widths:
\bea
\Gamma({D^*}^+\to D^0 \pi^+)&=&\dd\frac{g^2}{6\pi f^2} |{\vec p}_\pi|^3\nn\\
\Gamma({D^*}^+\to D^+ \pi^0)=\Gamma({D^*}^0\to D^0 \pi^0)
&=&\dd\frac{g^2}{12\pi f^2} |{\vec p}_\pi|^3
\label{wid0}
\eea
The decay ${D^*}^0\to D^+ \pi^-$ is also forbidden by the phase space.
The ${D^*}^+$ decay is dominated by the $D\pi$ channels (see table 
\ref{table:fer}, \cite{cleo,pdb}).
There is an experimental upper bound on the total ${D^*}^+$ width:
$\Gamma_{tot}(D^{*+}) < 131$ KeV \cite{acc}. By combining this bound
with the measured branching ratios of ${D^*}^+$ reported in table 
\ref{table:fer}, one obtains the following upper limit for $g$: 
\be
g^2 < 0.5~~~.
\ee
\begin{table}[htbp]
\caption{Experimental $D^*$ branching ratios (\%)}
\label{table:fer}
\centering
\begin{minipage}{5truecm}
\begin{tabular}{l c}
{\rm Decay mode}& {\rm Branching ratio}\\
\hline
${D^*}^0\to D^0 \pi^0$& $63.6\pm 2.8$\\
${D^*}^0\to D^0 \gamma$& $36.4\pm 2.8$\\
${D^*}^+\to D^0 \pi^+$& $68.1\pm 1.3$\\
${D^*}^+\to D^+ \pi^0$& $30.8\pm 0.8$\\
${D^*}^+\to D^+ \gamma$& $1.4\pm 0.8$\\
\end{tabular}
\end{minipage}
\end{table}

Also the radiative partial widths $\Gamma({D^*}^0\to D^0 \gamma)$ and
$\Gamma({D^*}^+\to D^+ \gamma)$ depend, via chiral loops, on the $g$
coupling constant \cite{abj,cho,cina1}. This dependence will be discussed in 
section \ref{sec:bepp1}.

A list of measurable quantities which depend on the $g$ coupling constant
either directly, or via chiral loop corrections, includes:
the rate for $B\to D(D^*) \pi l \nu$, the form factors
for the weak transitions between  heavy and  light pseudoscalars,
the chiral corrections to
the ratios $f_{D_s}/f_{D^+}$, $B_{B_s}/D_{D^+}$, to the Isgur-Wise
function $\xi(v\cdot v')$, to the double ratio 
$(f_{B_s}/f_{B^0})/(f_{D_s}/f_{D^+})$, and to several mass splittings in the 
$P^*_a, P_a$ system. The discussion of these observables will be presented
below.

\subsection{Theoretical estimates of $g$}

In this section we review some theoretical estimate of the strong coupling 
constant 
$g$
defined in eq. \ref{wise}.

\subsubsection{Constituent quark models}
\label{sub:cqm}

In the constituent quark model, one finds $g\simeq 1$ \cite{yan}. As a matter
of fact
the axial-vector current ${j_5^A}_\mu$ associated to the lagrangian 
of eq. (\ref{wise}) reads:
\be
{j_5^A}_\mu = - g Tr[{\bar H}_a  H_b \gamma_\mu\gamma_5] (T^A)_{ba} +...
\label{axi}
\ee
where $T^A~~(A=1,...8)$ are $SU(3)$ generators and the dots stand for
terms containing light pseudoscalar fields.
The matrix element of the combination ${j_5^1}_\mu+i~ {j_5^2}_\mu$ between
the ${D^*}^-$ and the $D^0$ states can be easily evaluated.
By working in the ${D^*}^-$ rest frame and by selecting the 
longitudinal helicity, one obtains:
\be
\langle {D^*}^- |({j_5^1}+i {j_5^2})_{\mu=3}|D^0 \rangle = -g
\label{uno}
\ee
On the other hand, if one identifies, without further 
renormalization, the partially conserved axial currents 
of eq. (\ref{axi}) with the corresponding
currents of QCD:
\be
({j_5^A})^{QCD}_\mu={\overline q}a (T^A)_{ab} \gamma_\mu \gamma_5 q_b
\ee
one can evaluate the same matrix element within the non-relativistic
constituent quark model, obtaining: 
\be
\langle {D^*}^- |({j_5^1}+i {j_5^2})_{\mu=3}|D^0 \rangle = - 1~~.
\label{due}
\ee
The comparison between the eq. (\ref{uno}) and (\ref{due}) leads  to:
\be
g=1~~~.
\label{gnr}
\ee
A similar argument provides $g_A=5/3$ for the nucleon, to be 
compared with the experimental result $g_A\simeq 1.25$ (analogous result
was previously obtained in the constituent quark model \cite{suzuki}).
The authors in ref. \cite{isgur} find a 
slightly different 
value: $g \simeq 0.8$, obtained in a calculation considering mock mesons (see 
references therein). A similar value $g \simeq 0.8$ is obtained in 
\cite{pham} using PCAC (see also \cite{nussi} and \cite{dompav88}).

In ref. \cite{defazio2} it has been suggested that
a departure from the naive constituent quark model might arise 
as a consequence of the 
relativistic motion of the light antiquark ${\bar q}$ inside the 
heavy meson. The model adopted in \cite{defazio2} is based 
on a constituent quark picture of the hadrons; the strong interaction between 
the quarks is described by a QCD inspired potential \cite{richardson}
and the relativistic effects due to the kinematics are included 
by considering as wave equation the 
Salpeter equation \cite{salpeter} (for more details see \cite{pietroni}).

In this model one finds \cite{defazio2}:

\be g =
{1 \over 4 M_D} \int_0^\infty \frac{dk}{2 \pi^2} k^2 |\psi|^2
 {E_q + m_q 
\over E_q} \left[ 1-{k^2 \over 3 (E_q+m_q)^2} \right] \hskip 3 pt . 
\label{eq : 28} \ee
where $E_q \; = \; \sqrt{ \vec{k}^2 + m_q^2}$ is the light quark energy,
and $\psi$ is the wave function.
By considering the non-relativistic limit ($E_q \simeq 
m_q \gg k$) one obtains $g=1$,
because of the normalization condition 

\be {1 \over (2\pi)^3} \int d \vec{k} |\psi|^2=2M_D 
\hskip 3 pt , \label{eq : 8} \ee 
This reproduces the constituent quark model result of eq. (\ref{gnr}).

Let us now take in (\ref{eq : 28}) the limit of very small light quark masses
(we note that there is 
no restriction to the values of $m_q$ in the Salpeter equation
and $m_q=0$ is an acceptable value). In this case, we obtain:

\be g={1 \over 3} \hskip 3 pt . \label{eq : 30} \ee

\noindent It is worth to stress that the strong reduction of the value of $g$ 
from the naive non relativistic quark constituent model value $g=1$ 
 (eq. \ref{gnr}) to the 
result (\ref{eq : 30}) has a simple explanation in the effect of the 
relativistic kinematics taken into account by the Salpeter equation. 
Similar results have been obtained in \cite{donnell}.

Including finite mass effects 
($m_u=m_d=38 \hskip 3 pt MeV$; $m_s=115 \hskip 3 pt  MeV$, $m_c=1452 
\hskip 3 pt MeV$, $m_b=4890 \hskip 3 pt MeV$ are used in this fit)
one obtains the  numerical results:

\be 
g = 0.40 \hskip 2 cm (D  \; \; case)  
\label{eq : 24} 
\ee
\be 
g = 0.39 \hskip 2 cm (B  \; \; case)~~.  
\label{eq : 25i} 
\ee

\subsubsection{QCD sum rules}
\label{subsec:QCD}

The coupling constant $g$ has also been determined within
the QCD sum rule approach \cite{ovc,qcdsr,bel,alievs}
\footnote{For a complete list of earlier references, see ref. \cite{bel}.}.
The starting point of this approach is the QCD correlation function:
\be
A_{\mu}(P,q) = i \int dx <\pi^-(q)| T(V_{\mu}(x) j_5(0) |0> e^{-iq_1x} = A
q_{\mu} + B P_{\mu}
\label{corr}
\ee
where, considering the case of the $B$ system, 
$V_{\mu}={\overline u} \gamma_{\mu} b$, $j_5=i{\overline b} \gamma_5 d$,
$P=q_1+q_2$, $q = q_1 - q_2$ and $A$, $B$ are scalar functions of $q_1^2$, $q_2^2$, $q^2$.

Both $A$ and $B$ satisfy dispersion relations and are computed, according to
the QCD sum rules method, in two ways: either by means of the operator product
expansion (OPE), or by writing a dispersion relation and
saturating the associated spectral function by physical hadronic states.

The OPE can be performed in the soft-pion limit
$q_\mu\to 0$, for large Euclidean momenta ($q_1^2=q_2^2\to -\infty$).
The various contributions  
come from the expansion of the heavy quark propagator and of the vector 
current $V_\mu$. 
This leads to a combination of matrix elements of local operators
 bilinear in the light quark fields, taken between the vacuum and 
the pion state.
On the other hand, when considering the dispersion relation for 
the correlator of eq. (\ref{corr}), the  constant $g$ enters 
via the contribution of the $B$ and $B^*$ poles to the spectral density,
through the S-matrix element:
\be
<\pi^-(q)~{\bar B}^o(q_2) | B^{*-}(q_1,\epsilon )> \; = \; \gib \;
\epsilon^{\mu} \cdot q_{\mu}
\label{gi}
\ee
 From the lagrangian in eq. (\ref{ppm}), one immediately finds:
\be
\gib={\dd{2 M_B}\over{f}} g ~~.
\ee

The QCD sum rule approach allows to estimate directly the
strong amplitude of eq. (\ref{gi}), characterized by the
coupling constant $\gib$. 
This includes the full dependence on the heavy quark mass $m_b$,
not only its asymptotic, large $m_b$, behaviour.

On the other hand, by retaining only the leading terms in the 
limit $m_b\to \infty$, one obtains the following numerical results from
the sum rule:
\be 
{\hat F}^2 \; g = 0.040 \pm 0.005~~~{\rm GeV}^3~~~. 
\label{hat}
\ee
where
${\hat F}$ parametrizes the leading term in the decay constants
$f_B$ and $f_{B^*}$:
\be
f_B=f_{B^*}= \frac{\hat F}{\sqrt{M_B}}~~.
\label{17aa}
\ee 
$\hat F$ can be computed by QCD sum rules. For example,
for $\omega=0.625$ GeV ($\omega$ is the binding energy of the meson,
finite in the 
large mass limit), and the continuum threshold parameter
$y_0$ in the range $1.1 - 1.4$  GeV, and neglecting
QCD corrections \cite{neubert1}
the result is
\be
{\hat F}= 0.30 \pm 0.05 {\rm GeV}^{3/2}~~.
\label{eq:eqf}
\ee
By including radiative corrections one finds higher values (around 
0.4-0.5 GeV$^{3/2}$)
that are compatible with the results obtained by lattice QCD;
\bea
\hat F&=&0.55\pm 0.07 ~~~~~\cite{APE}\nn \\
\hat F&=&0.61\pm 0.08 ~~~~~\cite{UKQCD}\nn \\
\hat F&=&0.49\pm 0.05 ~~~~~\cite{MILC}~~.
\label{eq:eqf2}
\eea
Since one has neglected in (\ref{hat}) radiative corrections, a safer value
for $\hat F$  is given in eq. (\ref{eq:eqf}), which is also the value we 
shall use in the subsequent sections.
 From eqs. (\ref{hat}) and (\ref{eq:eqf}), one would obtain:
$g=0.44 \pm 0.16$.

An independent estimate of $g$ can be obtained by expanding
the correlator of eq. (\ref{corr})
near the light-cone in terms of non-local operators whose 
matrix elements  define pion wave functions of increasing
twist (this method is called light-cone sum rules). 
In this way, an infinite series of matrix elements of local operators is
effectively replaced by a universal, non-perturbative, wave-function 
whose high-energy asymptotic behaviour is dictated by the approximate 
conformal 
invariance of QCD. By using this technique, 
in \cite{bel}, the following result has been obtained:
$g=0.32\pm 0.02$. 

Our best estimate for $g$, based on the analyses of both QCD sum rules 
\cite{qcdsr,bel} and relativistic quark model \cite{defazio2} is
\be
g\approx 0.38
\label{eq:038}
\ee
with an uncertainty that we estimate around $\pm 20\%$. This is the value we 
shall use in the next sections. In section \ref{sec:bepp1} we will show that
also the results from radiative $D^*$ decays 
are compatible with (\ref{eq:038}).

\subsection{Chiral corrections to $g$}
\label{sec:chirg}

Due to the exact chiral symmetry of the interaction terms in
eq. (\ref{wise}), the coupling constant $g$ does not depend
on the light flavour species.  Chiral breaking effects
can be accounted for by adding breaking terms to the symmetric lagrangian.
The chiral breaking parameters are the light quark masses, 
and the lowest approximation consists in keeping all the terms 
of the first order in the quark mass matrix.

On the other hand, in a given process, corrections to the chiral limit
can arise in two ways: either via chiral loops, with
mesons propagating with their physical, non-vanishing mass,
or via counterterms which affect the considered quantity
at tree-level. The latter corrections exhibit an analytic
dependence on the quark masses and are typically unknown,
being related to new independent parameters of the chiral
lagrangian. On the contrary, the former terms contain
a non-analytic dependence on the quark masses
which is calculable via a loop computation.
The loop corrections in turn depend explicitly on an arbitrary 
renormalization point 
$\mu^2$ (e.g. the t'Hooft mass of the dimensional regularization). 
This dependence is cancelled by the $\mu^2$ dependence of the counterterms. 

Although the overall result is given by the sum of these two 
separate contribution, it is current practice to 
estimate roughly the chiral corrections by neglecting the analytic
dependence and by fixing to about 1 GeV the renormalization scale $\mu$ in
the loop computation.
The adopted point of view is that the overall effect of adding the counterterm
consists in replacing $\mu^2$ in the loop corrections with the physical scale
relevant to the problem at hand, $\Lambda_{\chi}^2$. 
Possible finite terms in the counterterm are supposed to be small 
compared to the large chiral logarithms due to the formal
enhancement of the non-analytic terms as $m_{\pi}^2 {\rm log} m_{\pi}^2$
over the analytic ones. In view of these uncertainties the results of this
method are more an indication of the size of the corrections
than a true quantitative calculation, since there are examples e.g. in kaon 
physics where the finite counterterms are not negligible (\cite{maiani}).

With this philosophy in mind, the chiral corrections to the 
$g$ coupling constant have been evaluated in ref. \cite{fle,fag,cheng}.
Neglecting the $u$ and $d$ quark masses in comparison 
to the strange quark mass and by using the Gell-Mann-Okubo formula to express
$m^2_\eta$ in terms of $m^2_K$ ($m^2_\eta=4/3 m^2_K$), the leading one-loop
logarithmic corrections can be expressed in terms of
\bea
\chi&=&\frac{1}{16 \pi^2}\frac{m^2_K}{f_\pi^2} 
      ~{\rm log}(\dd\frac{m^2_K}{\mu^2})\nn\\
    &\simeq&-0.125~~~(\mu=1~{\rm GeV})~~.
\eea
The one-loop coupling  constant, $g_{eff}$, is
given by:
\be
g_{eff}=g\left[1-(1+\frac{35}{9}g^2) \chi \right]
       \simeq 0.45~~~~~~~~~~(g=0.38)~~.
\label{gef}
\ee
In computing these class of corrections one may use the Feynman rules 
reported in appendix A.
The result (\ref{gef}) will be used in the evaluation of the loop
corrections to the matrix element of the weak current between 
$\pi$ and $P~(P=D,B)$, as discussed in section \ref{subsec-bpieff}.

\subsection{Hyperfine splitting}

As an example of application of the chiral perturbation theory 
to the calculation of physical observables relative to heavy $Q {\bar q}$
mesons, in this section we work out in some detail the hyperfine mass
splitting between $1^-$ and $0^-$ mesons. As a matter of fact,
the spectroscopy of heavy mesons is probably the simplest framework where the 
ideas and the 
methods of heavy quark expansion can be quantitatively tested.
As explained in section \ref{sec:chiral}, the splitting among the $1^-$ and 
$0^-$ 
heavy mesons masses is due, at the leading order, by the $1/M$ correction
of eq. (2.48):
\be
\Delta=(m_{P^*}-m_P)=-\frac{2\lambda_2}{M}~~.
\label{t}
\ee
The experimental data, listed in table \ref{table:fer2}, supports quite well 
the approximate scaling law suggested by eq. (\ref{t}).
\begin{table}[htb]
\caption{Experimental mass splittings between $1^-$ and $0^-$ mesons.}
\label{table:fer2}
\centering
\begin{minipage}{6truecm}
\begin{tabular}{l c}
&$\Delta~({\rm MeV})$\\
\hline
$M_{D^{*+}}-M_{D^+}$ & $140.64\pm 0.09$\\
\hline
$M_{D^{*0}}-M_{D^0}$ & $142.12\pm 0.07$\\
\hline
$M_{D_s^{*\pm}}-M_{D_s^\pm}$ & $141.6\pm 1.8$\\
\hline
$M_{B^{*}}-M_{B}$ & $46.0\pm 0.6$\\
\hline
$M_{B_s^{*}}-M_{B_s}$ & $47.0\pm 2.6$\\
\end{tabular}
\end{minipage}
\end{table}
These data can be used to estimate the parameter $\lambda_2$:
\be 
\lambda_2\simeq 0.10-0.11~GeV^{2} \label{tbis}~~. 
\ee
The second term in eq. (\ref{chircorr}), independent of the heavy quark
flavour, is responsible 
for the mass splitting between strange and 
non-strange heavy mesons:
\be
\Delta_s = 2 \lambda_1 m_s~~.
\label{ds}
\ee
Experimentally one has \cite{pdb}:
\bea
M_{D_s^{*\pm}}-M_{D^{\pm}}&=&99.1\pm 0.6~{\rm MeV}\nn\\
M_{B_s^{0}}-M_{B}&=&96\pm 6~{\rm MeV}~~,
\eea
leading to
\be
\lambda_1 \simeq 0.33~~.
\ee
Recently, attention has been focused on the combinations 
\cite{ros,ran,jen,cheng}:
\be
\Delta_D=(M_{D^*_s}-M_{D_s})-(M_{D^{*+}}-M_{D^+})
\label{a}
\ee
\be
\Delta_B=(M_{B^*_s}-M_{B_s})-(M_{B^{*0}}-M_{B^0})
\label{b}
\ee
which are measured to be \cite{pdb}:
\be
\Delta_D\simeq 1.0\pm1.8~MeV
\label{c}
\ee
\be
\Delta_B\simeq 1.0\pm2.7~MeV~~.
\label{d}
\ee
This hyperfine splitting is free from electromagnetic corrections and  
vanishes separately in the $SU(3)$ chiral limit and in the heavy quark limit. 
In the combined chiral and heavy quark expansion, the leading 
contribution is of order $m_s/m_Q$ and
one would expect the relation \cite{ros}:
\be
\Delta_B=\frac{m_c}{m_b}\Delta_D~~.
\label{e}
\ee
In our framework the
lowest order operator
contributing to $\Delta_{D,B}$ is:
\be
\eta~ {\cal O}_2=\frac{\eta}{8 M} 
Tr[{\bar H}_a \sigma_{\mu\nu} H_{b} \sigma^{\mu\nu}]
                     \frac{(m_{\xi})_{ba}}{\Lambda_{\chi}}~~.
\label{f}
\ee
The matrix $m_{\xi}$ is
\be
m_{\xi}=(\xi \hat m \xi +\xi^\dagger \hat m \xi^\dagger)
\label{g}
\ee
where $\hat m$ is the light quarks mass matrix and $\xi$ the coset variable
defined in eq. (2.33).
By taking $m_s/\Lambda_{\chi}\simeq 0.15$ and $\eta \simeq
\Lambda_{QCD}^2 \simeq 0.1~ GeV^2$ 
one would estimate:
\be
\Delta_D\simeq 20~MeV
\label{h}
\ee
\be
\Delta_B\simeq 6~MeV~~.
\label{i}
\ee
Given the present experimental accuracy,
the above estimate is at most acceptable, as an order of magnitude, 
for $\Delta_B$,
while it clearly fails to reproduce the data for $\Delta_D$.
If the contribution from ${\cal O}_2$ were the only one
responsible for the hyperfine
splittings, agreement with the data  would clearly require
a much smaller value for $\eta$.

In chiral perturbation theory, an independent contribution arises from 
one-loop corrections to the heavy meson self energies \cite{jen},
evaluated from an initial lagrangian containing, at the lowest order,
both the chiral breaking and the spin breaking terms of eq. (\ref{chircorr}) 
and (\ref{1/m}). These corrections can be computed by using the Feynman rules
given in appendix A; they depend 
on an arbitrary renormalization point $\mu^2$ (e.g. the
t'Hooft mass of dimensional regularization). This dependence is cancelled by
the $\mu^2$ dependence of the counterterm $\eta(\mu^2) {\cal O}_2$.
Following the discussion in section \ref{sec:chirg} one can use
$\mu=1$ GeV.

The possible sources of hyperfine splittings via chiral loops
are the light pseudoscalar masses $m_\pi$, $m_K$ and $m_\eta$, 
the mass splittings $\Delta_s$, $\Delta$ of eqs. (\ref{t}), (\ref{ds}) and,
finally, the difference between the $P^* P^* \pi$
and the $P^* P \pi$ couplings ($P=D,B$) induced by the last term 
of eq. (\ref{1/m}).

This splitting is 
of order $1/M$, and, from eq. (\ref{g:1/m}), one obtains: 
\be 
\Delta_g\equiv g_{P^*P^* \pi}-g_{P^* P \pi}=2~ \frac{g_2}{M}~~. 
\label{u}
\ee
The second term in (\ref{1/m}), proportional to $g_1$,
breaks only the heavy flavour symmetry, making the 
$B^* B^{(*)} \pi$ and $D^* D^{(*)}\pi$ couplings different.         
The third term, proportional to $g_2$,
breaks also the spin symmetry and contributes 
differently to the $P^* P \pi$ and to the $P^* P^* \pi$ couplings.
This is precisely the effect relevant to the hyperfine splitting. 

In terms of these quantities, one finds \cite{jen,ran,cheng}:
\bea
\Delta_{P}&=&\frac{g^2 \Delta}{16 \pi^2 f^2}
            \Bigl[4 m_K^2 \log(\frac{\Lambda_{\chi}^2}{m_K^2})+
                  2 m_\eta^2 \log(\frac{\Lambda_{\chi}^2}{m_\eta^2})-
                  6 m_\pi^2 \log(\frac{\Lambda_{\chi}^2}{m_\pi^2})\Bigr]\nn\\
          &+&\frac{g^2 \Delta}{16 \pi^2 f^2} [24 \pi m_K \Delta_s]\nn\\
          &-&\frac{g^2}{6 \pi f^2}\frac{\Delta_g}{g}
             (m_K^3+\frac{1}{2} m_\eta^3-\frac{3}{2} m_\pi^3)~~.
\label{split}
\eea
The dependence upon the heavy flavour $P=D,B$ is contained in the 
parameters $\Delta$ and $\Delta_g$. 

The first term in eq. (\ref{split}) 
is the so called chiral logarithm 
\cite{jen}.
In the ideal situation with pseudoscalar masses much smaller than 
$\Lambda_{\chi}$, it would represent the dominant contribution to $\Delta_P$. 
Calling $\Delta^0_D$ and $\Delta^0_B$ its value for
the $D$ and $B$ mesons, respectively, one finds:  
\be
\Delta^0_D\simeq +13~MeV,~~~~~~\Delta^0_B\simeq + 4~MeV,
\label{l}
\ee
where we are using the representative 
value $g =0.38$ (see eq. (\ref{eq:038})).
 
The second term in eq. (\ref{split}) represents a non analytic
contribution of order $m_s^{3/2}$ \cite{ran}, which, although formally
suppressed with respect to the leading one, is numerically more important, 
because of the large coefficient $24 \pi$. The separate contributions
to the $D$ and $B$ hyperfine splittings read:
\be
\Delta^1_D\simeq +30~MeV,~~~~~\Delta^1_B\simeq +9~MeV~~.
\label{m}
\ee

Finally, the last term in eq. (\ref{split}) 
\cite{cheng} is
also of order $m_s^{3/2}$. 
Its evaluation requires the estimate of the difference $\Delta_g/g$,
which is not directly related to other experimental data.
In ref. \cite{hyp}, this difference has been computed in the framework of
QCD sum rules. Using $m_b=4.6 \; GeV$ and 
$m_c=1.34 \; GeV$  one gets:
\bea  
f_{B^*}^2 \; \gibs & = &  0.0094 \pm 0.0018 \; GeV^2  \nn\\
f_{D^*}^2\; g_{D^*D^* \pi} & = &  0.017 \pm 0.004 \; GeV^2  \; ; \label{15} 
\eea
and for the $\gi$ coupling 
\bea 
f_B \; f_{B^*} \; \gib & = &  0.0074 \pm 0.0014 \; GeV^2  \nn \\
 f_D \; f_{D^*}\; \gid & = &  0.0112 \pm 0.0030 \; GeV^2  \; . \label{16} 
\eea
To derive the difference $\Delta_g$ at first order in $1/M$, 
one should expand the relevant sum rules in the parameter $1/M$,
keeping the leading term and the first order corrections which are given by
\be
\left(\frac{g_1+g_2}{g}\right)+2A'= -0.15 \pm 0.20 \; GeV \;\;\;~~ \; 
\left(\frac{g_1-g_2}{g}\right)+A'+A= -1.15 \pm 0.20 \; GeV
\label{22}                                      
\ee
and
\be
2~\frac{g_2}{g}+(A'-A)=0.99 \pm 0.02 \; GeV~~.
\label{splitc}
\ee
The couplings $g_1$ and $g_2$ have been defined in eq. (\ref{g:1/m}) and
the parameters $A$ and $A'$ are related to the 
$1/M$ corrections to the leptonic decay constants, $f_P$ and $f_{P^*}$:
\be
f_P= \frac{\hat F}{\sqrt{M}} \left( 1+ \frac{A}{M} \right) \; \;~~~
f_{P^*}= \frac{\hat F}{\sqrt{M}} \left( 1+ \frac{A'}{M} \right)~~.
\label{17a} 
\ee
Neglecting radiative corrections, $A$ and $A'$ are given by 
\cite{Ball,neubert1}:
\be
A= -\omega + \frac{G_K}{2} + 3 G_{\Sigma} \;\;\; \;~~ 
A'= -\frac{\omega}{3} + \frac{G_K}{2} - G_{\Sigma} 
\label{eq:aap}
\ee 
where $\omega$ represents the difference between the pseudoscalar 
meson and the heavy quark masses, at leading order in $1/M$.
The splitting of the couplings depends on the quantity 
$2~g_2/g$ that
contains only the difference $A'-A$ given by:
\be
A'-A = \frac{2}{3} \omega -4 G_{\Sigma}
\label{diff}
\ee
There is disagreement in the literature on the values of
the parameter $G_{\Sigma}$: at the $b$ quark
mass scale from ref.\cite{Ball} one gets $G_{\Sigma} = (0.042 \pm 0.034 \pm
0.023 \pm 0.030) \; GeV$, while in ref. \cite {neubert1}  the central value
$G_{\Sigma} \simeq -(0.052)\; GeV$ is quoted.
In view of this discrepancy, to provide an estimate of the difference 
(\ref{diff}), it is reasonable to approximate 
$A' -A \approx 2/3 \omega \approx 0.4 \; GeV$,
obtaining
\be
2~\frac{g_2}{g} \approx 0.6 \;\; GeV \; \; .
\label{ab}
\ee

 From (\ref{u}), (\ref{ab}) and
from the formula (\ref{split}) of the hyperfine mass splitting one finds:
\be
\Delta_B \approx g^2 ~(27.3 + 61.4 -75.8)~ MeV= 12.9~ g^2 ~MeV
\label{z}
\ee

We notice that we have used in eq. (\ref{split}) $f=f_{\pi}=130$ MeV
for all the light pseudoscalar mesons of the octet. 
In eq. (\ref{z}) 
we have detailed the contributions $\Delta^0$, $\Delta^1$ and the one from
$\Delta_g/g$ respectively. We have also taken $\Lambda_{\chi}=1~GeV$.
It is evident that there is a large cancellation among the last term and
the other ones.  
For the value $g=0.38$ we obtain
\be
\Delta_B \simeq 1.9 ~MeV~~.
\label{z1}
\ee
The application of this result to the 
charm case is more doubtful, in view of the large values of the $1/m_c$
correction $\Delta_g/g$. By scaling the result (\ref{z1}) to the charm case,
one would obtain
\be
\Delta_D = \frac{m_b}{m_c}\Delta_B \simeq 6.3 ~MeV~~.
\label{z2}
\ee

In conclusion we observe that the application of chiral perturbation theory
to the calculation of the heavy meson hyperfine mass splitting is rather
successful, even though, given the large cancellations in eq. (\ref{z}),
the results (\ref{z1}) and (\ref{z2}) should be considered as order of
magnitude estimates only.

\subsection{ Strong decays of positive parity states}
\label{sec:strong}

In this section we shall examine the applications of the effective lagrangian
approach to the strong decays of the positive parity heavy meson states. We 
shall
first review the experimental evidence for these states; next we shall give 
the 
formulas for the
decay rates into final states with one pion. Finally we shall present some
estimates of the couplings based on QCD sum rules and we shall apply them
to the calculation of the strong decay rates.

Strong transitions of the positive parity states, 
contained in the multiplets $S$ and $T$ introduced in section \ref{sec:2.1},
are described in the present formalism by the lagrangian ${\cal L}_3$
of eq. (\ref{ll3}), explicitly discussed in section \ref{sec:2.5}.
The experimental data concerning these states are still at a
preliminary stage. In the charm sector, the total widths of the
$D_2(2460)$ and $D_1(2420)$ states, have been measured \cite{pdb}:
\bea
\Gamma_{tot}(D_2(2460))&=&21\pm 5~{\rm MeV}\label{a1}\\
\Gamma_{tot}(D_1(2420))&=&18\pm 5~{\rm MeV}~~.
\label{a2}
\eea
As for the $B$ sector, evidence has been recently reported \cite{delphi,opal}
of a bunch of positive parity states $B^{**}$, with an average mass 
\be
m_{B^{**}} = 5732 \pm 5 \pm 20~{\rm MeV}
\label{mass}
\ee
and an average width 
\be
\Gamma(B^{**}) = 145 \pm 28 ~{\rm MeV}~~~.
\label{wid}
\ee 
The OPAL collaboration of LEP \cite{opal} has also reported evidence of a
$B^{**}_s$ state with mass 
\be
m_{B^{**}_s} = 5853 \pm 15~{\rm  MeV}
\ee
and width
\be
\Gamma(B^{**}_s) = 47 \pm 22~{\rm MeV}~~~.
\label{bs}
\ee

The decay widths of the states $(1^+, 0^+)$, belonging to multiplet
$s_{\ell}^P = (1/2)^+$, here referred to as $P_1$
and $P_0$, are expected to be saturated by the single pion 
channels \cite{kilian,falkluke}:
$P_0 \to P \pi$ and $P_1 \to P^{*} \pi$. 
Therefore these transitions are controlled by the coupling constant $h$
of eq. (2.71).
In the $m_Q \to \infty$ limit one obtains
\begin{equation}
\Gamma( P_0 \to P^+ \pi^-)= \Gamma( P_1 \to P^{*+} \pi^-)=
\frac{1}{2\pi} \left( \frac{h}{f_\pi } \right)^2 (\delta m_s)^3
\label{1fer}
\end{equation}
where $\delta m_s$ is the mass splitting of the states $S$ 
with respect to the ground state $H$.
 From estimates based on quark model \cite{isgur2,veseli} and QCD 
sum rules \cite{noi,cola92}
computations of the masses of these states, one has $\Delta\,=\, 500 \pm 100$ 
MeV.
We notice that this mass splitting 
agrees rather well with the experimental result in the
$B$ sector, given in eq. (\ref{mass}). 

On the other hand,
the formula (\ref{1fer}) is of limited significance,
especially for the case of charm, due to the large $1/M$ corrections 
coming from the kinematical factors.
Keeping $M$ finite, the formulas become:
\begin{equation}
\Gamma (P_0 \to P^+ \pi^-) = 
\frac{1}{8\pi} G_{P^{**}P\pi}^2 \frac{\left[ (M^2_{P_0} -
(M_P + m_{\pi})^2) (M^2_{P_0} - 
(M_P - m_{\pi})^2)\right]^{\frac{1}{2}}}{2 M_{P_0}^3} \; .
\label{3fer}
\end{equation}
\begin{eqnarray}
\Gamma (P_1 \to P^{*+} \pi^-) & = &\frac{G_{P_1 P^*\pi}^2}{8\pi} 
\frac{\left[ (M^2_{P_1} -
(M_{P^*} + m_{\pi})^2) (M^2_{P_1} - 
(M_{P^*} - m_{\pi})^2)\right]^{\frac{1}{2}}}{2 M_{P_1}^3} \times \nonumber \\
& \times & \frac{1}{3} \left( 2+ \frac{(M^2_{P_1} 
+ M^2_{P^*})^2}{4 M^2_{P_1} M^2_{P^*}}
\right) \; .
\label{5fer}
\end{eqnarray}
For the $B$ system
the coupling constants $G_{P^{**}P\pi}$ and $G_{P_1 P^*\pi}$
are defined by the strong amplitudes:
\begin{equation}
 G_{B^{**} B \pi} \; = \; <\pi^+(q)~B^o(q_2) | B^{**+}(q_1)>
\label{g1}
\ee
\begin{equation}
 G_{B_1 B^* \pi} \; = \; <\pi^+(q)~B^{o*}(q_2) | B_1^+(q_1)>
\label{g2}
\end{equation}
where $B^{**}$ and $B_1$ denote the $0^+$ and the $1^+$ states in the 
$s_{\ell}^P = (1/2)^+$ doublet.
Analogous definitions are understood for the $D$'s.
In the infinite mass limit, $G_{P^{**}P\pi}$ and $G_{P_1 P^*\pi}$ coincide.
The amplitude $G_{B^{**} B \pi}$ is related to 
the strong coupling constant $h$ appearing in the heavy-light chiral 
lagrangian
(\ref{eq:lls}) by the formula: 
\begin{equation}
G_{B^{**} B \pi} \;= - \; {\sqrt {M_B M_{B^{**}}}} \; \; 
{{ M^2_{B^{**}}- M^2_B}\over{M_{B^{**}}}}\; 
{{h}\over{f_\pi}} \; . \label{grel}
\end{equation}
In the  limit $m_b \to \infty$ one has:
\begin{eqnarray}
M_B & = & m_b + \omega + {\cal O} \left( {1\over{m_b}}\right) \nonumber \\
M_{B^{**}} - M_B & = &  \delta m_s \; + \; {\cal O} \left( {1\over{m_b}} 
\right) 
\label{para}
\end{eqnarray}
\begin{equation}
G_{B^{**} B \pi} \simeq  -  \frac{2 h}{f_{\pi}} \; m_b \; \delta m_s  
\label{gas}
\end{equation}
and one recovers eq. (\ref{1fer}).

Differently from the decays
of the positive parity states having $s_{\ell}^P=(1/2)^+$, 
the single pion transitions of the
$s_{\ell}^P=(3/2)^+$ particles, here denoted
$P_2$ and $P_{1}'$, occur with
the final pion in $D$-wave.
The decay rates for these transitions are given by \cite{kilian,falkluke}:
\begin{eqnarray}
\Gamma (P_2^0 \to P^+ \pi^-) & = & \frac{1}{15\pi} \frac{M_P}{M_{P_2}}
\frac{h'^2}{{\Lambda_\chi}^2} \frac{|\vec{p}_{\pi}|^5}{{f_\pi}^2} 
\label{unof}\\
\Gamma (P_2^0 \to P^{*+} \pi^-) & = & \frac{1}{10\pi} \frac{M_{P^*}}{M_{P_2}}
\frac{h'^2}{{\Lambda_\chi}^2}\frac{|\vec{p}_{\pi}|^5}{{f_\pi}^2} 
\label{duef}\\
\Gamma (P_1^{\prime 0} \to P^{*+} \pi^-) & = & \frac{1}{6\pi} 
\frac{M_{P^*}}{M_{P_2}}
\frac{h'^2}{{\Lambda_\chi}^2}\frac{|\vec{p}_{\pi}|^5}{{f_\pi}^2} \; ,
\label{tref}
\end{eqnarray}
where the strong coupling $h'$ is given by:
\be
h'=h_1+h_2~~~,
\ee
in terms of the parameters $h_1$ and $h_2$ of eq. (\ref{eq:lld}).
 From the previous equations, one finds the following prediction
in the $D$ system:
\be
\frac{\Gamma(D_2^0\to D^+\pi^-)}{\Gamma(D_2^0\to D^{*+}\pi^-)}=2.7
\ee
in good agreement with the experimental result \cite{pdb}, $2.4\pm 0.7$.

To get numerical results for the rates given in eqs. (\ref{3fer}), 
(\ref{5fer}) and (\ref{unof}-\ref{tref}),
one should specify the relevant coupling constants.
The parameter $G_{P^{**}P\pi}$ has been evaluated in the framework of QCD
sum rules, by means of two independent methods \cite{faz}.
The first method, based on the single Borel transform of an appropriate
correlator evaluated in the soft pion limit, gives the results:
\begin{equation}
G_{B^{**}B\pi}= 13.3 \pm 4.8 \; GeV \;
\end{equation}
\begin{equation}
G_{D^{**}D\pi} = 11.5 \pm  4.0 \; GeV \; .
\end{equation}
We observe substantial violations of the scaling law 
$G_{D^{**}D\pi}/G_{B^{**}B\pi} \approx m_c/m_b$. 

In the limit $M\to\infty$
 we obtain, from the asymptotic ($m_b \to \infty$) limit of the sum rule:
\begin{equation}
h=\, - \, 0.52 \pm 0.17 \; . 
\label{h1b}
\end{equation}

The second method is based 
on the light-cone sum rules
\cite{chern1,chern2,bel,fil}. One obtains \cite{faz}:
\be
G_{B^{**}B\pi}= 21 \pm 7 \;{\rm GeV}
\ee
\be
G_{D^{**}D\pi} = 6.3 \pm 1.2 \; {\rm GeV}
\ee 
A two-parameter fit of the above results in the form
\begin{equation} h(m)=h \; (1 + {\sigma \over m}) \label{fit}
\end{equation}
gives for $h$ (see eq. (\ref{grel}) the result:
\begin{equation}
h=-0.56 \pm 0.28 \label{h2b}
\end{equation}
 and for the parameter $\sigma$,
$\sigma=0.4 \pm 0.8 \; GeV$.

The values
of $h$ found by the two methods agree with each other. 
As for the finite mass results, the two methods
sensibly differ (almost a factor of 2) in the case
of the charm, while the deviation is
less important for the case of beauty (around $40\%$). These
differences should be attributed to
corrections to the soft pion limit that have been
accounted for by the sum rule based on a light-cone expansion.
 
Using $G_{D^*D\pi}=6.3 \pm 1.2 \; GeV$, 
$G_{B^*B\pi}=21\pm 7 \; GeV$, and
$\Delta_D = \Delta_B = 500 \; MeV$, from eq. (\ref{3fer}) one finds
\begin{eqnarray}
\Gamma(D_0 \to D \pi) & \simeq & 180 \; MeV \\
\Gamma(B_0 \to B \pi) & \simeq & 360 \; MeV \; .
\label{4g}
\end{eqnarray}

There is no direct information on the coupling $G_{P_1 P^* \pi}$.
In the infinite-mass limit it coincides with
$G_{P^{**}P \pi}$ and, in order to
estimate the widths of the $1^+$ states, we assume that this equality
holds for finite mass as well. From eq.(\ref{5fer}) we obtain:
\begin{eqnarray}
\Gamma(D_1 \to D^* \pi) & \simeq & 165 \; MeV \\
\Gamma(B_1 \to B^* \pi) & \simeq & 360 \; MeV \; .
\label{6g}
\end{eqnarray}
Also in this case we have taken $M_{P_1} - M_{P^*} = 500 \;$ MeV ($P = B,D$)
as suggested by HQET considerations.

\vspace*{0.5cm}
To estimate the strong coupling constant $h'/\Lambda_\chi$ ,
one can make use of the total decay width of eq. (\ref{a1}),
$\Gamma_{tot}(D_2(2460))=21 \pm 5$ MeV. 
Assuming that only two body
decays are relevant, one gets $h'/\Lambda_\chi \approx 
0.55$ GeV$^{-1}$. From this
result and from eq. (\ref{tref}) one obtains for the state
${\tilde D}_1^{0}$
the total width $\Gamma_{tot}\approx 6$ MeV to be compared with
the experimental width of the other narrow state observed
in the charm sector, $\Gamma_{tot} (D_1(2420)) = 18 
\pm 5$ MeV, also given in eq. (\ref{a2}).
This discrepancy could be attributed to a mixing between 
the $D_1$ and the ${\tilde D}_1$ states \cite{iswi}.
If $\alpha$ is the mixing angle,
we have 
\begin{equation}
\sin^2 (\alpha) \approx \frac{12 \; MeV}{\Gamma(D_1) - \Gamma({\tilde D}_1)} \simeq 
0.08
\label{7}
\end{equation}
and therefore one gets the estimate $\alpha \approx 16^o$ \cite{faz}. This 
determination agrees with the result of Kilian et al. in ref.\cite{kilian}.

In ref. \cite{falkluke} the decay rates for the transitions of the 
$(1^+,2^+)$ states
with the emission of two pions, $D^{**}\to D^{(*)}\pi\pi$, 
have also been estimated.
They appear to be suppressed with respect to the single pion rates.

In the $B$ sector, the recently observed positive parity states $B^{**}$,
whose average mass and width are given in eqs. (\ref{mass}) and (\ref{wid}),
can be identified with the two doublets $(2^+, 1^+)$ and $(1^+, 0^+)$.
To compare previous estimates with the data, we average the widths
of the $(1^+, 0^+)$ multiplet, eqs. (\ref{4g}) and (\ref{6g}),
with those of the $2^+$ and $1^+$ states, obtained from eq. 
(\ref{unof}-\ref{tref}):
\bea
\Gamma_{tot}(B_2)&\simeq& 12 ~{\rm MeV}\nn\\
\Gamma_{tot}({\tilde B}_1)&\simeq& 10 ~{\rm MeV}\nn\\
\eea
It is difficult to perform a detailed comparison of 
these results with the yet incomplete experimental outcome.
However, assuming that
the result obtained by LEP collaborations in the $B$ system represents an
average of several states, and neglecting a possible
mixing between the states $B_1$ and ${\tilde B}_1$ (a $1/M$ effect), 
the experimental width is compatible with the previous estimate.

Finally, the total width in eq. (\ref{bs}) can be interpreted
as connected to the decay $B^{**}_s \to B K, \; B^* K$.
Assuming again that
the width is saturated by two-particle final states, and
using $M_{B^{**}_s} = 5853 \; MeV$, 
we obtain:
\begin{eqnarray}
\Gamma(B^{**}_s(0^+)) & \simeq & 280 \; MeV \\
\Gamma(B^{**}_s \to B^* K) & \simeq & 200 \; MeV \; \; \; \; \; (s_{\ell})^P=
(1/2)^+ \\ 
\Gamma(B^{**}_s(1^+))  & \simeq & 0.45 \; MeV \; \; \; \; \; (s_{\ell})^P=
(3/2)^+ \\
\Gamma(B^{**}_s(2^+)) & \simeq & 1.4 \; MeV \; .
\end{eqnarray}
Also in this case a detailed comparison with the experimental results 
cannot be performed without more precise measurements;
we  observe, however, that the 
computed widths of the different
$B^{**}_s$ states are generally smaller than the corresponding
quantities of the $B^{**}$ particles, a feature which is reproduced by the 
experiment.

\section{ $B \to D$ decays and chiral dynamics}
\label{sec:BD}

One of the most important applications
of the heavy quark symmetry is the analysis of the exclusive 
semileptonic decays $B \to D l \nu_l$ and $B \to D^* l \nu_l$. 
We shall here give a brief summary of this extensively studied subject:
for more details see for instance \cite{neubert} and references
therein.

In the symmetry limit, i.e. infinite $D$ and $B$ masses, the six form factors
generally needed to parameterize the matrix elements 
$<D^{(*)}(v')| J_\mu |B(v)>$ ($v$, $v'$ velocities)
reduce to a single function $\xi(v \cdot v')$, the Isgur-Wise function.
One finds \cite{hq1}:
\bea
<D(v')|{\bar c} \gamma_\mu b |B(v)> & =& 
\sqrt{M_B M_D} \xi (v \cdot v') (v+v')_\mu \nn \\
<D^* (v', \epsilon)|{\bar c} \gamma_\mu b |B(v)> & =& 
\sqrt{M_B M_D}
i \xi (v \cdot v')\epsilon_{\mu\nu\alpha\beta} \epsilon^{*\nu} v'^{\alpha}
v^\beta \nn \\
<D^* (v', \epsilon)|{\bar c} \gamma_\mu \gamma_5 b |B(v)> & =& 
\sqrt{M_B M_D}
\xi (v \cdot v') \left[ (1 + v \cdot v') \epsilon^*_\mu - (\epsilon^* \cdot v)
v'_\mu \right] \; .
\label{wisgur}
\eea

Various calculations of the Isgur-Wise function exist in the literature;
they use different non-perturbative approaches, such as QCD sum rules 
\cite{iwsumrul} or lattice QCD \cite{latiw}.
A review of these results would be outside the scope of the present report,
and we refer the interested reader to the literature.

At the symmetry point, i.e. $v=v'$,  the normalization of the Isgur-Wise
function is known: $\xi (1) =1$.
This is a consequence of the conservation of the vector current 
$J^\mu = {\bar h}'_v \gamma_\mu h_v = {\bar h}'_v v_\mu h_v$
and allows a model independent
determination of the CKM matrix element $V_{cb}$ from semileptonic heavy to
heavy decays by extrapolating the lepton spectrum to the endpoint $v=v'$.

Of special interest  for this determination
is the decay $B \to D^* l \nu$, since there are no $1/m_Q$
corrections for the axial form factor $A_1$, dominating the decay rate, 
at the symmetry point. This is the content of the Luke's theorem
\cite{luke90}. A simple proof of this important result has been presented by 
Lebed and Suzuki \cite{luke90}. Luke's theorem is an extension
to the spin-flavour symmetry of the Ademollo-Gatto
theorem \cite{ag}, which was originally stated for the 
$SU(3)$ flavour symmetry of
light quarks and refers to the matrix element of the vector current between
states belonging to the same $SU(3)$ multiplet at $q^2 = 0$.
 The statement is that matrix
elements of a charge operator, i.e. a generator of the symmetry, can deviate
from their symmetry values only for corrections of the
 second order in symmetry breaking.
In the case of semileptonic decays $B \to D (D^*) l \nu$, the only form
factor protected by this theorem against $1/m_Q$ corrections at the symmetry
point $v=v'$ is $A_1$, dominating the decay $B \to D^* l \nu$ at $v=v'$. 
In practice, Luke's theorem reduces to the result:
\be
\frac{M_B + M_{D^*}}{2 \sqrt{M_B M_{D^*}}} A_1 (q^2_{\rm max}) = \eta_A +
\delta_{1/m^2}
\label{luke}
\ee
where $\eta_A=1$ if strong radiative corrections are neglected.

The $1/m_Q^2$ corrections at the point $v=v'$
have been estimated 
\cite{2corr}: a combined analysis \cite{neub94} gives a correction to $\xi 
(1)$:
$\delta_{1/m^2}= -(5.5 \pm 2.5 )\%$. Also leading and subleading QCD
corrections arising from virtual gluon exchange 
have been computed:  see for instance
\cite{neubert}. 

Experimental measurements close to $v=v'$ suffer of large errors, 
due to the smallness of the phase space: high statistics is needed to reduce
the uncertainty in the extrapolation of the lepton spectrum to this point: 
nevertheless, the exclusive
semileptonic decay $B \to D^* l \nu_l$ can provide 
a rather precise measurement of
the element $V_{cb}$ of the CKM matrix, complementary to the analysis of the
inclusive  semileptonic decay rate.

The HQET has also been used to investigate 
the semileptonic decay  of a $B$ meson into an excited charm meson
$D^{(s_\ell,l)}$ \cite{manwin94}, where $s_\ell$ is the total angular momentum 
of the light degrees of freedom
and $l$ the corresponding orbital angular momentum 
 of the charm meson ($ s_\ell = l \pm 1/2$).
At the leading order, the matrix element
\be
<D^{(s_\ell,l)} (v' )| J_\mu | B(v)>
\ee
appearing in the semileptonic transition is described by a single form factor
$\xi^{(s_\ell,l)} (v \cdot v' )$: the Isgur-Wise function for the $B \to 
D^{(*)}$
transitions is the function $\xi^{(1/2,0)}\equiv\xi$; for the $P$-wave heavy 
mesons
we have $\xi^{(3/2,1)}\equiv\tau_{3/2}$ and $\xi^{(1/2,1)}\equiv\tau_{1/2}$; 
they have been computed by QCD sum rules in \cite{cola92}, and by constituent 
quark models in \cite{veseli}.

\subsection{Chiral corrections}
\label{subsec:chircorr}

Violations to $SU(3)$ symmetry can be computed by means of
the effective heavy meson chiral lagrangian. 

To estimate the size of the chiral corrections, it is common practice, as we 
have stressed already, to retain only the non-analytic terms
arising from chiral loops. 
Moreover when the subtraction scale $\mu$ is of order of the chiral symmetry
breaking scale $\Lambda_{\chi} \approx 1 \; GeV$, 
the coefficients of the higher order terms do
not contain large logarithms, and therefore the numerical estimates are 
carried
out at this scale.

Chiral perturbation theory has been used to compute the leading corrections 
to the form factors for $B \to D (D^* )$ semileptonic decays, arising from
the chiral loops of figure \ref{fig:iwcorr}.
\begin{figure}[htbp]
\epsfxsize=12cm
\epsfysize=7cm
\epsfbox{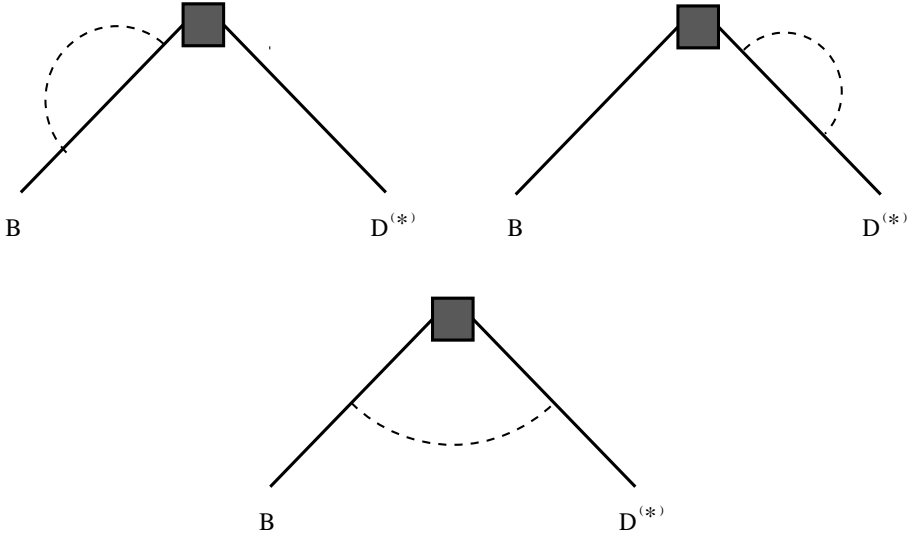}
\vspace{3mm}
\caption{Diagrams for the calculation of  
one-loop chiral corrections to the $B \to D^{(*)}$
transition matrix element. The box represents the $b \to c$ weak current, the
dashed line a light pseudoscalar.}
\label{fig:iwcorr}
\end{figure} 
The dominant corrections at zero recoil, i.e. 
$v = v'$, are of special interest and have been computed in 
\cite{goity92,randall93}.
According to Luke's theorem, these corrections appear at the order 
$1/m_Q^2$. 
This class of corrections should not be confused with those
coming from the $1/m_Q^2$ terms present in the
effective lagrangian or in the current: the $1/m_Q$ terms in the lagrangian,
in particular the one 
responsible for the hyperfine mass splitting $P^* - P$
and the one giving the splitting between the couplings $g_D$ and $g_B$, 
generate at one-loop $1/m_Q^2$ non-analytic corrections. The effect of the
$g_B$ and $g_D$ splitting has been neglected in \cite{goity92,randall93}.
For instance, the $B \to D^{(*)}$ matrix element at the recoil point $v = v'$,
as computed by the formulas of appendix A,
is \cite{randall93}
\bea
<D(v) | J^{\bar c b}_\mu|B(v)> &=& 2v_\mu\left[ 1 + C(\mu)/m_c^2 +\right.
 \nonumber \\
& - & \left. {3g^2\over2}\left({\Delta_c\over4\pi f}\right)^2\left[ 
f(\Delta_c/m_\pi)
+\log(\mu^2/m_\pi^2)\right] \right] \nonumber
\\
<D^*(v,\epsilon) | J^{\bar c b}_\mu|B(v)> &=&
2\epsilon^*_\mu\left[ 1 + C'(\mu)/m_c^2 + \right. \nonumber \\
& - & \left.  {g^2\over2}\left({\Delta_c\over4\pi
f}\right)^2\left[ f(-\Delta_c/m_\pi) +\log(\mu^2/m_\pi^2)\right] 
\right]
\label{bdcorr}
\eea
where $C$ and $C'$ stand for tree level counter-terms and
\be
f(x)=\int_0^\infty dz {z^4\over (z^2+1)^{3/2}}
\left(\frac{1}{[(z^2+1)^{1/2}+x]^2}-
\frac{1}{z^2+1}\right)~~.
\ee

In (\ref{bdcorr}) only the dependence on $\Delta_c = M_{D^*} - M_D$
has been kept, discarding the $\Delta_b$ terms and those
 proportional to the $g_B - g_D$ splitting. 
Numerically, for $\mu = 1 \; GeV$ and $g = 0.38$, the correction from
the logarithmically enhanced term in (\ref{bdcorr}) is $ -0.6 \%$, and the
correction from $f(x)$ is $0.3\%$.

A complete calculation of the $1/m_Q$ and $SU(3)$ breaking corrections
to the $B_a \to D^{(*)}_a l \nu_l$ process has been performed in 
\cite{boyd95}.
This analysis includes non-analytic terms arising from chiral loops and the
analytic counterterms, but it lacks predictive power  due to the introduction
of many unknown effective parameters. 

In the $SU(3)$ limit, the Isgur-Wise function is independent of light quark
flavor of the initial and final mesons, i.e. 
\be
\xi_u =\xi_d = \xi_s~~, 
\label{idfun}
\ee
 where
$\xi_{u,d,s}$ is the Isgur-Wise function occurring respectively in
$B_{u,d,s}$ decays.
In \cite{jenkins92,goity92}
the leading corrections to the
equality (\ref{idfun})    have been computed in chiral perturbation theory,
giving \cite{goity92}
\bea
\frac{{\xi_{s}(v\cdot v')}}{{\xi_{u,d}(v\cdot v')}} & = &
1+ \frac{g^2\Omega(v\cdot v')}{16 \pi^2 f_{\pi}^2} \left[ m_K^2 
 \log\left(m_K^2/ \mu^2 \right) \right. \nonumber \\
& + & \left. \frac{1}{2} m_\eta^2 
\log\left(m_\eta^2/ \mu^2 \right) - \frac{3}{2} m_{\pi}^2
\log\left(m_\pi^2/ \mu^2 \right)\right]
\label{fraciw}
\eea
where
\bea
\Omega(x)& = &
 -1+\frac{2+x}{2\, \sqrt{x^2-1}} \;\;
 \log \left( \frac{x+1+\sqrt{x^2-1}}{x+1-\sqrt{x^2-1}}
\right)\nonumber \\
& + &  \frac{x}{4\,\sqrt{x^{2}-1}}\;\;
\log\left( \frac{x-\sqrt{x^2-1}}{x+\sqrt{x^2-1}} \right)~~.
\eea

In (\ref{fraciw})  the analytic counterterms are neglected. Numerically, 
the nonanalytic chiral correction is  a few percent.

We mention here another calculation in the framework of the heavy meson
chiral perturbation theory, the ratio of the parameters $B_{B_s}$ and
$B_{B}$, entering in the analysis of $B_{(s)} - {\bar B_{(s)}}$ mixing
and defined as:
\bea
<\bar B(v)| \bar b \gamma_\mu (1-\gamma_5) d
\ \bar b \gamma^\mu (1-\gamma_5) d | B(v)> &=& {8\over 3} f_B^2 B_B \\
<\bar B_s(v)| \bar b \gamma_\mu (1-\gamma_5) s
\ \bar b \gamma^\mu (1-\gamma_5) s | B_s(v)>&=&
 {8\over 3} f_{B_s}^2 B_{B_s} 
\eea

 In the chiral symmetry limit
$B_{B_s}/B_{B}=1$.  For non-zero strange quark mass, the ratio is no
longer equal to $1$, and the one loop chiral corrections, arising from
the diagrams of figure \ref{fig:bbar}, are \cite{lepratio}:
\bea
{B_{B_s} \over B_{B} } & = & 1 - {2\over 5}\frac{\left(1 - 3g^2\right)}{16 
\pi^2
f_\pi^2}
\left[ m_K^2 \log\left(m_K^2/ \mu^2 \right) + \right. \nonumber \\
&  + & \left. \frac{1}{2} m_\eta^2 
\log\left(m_\eta^2/ \mu^2 \right) - \frac{3}{2} m_{\pi}^2
\log\left(m_\pi^2/ \mu^2 \right) \right]
\eea
 
Using $\mu = 1 \; GeV$ and $g\simeq 0.38$ the previous formula gives
$B_{B_s}/B_B \simeq 1.03$.
\begin{figure}[htb]
\epsfxsize=12cm
\epsfysize=2cm
\epsfbox{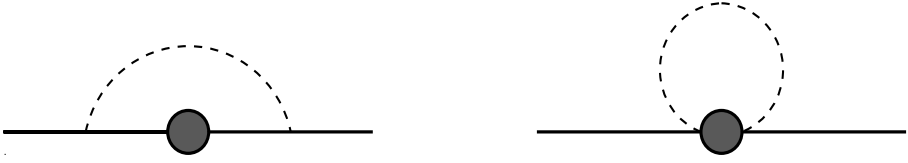}
\vspace{3mm}
\caption{Diagrams appearing in the calculation of
one-loop chiral corrections to the $B-{\bar B}$
mixing. The self-energy diagrams are not shown. The dot represents the $\Delta 
B =2$
operator}
\label{fig:bbar}
\end{figure}

\subsection{The $B \to D^{(*)} \pi l \nu_l$ decay}

Another application of the chiral lagrangian can be found in the semileptonic 
decays of $B$  into a charmed meson with the emission of a single soft pion, i.e.
$B \to D^{(*)} \pi l \nu_l$.
The phenomenological heavy-to-heavy leading current \cite{hq5}:
\be
J_{\mu}^{cb} = -\xi (v \cdot v' ) < {\bar H}^{(c)}_a \gamma_{\mu}
(1 - \gamma_5 ) H^{(b)}_a >
\label{effcurrf}
\ee
does not depend on the pion field, and therefore the amplitude with emission 
of
a single pion is dominated by pole diagrams, where the pion is emitted by
the initial $B$ or the final $D^{(*)}$, and is proportional to the 
coupling $g$. The two diagrams are shown in figure \ref{fig:bdpi}.
\begin{figure}[htb]
\epsfxsize=12cm
\epsfysize=2.3cm
\epsfbox{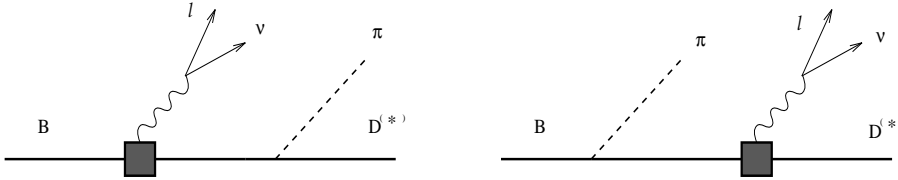}
\vspace{3mm}
\caption{Pole diagrams for the $B \to D^{(*)}\pi l \nu_l$ decay. The square 
represents the $b \to c$ current, the dashed line is the pion}
\label{fig:bdpi}
\end{figure} 
These decays might be used to determine the value of $g$,
or even to test the heavy quark flavour symmetry prediction 
for the $D^* D \pi$ and $B^* B \pi$ vertices $g_B = g_D =g$.
Moreover these processes may give indications on resonance effects. 
 
 The chiral calculation is reliable only in the kinematical
region of soft pions. 
In  the decay $B \to D \pi l \nu_l$, the soft pion domain is a large fraction 
due to the inclusion of the cascade decay $B \to D^* l \nu_l \to
D \pi l \nu_l$. This process has been treated by various authors 
\cite{lee92,cheng93,kramer93}, and the analysis has been extended to
$B \to D^* \pi l \nu_l$ in \cite{lee93,cheng93,goity95}: in \cite{goity95},
in addition to the ground state mesons $D, D^*, B$ and $B^*$,
also the contribution of the low-lying positive parity $0^+$ and $1^+$
resonances and some radially excited states is estimated.

\subsection{The heavy-to-light effective current}
\label{sec-effcurr}

The weak current for the transition from a heavy to a light quark, $Q \to 
q_a$, is given at the quark level by ${\bar q}_a \gamma_\mu (1 - \gamma_5) Q$; 
when written in terms of a heavy meson and light pseudoscalars \cite{wise}, it 
assumes the form, at the lowest order in the light meson derivatives,
\be
L^{\mu}_a = \frac{ i {\hat F}}{2} <\gamma^{\mu} ( 1 - \gamma_5 )H_b
\xi^{\dagger}_{ba}>~~.
\label{c1}
\ee

This operator transforms as $({\bar 3}_L , 1_R )$ under $SU(3)_L \times
SU(3)_R$, i.e. analogously to the quark weak current, 
and is uniquely defined at this order
in the chiral expansion. 

 From the definition of decay constant of a heavy meson $P$
\be
<0| {\bar q}_a \gamma^\mu \gamma_5 Q|P_b (p)>  = i p^\mu f_{P_a} \delta_{ab}
\label{c2}
\ee
one gets
\be
f_{P_a} = \frac{{\hat F}}{\sqrt{M_P}}~~.
\label{c3}
\ee

We note that in the infinite quark mass limit, $M_{P_a} \to m_Q$, and there is
no dependence on the light flavour.
The previous formula shows the $1/\sqrt{m_Q}$ scaling of the heavy meson
leptonic decay constant in the $m_Q \to \infty$ limit, and its light-flavour
independence in the chiral limit (we neglect the small
logarithmic dependence of $\hat F$ on $m_Q$). In section \ref{subsec:QCD}
 we have already discussed the various determinations of ${\hat F}$, 
see eqs. (\ref{eq:eqf}), (\ref{eq:eqf2}).

Higher derivative, spin breaking, and $SU(3)$ breaking current operators
are written explicitly in \cite{Boyd-Grin}: their introduction adds many
unknown effective parameters, and they correct the leading behaviour
(\ref{c3}). Lattice calculation \cite{latfp} and QCD sum rules
\cite{neubert1,Ball} indicate that the $1/m_Q$ corrections 
are sizeable at least for $f_D$.

The current describing  weak interactions 
between pseudoscalar Goldstone bosons and the positive parity $S$ fields
is introduced in a similar way:
\be
\hat {L}^{\mu}_a = \frac {i\hat{F}^+} {2} <\gamma^{\mu} (1-\gamma_5) S_b
\xi^{\dagger}_{ba}>
\label{c4}
\ee
The analysis done in \cite{cola92}, based on QCD sum rules, gives for ${\hat
F}^+$:
\be
{\hat F}^+ \simeq 0.46 \; GeV^{3/2}~~.
\label{c4bis}
\ee

The current describing the interaction of the  $H$ fields  with the light 
vector 
mesons, is, at
the lowest order in the derivatives:
\bea
L_{1 a}^{\mu} & = & 
\alpha_1 <\gamma^{\rho} (1 -\gamma_5) H_b v_{\rho}
 (\rho^{\mu}-V^{\mu})_{bc} \xi^{\dagger}_{ca}> + \nn \\
& + & \alpha_2 <\gamma^{\mu} (1 -\gamma_5) H_b v_{\rho}
 (\rho^{\rho}-V^{\rho})_{bc} \xi^{\dagger}_{ca}> + \nn \\
& + & \alpha_3 <\gamma^{\rho} (1 -\gamma_5) H_b v^{\mu}
 (\rho_{\rho}- V_{\rho})_{bc} \xi^{\dagger}_{ca}>~~. 
\label{c5}
\eea

The current (\ref{c5}) is of the next order as compared
 to the currents
(\ref{c1}) and (\ref{c4}), and does not contribute to the leptonic decay
constant $f_P$. As we shall see below, the term in (\ref{c5})
proportional to $\alpha_1$ contributes 
in a leading way to the $A_1$ form factor in
the $P \to \rho$ semileptonic matrix element, while the terms proportional
to $\alpha_2$ and $\alpha_3$ contribute to the $A_2$ form factors, but they 
are
subleading with respect to the pole diagram contribution. 

We  observe that there is no similar coupling between the fields 
$T^{\mu}$, defined in (\ref{r:tmu}) and $\xi$. 
Indeed (\ref{c1}) and (\ref{c4}) 
also describe the matrix element 
between the meson and the vacuum, and this coupling vanishes for the $1^+$ and 
$2^+$ states having $s_{l}=3/2$. This can be proved explicitly by 
considering the current matrix element ($A^{\mu}=\overline{q}_a \gamma^{\mu} 
\gamma_5 Q$):
\be
<0|A^{\mu}|\tilde{D}_1> = \tilde{f} \epsilon^{\mu}~~,
\label{c6}
\ee
where $\tilde{D}_1$ is the $1^+$ partner in the $s_l = 3/2$ multiplet.
Using the heavy quark spin symmetry, 
(\ref{c6}) turns out to be proportional to the matrix element of the vector 
current between the vacuum and the $2^+$ state, which vanishes.

\subsection{Chiral corrections for $f_{P_s}/f_P$}
\label{sec-fpsfp}

In the chiral limit, the leptonic decay constant does not depend on the light
flavour, i.e.
\be
\frac{f_{P_s}}{f_P} =1~~.
\label{c7}
\ee

As discussed in section (\ref{sec:chirg}), one can obtain an estimate of the 
$SU(3)$ violations by computing the non-analytic terms arising from the chiral 
loops.

The one-loop diagrams contributions to the leptonic decay constant $f_P$ are 
shown in fig. \ref{fig:fp}, and have been computed in
\cite{lepratio,goity92}, keeping only the ``log-enhanced'' terms of
the form $m^2 \log (m^2/\mu^2 )$. For the ratio $f_{D_s}/f_D$ one has:
\be
\frac{f_{D_s}}{f_D}  =  1 - \frac{1}{32 \pi^2 f_\pi^2} \left[
m_K^2 \log (\frac{m_K^2}{\mu^2}) +\frac{1}{2}
m_\eta^2 \log (\frac{m_\eta^2}{\mu^2}) -\frac{3}{2}
m_\pi^2 \log (\frac{m_\pi^2}{\mu^2}) \right] 
( 1 + 3 g^2)~~. 
\label{c8}
\ee

The corrections proportional to $g^2$ arise from the self-energy diagrams, 
fig.
\ref{fig:fp}b, while  the diagram \ref{fig:fp}c gives the $g$-independent
corrections. The diagram \ref{fig:fp}d, linear in $g$, vanishes at the leading
order in $1/m_Q$. 
Using $g \simeq 0.38$ in (\ref{c8}), one gets $f_{D_s}/f_{D} \simeq 1.11$.
\begin{figure}[htb]
\epsfxsize=12cm
\epsfysize=6.3cm
\epsfbox{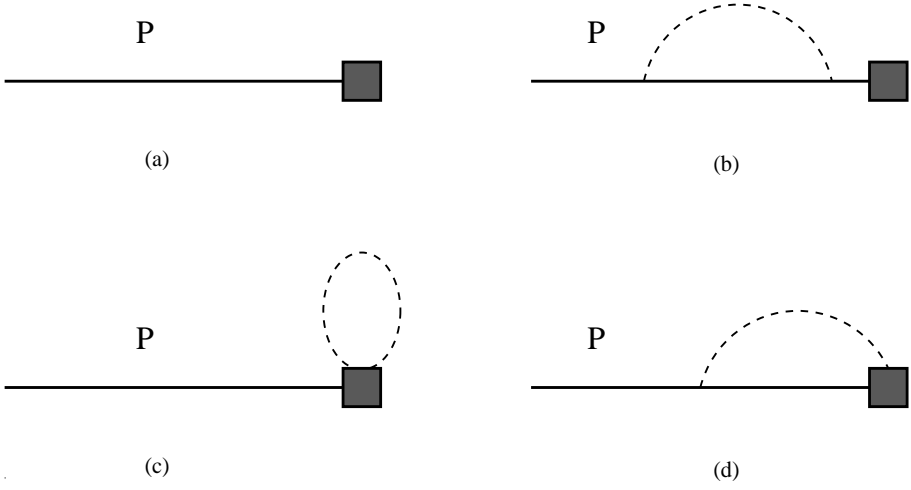}
\vspace{3mm}
\caption{Diagrams for the calculation of 
one-loop chiral corrections to the heavy meson
leptonic decay constant $f_P$. The box represents the $b \to q$ weak current,
 the dashed line a light pseudoscalar.}
\label{fig:fp}
\end{figure} 
The excited positive parity heavy mesons  contribute
to $SU(3)$ violating effects as virtual
intermediate states in chiral loops.
In Ref.\cite{falk93} the ``log-enhanced'' 
terms due to these excited-state loops
have been computed: some of them are proportional
to $h^2$ and others depend linearly on $h$, $h$ being the coupling
of the vertex $P^{**} P \pi$.
 In \cite{falk93} it has been  pointed out  that these
terms could be numerically relevant and could invalidate the chiral
estimate based only on the states $D$ and $D^*$; 
as discussed in section \ref{sec:strong} 
the coupling $h$ is estimated by QCD sum rules 
in \cite{faz}, with the result $h \simeq -0.5$, see eqs. (\ref{h1b}, \ref{h2b}),
where a more accurate chiral computation of the ratio $f_{D_s}/f_D$ is 
performed. 
We present here some details of the calculation: the vertices and 
the integrals needed for the
loop integration can be found respectively in appendices A and B. 

The self-energy diagrams \ref{fig:fp}b give the following 
wave function renormalization
factors:
\begin{eqnarray}
Z_D & =& 1 - \frac{3 g^2}{16 \pi^2 f_\pi^2} \left[3/2 
C_1(\Delta_{D^*D},\Delta_{D^*D},m_\pi ) +
C_1(\Delta_{D^*_s D},\Delta_{D^*_s D},m_K ) + 
\frac{1}{6} C_1(\Delta_{D^*D},\Delta_{D^*D},m_\eta ) \right] \nonumber \\
& + &  \frac{ h^2}{16 \pi^2 f_\pi^2} \left[3/2 
C(\Delta_{P_0 P},\Delta_{P_0 P},m_\pi ) +
C(\Delta_{P_{0s} P},\Delta_{P_{0s} P},m_K ) + \right. \nonumber \\
& + & \left. \frac{1}{6} 
C(\Delta_{P_0 P},\Delta_{P_0 P},m_\eta ) \right] 
\end{eqnarray}
\begin{eqnarray}
Z_{D_s} & = & 1 - \frac{3 g^2}{16 \pi^2 f_\pi^2} \left[
2 C_1(\Delta_{D^* D_s},\Delta_{D^* D_s},m_K ) 
+ \frac{2}{3} C_1(\Delta_{D^*D},\Delta_{D^*D},m_\eta ) \right] \nonumber \\
& + &  \frac{ h^2}{16 \pi^2 f_\pi^2} \left[ 
2 C(\Delta_{P_{0} P_s},\Delta_{P_{0} P_s},m_K ) + \frac{2}{3} 
C(\Delta_{P_0 P},\Delta_{P_0 P},m_\eta ) \right] 
\end{eqnarray}
where the mass splittings $\Delta_{P^* P}= M_{P^*}- M_{P}$, 
$\Delta_{P^* P_s}= M_{P^*}- M_{P_s}$, and
$\Delta_{P_s^* P}= M_{P_s^*}- M_{P}$  are  ${\cal O}(1/m_Q)$, while the mass
splittings
$\Delta_{P_0 P}= M_{P_0}- M_{P}$,
$\Delta_{P_{0s} P}= M_{P_{0s}}- M_{P}$, and
$\Delta_{P_0 P_s}= M_{P_0}- M_{P_s}$ between excited and ground states 
are finite in the limit $m_Q \to \infty$.
 
The functions 
$C_1$ and $C$ come from the loop integration and are defined in appendix B 
(here
we use $\hat\Delta=0$).

The diagram \ref{fig:fp}c gives the same contribution as in (\ref{c8}),
while
the diagram \ref{fig:fp}d is linear in $h$ (the analogous term proportional to 
$g$
vanishes), and proportional to ${\hat F}^+$:
combining all the diagrams one obtains \cite{faz}:
 
\begin{eqnarray}
 f_{D} & =& \frac{{\hat F}}{\sqrt{M_D}} \left[ 1 -
 \frac{1}{32\pi^2 f_\pi^2} \left[\frac{3}{2} m_\pi^2 
\log (\frac{m_\pi^2}{\mu^2}) + 
 m_K^2 \log (\frac{m_K^2}{\mu^2})+ \frac{1}{6} m_\eta^2 
\log (\frac{m_\eta^2}{\mu^2})\right] \right.\nonumber \\ 
 & - &  \frac{3 g^2}{32 \pi^2 f_\pi^2} \left[\frac{3}{2} 
C_1(\Delta_{D^* D},\Delta_{D^* D},m_\pi ) +
C_1(\Delta_{D^*_s D},\Delta_{D^*_s D},m_K ) +
 \frac{1}{6} C_1(\Delta_{D^* D},\Delta_{D^* D},m_\eta ) \right] \nonumber \\
& + &  \frac{ h^2}{32 \pi^2 f_\pi^2} \left[
\frac{3}{2} C(\Delta_{P_{0} P},\Delta_{P_{0} P},m_\pi ) + 
 C(\Delta_{P_{0s} P},\Delta_{P_{0 s} P},m_K ) + \frac{1}{6} 
C(\Delta_{P_0 P},\Delta_{P_0 P},m_\eta ) \right] \nonumber \\
& + & \left.\frac{{\hat F}^+}{{\hat F}} \frac{h}{16\pi^2 f_\pi^2} \left[
\frac{3}{2} 
C(\Delta_{P_0 P},0,m_\pi ) +
C(\Delta_{P_{0s} P},0,m_K ) + \frac{1}{6} 
C(\Delta_{P_0 P},0,m_\eta )\right] \right]
\label{fd}
\end{eqnarray}

\begin{eqnarray}
 f_{D_s} & =& \frac{{\hat F}}{\sqrt{M_D}} \left[ 1 -
 \frac{1}{32\pi^2 f_\pi^2} \left[ 
2 m_K^2 \log (\frac{m_K^2}{\mu^2})+ \frac{2}{3} m_\eta^2 
\log (\frac{m_\eta^2}{\mu^2})\right] \right.\nonumber \\ 
 & - &  \frac{3 g^2}{32 \pi^2 f_\pi^2} \left[
2 C_1(\Delta_{D^* D_s},\Delta_{D^* D_s},m_K ) +
 \frac{2}{3} C_1(\Delta_{D^* D},\Delta_{D^* D},m_\eta ) \right] \nonumber \\
& + &  \frac{ h^2}{32 \pi^2 f_\pi^2} \left[ 
2 C(\Delta_{P_{0} P_s},\Delta_{P_{0} P_s},m_K ) + \frac{2}{3} 
C(\Delta_{P_0 P},\Delta_{P_0 P},m_\eta ) \right] \nonumber \\
& + & \left.\frac{{\hat F}^+}{{\hat F}} \frac{h}{16\pi^2 f_\pi^2} \left[
2 C(\Delta_{P_{0} P_s},0,m_K ) + \frac{2}{3} 
C(\Delta_{P_0 P},0,m_\eta )\right] \right] \; .
\label{fds}
\end{eqnarray}

 From the previous formulas, using $\Delta_{P_0 P} = 0.5 \; GeV$, $\mu = 1$,
${\hat F}^+ = 0.46 \; GeV^{3/2}$ and ${\hat F} = 0.30 \; GeV^{3/2}$,
one gets numerically:
\begin{eqnarray}
f_D & = & \frac{{\hat F}}{\sqrt{M_D}} \left( 1 + 0.09 + 
0.003 g^2 -0.33 h^2 -1.00 h \right)
\label{fdnum}
\\
f_{D_s}& = &\frac{{\hat F}}{\sqrt{M_D}} \left( 1 + 0.17 + 0.59 g^2 - 
0.66 h^2 - 1.15 h \right) \; .
\label{fdsnum}
\end{eqnarray}
In the previous formulas we have kept only the leading order in the $1/m_Q$,
i.e. we have put $\Delta_{D^*D}=0$ in (\ref{fd}) and (\ref{fds}).

It is found that the terms ${\cal O}(h^2 )$ and ${\cal O}( h )$, 
while important, tend to cancel out in (\ref{fdnum}, \ref{fdsnum})
and that the ratio of leptonic decay constants  is numerically the same
as obtained from (\ref{c8}):
\be
\frac{f_{D_s}}{f_D} \simeq 1.10~~.
\label{c13}
\ee
These values are obtained by using $g=0.38$ and $h = -0.5$.

The formula (\ref{c8}) is valid at the leading order in $1/m_Q$, and
in this limit it is the same for $B$ and $D$ systems. In other terms,
the double ratio $R_1$
\be
R_1 = \frac{f_{B_s}/f_B}{f_{D_s}/f_D}
\label{c9}
\ee
is equal to $1$ in the chiral limit and in the heavy quark limit, separately.
To see  how $R_1$ deviates from unity one has to take into account the
$1/M$ terms in the chiral effective lagrangian and in the
effective current. As discussed in
\cite{Boyd-Grin}, four new  parameters contribute at the order $1/m_Q$ to 
the leptonic decay constants: two of them, $\rho_1$ and $\rho_2$,
come from the $1/m_Q$ terms in the current as 
\bea
L^{\mu}_a & = & \frac{ i {\hat F}}{2} (1 + \frac{\rho_1}{M_P} )
 <\gamma^{\mu} ( 1 - \gamma_5 )H_b\xi^{\dagger}_{ba}> + \nn \\
& + &  \frac{ i {\hat F}}{2} \frac{\rho_2}{M_P} 
 <\gamma_{\alpha} \gamma^{\mu} ( 1 - \gamma_5 )\gamma^{\alpha}
H_b\xi^{\dagger}_{ba}> 
\label{c10}
\eea
and they modify the leptonic decay constants as follows
\bea
\sqrt{M_P} f_P & = & {\hat F} \left( 1 + \frac{\rho_1 + 2 \rho_2}{M_P} \right)
\nn \\
\sqrt{M_P} f_{P^*} & = & {\hat F} \left( 1 + \frac{\rho_1 - 2 \rho_2}{M_P} 
\right)~~.
\label{c11}
\eea

The two parameters $\rho_1$ and $\rho_2$ can be related to the  HQET matrix
elements $G_K$ and $G_\Sigma$ defined in eq. (\ref{eq:aap}), 
and estimated by QCD sum rules in \cite{neubert1,Ball}.

The other two couplings, $g_1$ and $g_2$, 
have been already introduced in (\ref{1/m}), and they
 parameterize the $1/M$ corrections
to the couplings $g_{P^*P\pi}$ and $g_{P^* P^* \pi}$:
\bea
g_{P^*P^*\pi} & = & g + \frac{1}{M_P} (g_1 + g_2) \\
g_{P^*P\pi} & = & g + \frac{1}{M_P} (g_1 - g_2)~~. 
\eea

For the chiral correction to the double ratio $R_1$,
neglecting as usual the analytic counterterms, only the quantity
$g_1 - g_2$, i.e. the $1/M_P$ correction to $g_{P^*P\pi}$,
 is relevant, and one gets \cite{Boyd-Grin}
\be
R_1 -1 = -0.11 g^2 - 0.06\;  g (g_1 -g_2 )~GeV^{-1}~~.
\label{c12}
\ee

\section{Heavy-to-light semileptonic exclusive decays}
\label{chap-semilep}

Most of the known CKM matrix elements have been determined using semileptonic
decays. In particular, from semileptonic $B$ decays one can extract  
$V_{cb}$ and $V_{ub}$. 

The extraction of the value of $V_{cb}$ from the exclusive process $B \to D^* 
l \nu_l$
has been studied in the HQET context, and we mentioned it
before. 
The situation for $V_{ub}$ is apparently more uncertain, both for 
the inclusive and the exclusive semileptonic rates. Its determination
is one of the most important goals in $B$ physics, but it involves great 
experimental and theoretical difficulties.
At the moment there is a safe experimental evidence for $b \to u$
transitions, and  experimental data on the $B \to \pi l \nu_l$ and
$B \to \rho l \nu_l$ exclusive
 processes have been presented by the CLEO II collaboration
\cite{cleobrux}. 

The interpretation of the inclusive $ b \to u $ semileptonic rate is difficult
because of the dominant $b \to c $ background: to eliminate it, one works
beyond the end-point region of the lepton momentum spectrum for $b \to c $
processes. This is a very small fraction of the phase space, where theoretical
inclusive models have relevant uncertainties. 

Predictions for the exclusive channels $B \to X_u l \nu_l$ are also model 
dependent,
and here HQET is much less useful than in the $B \to D$ process,  
because of the presence of a light
meson in the final state. 
We shall show in the following sections how to relate 
$B \to \pi l \nu_l$ to  $D \to \pi l \nu_l$, and 
$B \to \rho l \nu_l$ to  $D \to \rho l \nu_l$, in the $b$ and $c$ infinite 
mass
limit. The main problem of this approach are the $1/m_c$ corrections, 
potentially relevant and not under control.

The effective lagrangian approach can shed light on these semileptonic decays 
and
can give indications on the values of the relevant form factors at the zero 
recoil 
point. In order to extract information from the experimental data the complete 
$q^2$ dependence of the form factors is required, which goes beyond the 
chiral lagrangian approach.
For this reason external inputs, either phenomenological or purely 
theoretical, are
required, and, in the next section, we shall discuss this issue in some 
details.

\subsection{Form factors}
\label{sec-formfact}

We introduce now form factors that parameterize the hadronic matrix
elements of the weak currents.

In the case of  semileptonic decays, $P \to P' l \nu_l$ ($P$, $P'$ 
pseudoscalar
mesons)
there is no contribution from the axial-vector part of the current and 
the matrix element can be written as
\be
<P'(p')|V^{\mu}|P(p)> =
 \big[ (p+p')^{\mu}+\frac{M_{P'}^2-M_P^2}{q^2} q^{\mu}\big] 
F_1(q^2) -\frac {M_{P'}^2-M_P^2}{q^2} q^{\mu} F_0(q^2)
\label{scalscal}
\ee
where $V_{\mu} = {\bar q}' \gamma_{\mu} Q$ and $q = p - p'$.
There is no singular behaviour at $q^2 = 0$ because $F_1 (0) = F_0 (0)$.

The form factor $F_1(q^2)$ can be associated, in a dispersion relation
approach, to intermediate states
with quantum numbers $J^P = 1^-$, and $F_0 (q^2)$ to states with
$J^P = 0^+$. In the limit of massless lepton, the terms proportional
to $q_\mu$ in (\ref{scalscal}) do not contribute to the rate, so that only
the form factor $F_1 (q^2)$ is relevant.

For the pseudoscalar to vector matrix elements also the axial-vector current
 contributes and four form factors are required: 
\bea
<V(\epsilon,p')| &(&V^{\mu}-A^{\mu})|P(p)> = 
\frac {2 V(q^2)} {M_P+M_V} 
\epsilon^{\mu \nu \alpha \beta}\epsilon^*_{\nu} p_{\alpha} p'_{\beta} \nn\\ 
&+& i  (M_P+M_V)\left[ \epsilon^{*\mu} -\frac{\epsilon^* \cdot q}{q^2}
q^\mu \right] A_1 (q^2) \nn\\  
& - & i \frac{\epsilon^* \cdot q}{(M_P+M_V)} \left[ (p+p')^\mu -
\frac{M_P^2-M_V^2}{q^2} q^\mu \right] A_2 (q^2) \nn \\
& + & i \epsilon^* \cdot q \frac{2 M_V}{q^2} q^\mu A_0 (q^2)~~, 
\label{v1}
\eea
where
\be
A_0 (0)=\frac {M_V-M_P}{2 M_V} A_2(0) + \frac {M_V+M_P} 
{2M_V} A_1(0)~~.
\label{v2}
\ee

Neglecting the lepton mass, only the form factors $V(q^2)$, $A_1 (q^2)$ and
$A_2(q^2)$ contribute to the decay rate. The form factors $A_1$ and $A_2$ can
be associated to $J^P = 1^+$ intermediate states, and $V$ to 
$J^P = 1^-$ states.

The form factor dependence on $q^2$ is still an open question, and, at
the present time, there is
no general  theoretical agreement.
Quark model calculations are based on meson wave functions,
generally derived by some wave equation, and make use of them  to compute 
hadronic matrix elements. These calculations are normally reliable only at 
some 
specific value of $q^2$, and the dependence on the variable $q^2$ has to be 
assumed 
as an additional hypothesis.
The physical region for semileptonic decays covers the range
$0 \leq q^2 \leq q^2_{max}= (M_P - m_\pi )^2$
(in the limit of massless leptons). Close to $q^2_{max}$, 
the form factors should be dominated by the nearest
$t$-channel pole, located at the mass of the lightest heavy meson exchanged in
that channel.  

With decreasing $q^2$, the influence of the pole becomes weaker:
in a dispersion relation the form factor can be written as a $P^*$ pole
contribution ($P^*$ nearest pole for that channel)
 plus a continuum contribution, that, in the narrow width approximation,
reduces to a sum over higher resonances $P_n$. We can therefore 
write,
e.g., for the form factor $F_1$:
\be
F_1 (q^2)= \frac{f_{P^*} g_{P P^*\pi}}{q^2 - M^2_{P^*}} +
\sum_{n} \frac{f_{P_n} g_{P P_n\pi}}{q^2 - M^2_{P_n}}~~,
\label{n25i}
\ee
where $g_{P P^* \pi}$ is the trilinear coupling among $P$, $P^*$ and $\pi$.
In the combined limit $m_\pi \to 0$, $M_P \to \infty$, the $P^*$ pole
contribution goes like $M_P^{1/2}$ when $q^2 \to M^2_P$ ($g_{P P^* \pi} \sim
M_P$), while the higher resonances contributions go only like
$M_P^{-1/2}$, as they do not become degenerate with the $P$ in the heavy
mass limit. But far away from the kinematical end point $q^2_{max}$, many
resonances can in general contribute to the form factors. This observation
leads to two-component models for the form factors \cite{burdman92}.
In \cite{grinstein94} it is shown that the form factors for $B \to \pi$
semileptonic decay are dominated by the $B^*$ pole at all $q^2$ in
two-dimensional planar QCD; it remains to be seen if such dominance hold
also in four dimensions.

The nearest pole-dominance on the whole
$q^2$ range, i.e.
\be
F (q^2)  =  \frac{F(0)}{1 - q^2/M^2_{pole}}
\label{n7}
\ee
should therefore be taken as an additional assumption in building models,
as done for instance in the popular BSW model \cite{wsb85}. Other dependences
can be found in the literature; for example in
the ISGW model \cite{isgw89}, which  is expected to work well close to
$q^2_{\rm max}$, the extrapolation to lower $q^2$ is done by
an exponential dependence for the form factors.
It is important to stress that in general predictions for the widths
are not  sensitive to the assumed dependence only 
when the available range in $q^2$ is not
large, as in $D$ decays: for $B$ decays into light mesons, 
as $B \to \pi (\rho) l \nu_l$, different form factor behaviours can
lead to  different predictions.

Experimentally, only the
decay $D^0 \to K^- e^+ \nu_e$ allows at the moment 
a study of the $q^2$ dependence of the
form factor $F_1$. The data are compatible with the pole form (\ref{n7}),
but the precision is still poor. The value of the pole mass, as fitted by the
data, is compatible  with the $D^*_s$ mass, and 
the intercept $F_1 (0)$ is \cite{pdb}
\be
F_1^{D K} (0) = 0.75 \pm 0.03~~. 
\label{n8}
\ee

The Cabibbo-suppressed decay $D \to \pi l \nu_l$ suffers from poor statistics:
MARK III and CLEO II data, extracted with the pole-dominance
hypothesis, give\cite{pdb}:
\bea
\frac{F_1^{D \pi} (0)}{F_1^{D K} (0)} & = & 
1.0^{+ 0.3}_{- 0.2} \pm 0.04 \; \; {\rm MARK~III} \nn \\
& = & 1.3 \pm 0.2 \pm 0.1 \; \; {\rm CLEO~II}~~.
\label{n9}
\eea

Theoretically, QCD sum rules allow to compute the $q^2$ dependence of the
form factors, except when close to $q^2_{max}$. The analyses performed in 
\cite{bbd91,ball93,belyaev93,ali94,santo94,narison95} are  generally
compatible with nearest pole dominance for the vector current form
factor, $F_1$ and $V$: for the axial form factors, $A_1$ and $A_2$, there are
discrepancies among the different calculations.
In \cite{ball91,ball93}, the form factor $A_1^{B \to \rho}$ has an
unexpected behaviour, decreasing from $q^2 = 0$ to $q^2 = 15 \; GeV^2$:
for $A_2^{B \to \rho}$, a moderate increase in $q^2$ is found, at least
up to  $q^2 \simeq 15 \; GeV^2$. For higher values of $q^2$ the estimate
is unreliable.
Such a behaviour is the result of cancellations among large terms, and 
therefore
can suffer from relevant uncertainties.
Also, in \cite{narison95} the form factor $A_1$ decreases with $q^2$, while 
$A_2$ can be fitted by a pole formula.
Light-cone sum rules \cite{ali94} show, on the contrary, an increasing $A_1$, 
with a dependence close to the pole behaviour, and a steeper increase for $V$.

Current lattice QCD simulations cannot study directly the $b$ quark,
because its mass is above the UV cut-off. 
Quantities are computed around the charm scale, and then extrapolated
up to the $b$ mass using the Isgur-Wise scaling relations \cite{isgur}.
With this strategy, suggested in \cite{abada94}, one is forced to make
assumptions on the $q^2$-dependence at the $b$ scale, because the 
extrapolation
pushes the $q^2$ value towards $q^2_{max}$. 
For  $D$ meson form factors, the determination of  $q^2$ dependence is
still poor, but compatible with pole dominance \cite{uk95,ape95}.
Preliminary lattice computations of the $q^2$ dependence of the form factors 
$F_1$ and $F_0$ in $B \to \pi l \nu_l$ \cite{ukqcd95a}, and 
$A_1$ in $B \to \rho l \nu_l$ \cite{ukqcd95}, 
seem to favor a dipole/pole fit for $F_1$ and $F_0$ respectively,
and a pole behaviour for $A_1$: the data have 
however large uncertainties, and will be improved by
working with heavier quark masses and by using larger lattices. 

An additional constraint to the form factor $q^2$ dependence is
provided by nonleptonic heavy meson decays.
It has been shown that the commonly used form factors, when used together
with the additional hypothesis of factorization to evaluate non-leptonic decay
amplitudes,
do not agree with the data on $B \to J/\psi K (K^*)$ transitions
\cite{gourdin94,aleksan94}.
These non-leptonic decays can be computed, using the factorization
approximation, as functions of the leptonic decay constant $f_{J/\psi}$ and
the form factors $F_1^{B K}$, $A_1^{B K^*}$, $A_2^{B K^*}$, and
$V^{B K^*}$, at $q^2 = M^2_{J/\psi}$. The problem is to fit
simultaneously the rather small ratio of vector to pseudoscalar decay rates,
$\Gamma( B \to K^* J/\psi )/\Gamma( B \to K J/\psi )$, and the large fraction
of longitudinal polarization in $B \to J/\psi K^*$.

In \cite{gourdin95}, it has been shown that the discrepancy can be eliminated
allowing for a non-polar behaviour of some form factors. 
Using the Isgur-Wise scaling laws,
these authors compute the relevant form factors from the experimental data
on the semileptonic transitions $D \to K (K^*)$ at $q^2 =0$.
Subsequently they adopt for the form factors a generic dependence 
$(1- q^2/\Lambda^2)^{-n}$, with $n= -1,0,1,2$. Simple-pole
dominance corresponds to $n=1$. Three scenarios survive to the 
phenomenological
analysis:  the form factors $F_1$, $A_1$, $A_2$ and $V$
can only have a dependence $[n_F,n_1,n_2,n_V]=[+1,-1,n_2,+2]$ respectively,
and $n_2$ can be equal to $2,1,0$ (with a preference for $n_2 =2$).
Notice that $A_1(q^2)$ is linearly decreasing in these scenarios:
a similar behaviour is found also in a theoretical analysis \cite{ball93},
at least for $q^2 < 15 \; GeV^2$.

The hypothesis that $A_1$ is linearly decreasing on the whole range in
$q^2$ is certainly not valid: the form factor should have a pole at
$q^2 =M^2_{1^+}$, which should affect the $q^2$ dependence
at least close to the zero-recoil point.
In \cite{aleksan94} it is argued that $A_1$ should have a flatter $q^2$ 
dependence than the one predicted by the pole dominance.

In section \ref{sec-brho} we shall present a phenomenological analysis of the
$B \to V$ ($V$ light vector meson) processes, showing that a constant 
$A_1$ behaviour leads to discrepancies with the available data, when scaling
laws are used to scale the form factors from $D$ to $B$ systems:
the situation would be even worse for a decreasing $A_1$. There is thus
some rough suggestions for an increasing $A_1$ with $q^2$, and in section 
\ref{subsec-brhoeff}
a two-component model for $A_1$, a constant term plus a pole term, is
used with satisfactory phenomenological agreement. 

In any event, further theoretical and  experimental studies are needed to
clarify the problem of $q^2$ dependence in the form factors.

\subsection{$B \to \pi$ semileptonic decays}
\label{sec-bpi}

In this section we shall analyze the 
semileptonic exclusive decays of a heavy meson $P=B,\;D$ into a 
light particle belonging to the pseudoscalar octet, e.g. $\pi$.

The relevant hadronic matrix element is:
\be
<\pi| {\bar q} \gamma_\mu Q |P>~~.
\label{n1}
\ee
In absence of a knowledge of the hadronic current in terms of hadrons,
various theoretical approaches to the evaluation of (\ref{n1})
have been developed:  potential
models, for instance in \cite{wsb85,isgw89},  lattice QCD 
\cite{abada94,uk95,ape95,ukqcd95a}, 
QCD sum rules 
\cite{sumrul}, and the method based on the chiral and heavy quark symmetries,
that we shall review here. 

In general two attitudes are possible. In the first one, that we shall call the
 scaling approach, one relates the different hadronic matrix elements 
 (\ref{n1}) using the spin and flavour symmetries. In this way for instance
it is possible to relate the matrix elements  $<\pi| j_\mu|B>$ and
$<K| j_\mu|D>$, $\pi$ and $K$ belonging to the same chiral multiplet, and 
$B$ and $D$ being related by heavy flavour symmetry. 
In the second approach, one builds up an effective lagrangian incorporating 
chiral
and heavy quark symmetries, and  computes the form factors  within such 
framework.
The advantage of the second approach is the possibility to include in a
rigorous way symmetry breaking corrections, at least formally.
Both approaches lead to the same results at the leading order, and 
require some experimental or external 
inputs: in the scaling approach one starts with a known 
matrix element, while the effective lagrangian contains
unknown couplings that are determined by the data. 
As we shall see, some of these couplings have also
been estimated theoretically, and we shall review also these analyses.

\subsubsection{The scaling approach}
\label{subsec-bpiscal}

We shall discuss first the so-called Isgur-Wise scaling laws for the form
factors. Let us parameterize the hadronic matrix element as follows: 
\be
<\pi (p_{\pi})|J_\mu| P (p_P)> = (p_\pi + p_P )_\mu f_{+}^{P\pi}(q^2) +
(p_P - p_\pi )_\mu f_{-}^{P\pi}(q^2)
\label{n2}
\ee
where $q^2 = (p_P - p_\pi)^2$.
The Isgur-Wise relations, following from the $SU(2)$ flavour symmetry between
$b$ and $c$ quarks, give \cite{isgur}:
\bea
(f_+ + f_{-})  & \sim &  m_Q^{-1/2} \nonumber \\
(f_+ - f_{-})  & \sim &  m_Q^{+1/2}~~. 
\label{n3}
\eea
We have neglected here the logarithms of $m_Q$ arising from perturbative
QCD corrections. These scaling laws are valid as long as $v \cdot p_\pi$
does not scale with $m_Q$, i.e. in the kinematical regime of soft $\pi$,
close to $q^2_{max} = (m_P -m_\pi )^2$. In this region $E_\pi \simeq m_\pi <<
m_Q$. $q^2_{max}$ is the no recoil point, where the final scalar and the
dilepton system are at rest.

Using the form factors $F_1, F_0$, the eqs. (\ref{n3}) become, including QCD 
corrections,
\bea
F_1^{B\pi} (q^2_{max,B}) & = & \left[ \frac{\alpha_s (M_B)}
{\alpha_s( M_D)}\right]^{-6/25} \sqrt{\frac{M_B}{M_D}} 
F_1^{D\pi} (q^2_{max,D})
\label{n4}
\\
F_0^{B\pi} (q^2_{max,B}) & = & \left[ \frac{\alpha_s (M_B)}
{\alpha_s( M_D)}\right]^{-6/25} \sqrt{\frac{M_D}{M_B}} 
F_0^{D\pi}(q^2_{max,D}).
\label{n5}
\eea

Eqs. (\ref{n4}) and (\ref{n5}) are valid in the $m_b, \; m_c \to \infty$ 
limit.
The $1/m_c$ corrections can be large, as we will discuss later, and can be 
estimated in the effective lagrangian approach.

The application of the 
chiral symmetry is straightforward, and, at the leading order gives:
\be
F^{P\pi} (q^2) = F^{P\pi'} (q^2)~~.
\label{n6}
\ee
Here $\pi$ and $\pi'$ are two arbitrary light pseudoscalar mesons, 
for instance $\pi$ and
$K$. We notice that (\ref{n6}) is valid for any value of $q^2$.

 From the knowledge of the form factors  
at any value of $q^2$ for a given
 decay mode one can compute a whole class of decays as follows:
\begin{itemize}
\item[i)]
Using (\ref{n6}), one computes all the chiral-related decays. We notice that
 all the form factors related by light flavour symmetry
have the same $q^2$-behaviour,    
but $F_0$ and $F_1$ can have a different
behaviour.
\item[ii)]
The Isgur-Wise scaling laws (\ref{n4}) and (\ref{n6}) allow to relate
$B$ and $D$ form factors, at least close to $q^2_{max}$.
\item[iii)] 
The strongest assumption concerns the evolution in $q^2$ of the scaled form 
factors. 
The $B$ decay rates are quite sensitive to the 
explicit $q^2$ dependence, while for $D$ decays the $q^2$-range is much 
smaller. 
\end{itemize}

We proceed to the computation of the form factors, widths and
branching ratios for the semileptonic decays of a heavy meson into a light
scalar, following the strategy we have  described.
Such an approach was followed in \cite{casa92,xu93} for the semileptonic
decays, and in
\cite{gourdin94,aleksan94} to compute the $B \to K (K^*)$ form factors.

For the form factor $F_1$ we shall assume a simple pole behaviour, which, as
we have already discussed, agrees with present experimental and theoretical
evidence.
As an input, we use the decay $D \to K l \nu_l$, i.e. the form factor 
at $q^2=0$ (\ref{n8}). For chirally-related decays, we impose the
same value to the form factors at $q^2=0$. 
For generic  $q^2$ we have
\be
F_1 (q^2) = \frac{F_1(0)}{1 - q^2/M^2_{P*}}~~.
\label{n10}
\ee
The $1^-$ pole $P^*$ can be a strange or a non-strange heavy meson, depending
on the decay mode: their mass difference $\Delta_s= M_{P_s} - M_P \simeq 100 
\; MeV$
is a chiral breaking effect, which we will not neglect.
 
Let us make some comments on the Isgur-Wise scaling law, which follows from
the observation that the matrix element $<\pi (p_\pi)|J_\mu| P(v)>$ behaves as
$\sqrt{m_Q}$, in the limit $v\cdot 
p_\pi << m_Q \to \infty$. 
These asymptotic scaling laws have $1/m_Q$ corrections
\be
<\pi (p_\pi)|J_\mu| P(v)> \sim \sqrt{m_Q} \left( 1 + {\cal O} \left(
\frac{\Lambda_{QCD}}{m_Q} \right) + {\cal O} \left(
\frac{v \cdot p_{\pi}}{m_Q} \right) + {\cal O} \left(
\frac{m_{\pi}}{m_Q} \right)\right)~~,
\label{n11}
\ee
 and a number of  different choices for the scaling
relations are possible. They are all compatible with the asymptotic behaviour, 
but
differ for corrections of the type indicated in (\ref{n11}).
Eqs. (\ref{n4}) and (\ref{n5}) is a possible one, but other choices might be
done.
For instance we could have taken, for $F_1$, neglecting QCD corrections:
\be
\frac{F_1^{B\pi}}{F_1^{D\pi}}\Big|_{(q^2_{max})} = \sqrt{\frac{M_D}{M_B}}
\frac{M_B + m_\pi}{M_D + m_\pi}         
\label{n12}
\ee
which reduces to (\ref{n4}) in the $M_B,M_D \to \infty$ limit.
Notice that the ``soft scaling'' (\ref{n12}) can lead to results 
numerically different when the mass of the light final meson is not so small,
as for instance in the case of  $K$ or  $K^*$. In these cases the 
differences between (\ref{n4}) and (\ref{n12}) are of the order $m_K/M_D$,
or $m_{K^*}/M_D$. Other possible scaling forms can be found  in
\cite{gourdin94,aleksan94}.
QCD radiative corrections, written in (\ref{n4},\ref{n5}),
are of the order of 10\%: and 
therefore they can be neglected within our approximations.

The chiral limit, $m_\pi \to 0$, presents some subtleties.
In order to examine them, let us find the scaling laws at $q^2 =0$, as arising
from the  simple-pole behaviour (\ref{n10}) and the asymptotic scaling
at $q^2_{max}$. One finds:
\be
\frac{F_1^{B\pi}}{F_1^{D\pi}}\Big|_{q^2=0} \simeq \frac{M_D}{M_B}
\frac{\Delta_B + m_\pi}{\Delta_D + m_\pi}   
\frac{F_1^{B\pi}}{F_1^{D\pi}}\Big|_{q^2_{max}}\; .
\label{n13}
\ee
In the $m_b,m_c \to \infty$ limit, $\Delta_P = M_{P^*} - M_P \to 0$, and one
gets:
\be
\frac{F_1^{B\pi}}{F_1^{D\pi}}\Big|_{q^2=0} \simeq  \sqrt{\frac{M_D}{M_B}}~~.
\label{n14}
\ee
On the other hand, performing first the  $m_\pi \to 0$ limit in (\ref{n13}),
 one finds:
\be
\frac{F_1^{B\pi}}{F_1^{D\pi}}\Big|_{q^2=0} \simeq  \left( 
\frac{M_D}{M_B}\right)^{
\frac{3}{2}}\; .
\label{n15}
\ee

The contradiction between (\ref{n14}) and (\ref{n15}) means a breaking of the
naive scaling laws at $q^2_{max}$ in the chiral limit: as shown by
\cite{isgur}, in the limit $m_\pi \to 0$ (\ref{n4}) becomes:
\be
\frac{F_1^{B\pi}}{F_1^{D\pi}} \Big|_{q^2_{max}} \simeq  
\left( \frac{M_B}{M_D}\right)^{\frac{3}{2}}~~.
\label{n16}
\ee
This behaviour can be explained, as we shall see, from the polar diagram with
exchange of the $P^*$.
Therefore, combining (\ref{n13}) and (\ref{n16}), 
we find also in the chiral limit the scaling law (\ref{n14}), which we shall
use in the subsequent analysis.

Using (\ref{n8}) as an input, we get:
\be
F_1^{B K} (0) \simeq 0.45~~.
\label{n17}
\ee
As previously discussed, the uncertainties due to scale corrections
are expected to be of the 
order $m_K/M_D$,
i.e. about 30 \%.
The uncertainties due to deviations from the 
polar behaviour of the form factor are hard to
estimate and essentially unknown.

We could  use as an input 
the Cabibbo-suppressed decay $D \to \pi$, which has however larger
experimental errors. The scaling uncertainties, moreover, even if probably
smaller than in the case of a final $K$, are expected to
 be of order ${\bar \Lambda}/M_D$ (${\bar \Lambda} = M_P - m_Q$) and 
not of order $m_\pi/M_D$ only; therefore they could be really significant. 

The  prediction for widths and branching ratios following from (\ref{n17})
and chiral symmetry are reported in table \ref{bpi1}.
\begin{table}[htb]
\caption{Predictions for semileptonic $D$ and $B$ decays in a pseudoscalar 
meson, in the scaling approach. We have neglected the $\eta ~- \eta '$ mixing.
The branching ratios and the widths for $B$ must be multiplied for 
$|V_{ub}/0.0032|^2$. We assume $\tau_{B_s}=\tau_{B^0} =\tau_{B^+}= 1.55~ps.$}
\label{bpi1}
\centering
\begin{minipage}{10truecm}
\begin{tabular}{l c c c}
Decay & $F_1 (0)$  & BR & exp. BR
 \\ \hline
$D^0 \to \pi^{-}$ & 0.75 &   $3.4 \cdot 10^{-3}$ & 
$(3.9^{+ 2.3}_{ - 1.2})\cdot 10^{-3}$ \cite{pdb}\\
$D^+ \to \eta$ & 0.31 & $6.9 \cdot 10^{-4}$\\
$D_s \to \eta$ & 0.61 &  $3.2 \cdot 10^{-2}$ & \\
$D_s \to K^0$ & 0.75 &  $4.2 \cdot 10^{-3}$ & \\
$B^0 \to \pi^- $ & 0.45 &  $2.2 \cdot 10^{-4}$ &
$(1.63 \pm 0.46 \pm 0.34)\cdot 10^{-4}$\cite{cleobrux} \\
$ B_s \to K $ & 0.45 &  $2.2 \cdot 10^{-4}$ & \\
\end{tabular}
\end{minipage}
\end{table}

The available experimental data shown in  table \ref{bpi1} 
have large uncertainties.
Concerning the $B^0 \to \pi^- l^+ \nu_l$ decay, the CLEO II collaboration
\cite{cleobrux}
quotes  for the branching ratio two different values, depending on the model 
used for the  detector efficiency: in the previous table \ref{bpi1}
we have put the
number corresponding to the BSW model \cite{wsb85}
\be
BR(B^0 \to \pi^- l^+ \nu_l) = (1.63 \pm 0.46 \pm 0.34)\cdot 10^{-4}
\ee
 while the value corresponding
to the ISGW model \cite{isgw89} is
\be
BR(B^0 \to \pi^- l^+ \nu_l) = (1.34 \pm 0.35 \pm 0.28)\cdot 10^{-4}~~.
\ee

\subsubsection{Effective lagrangian approach}
\label{subsec-bpieff}
 
We now discuss  the effective chiral lagrangian approach to the 
semileptonic heavy-light form factors. 

We have already presented the effective lagrangian that combines heavy quark
and chiral symmetry and describes the low-momentum interactions
of heavy mesons with light pseudoscalars, and the chiral representation in
terms of meson fields of the weak current 
${\bar q} \gamma_\mu (1 - \gamma_5)Q$. 
In this framework one can compute the hadronic matrix elements 
$<\pi|J_\mu|P(v)>$
in terms of
the effective couplings of the lagrangian and of the weak current, at least in 
the
soft-pion region, i.e. close to $q^2_{max}$. 
Two diagrams contribute to the form factors, at least in the leading order:
the $P^*$ pole diagram, proportional to the strong coupling constant $g$, and 
a 
direct diagram, as shown in fig. \ref{fig:semilep}.
\begin{figure}[htb]
\epsfxsize=12cm
\epsfysize=3cm
\epsfbox{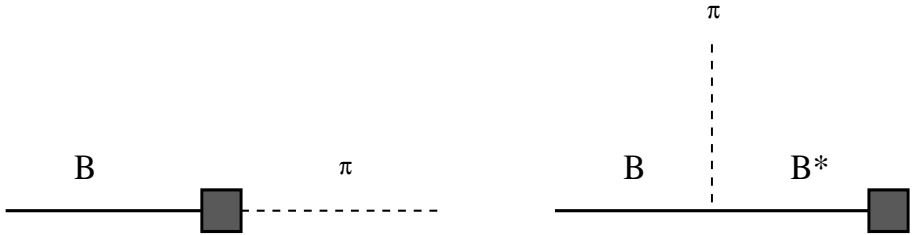}
\vspace{3mm}
\caption{Tree diagrams for the $B \to \pi$
transition matrix element. The box represents the $b \to u$ weak current, the
dashed line the pion.}
\label{fig:semilep}
\end{figure}   
At the leading order in $1/m_Q$ they give \cite{wise,burdman92}:
\bea
F_1 & = & \frac{g {\hat F} \sqrt{M_P}}{2 f_{\pi} ( v \cdot p_\pi + \Delta_P)} 
\label{n18}
\\
F_0 & = & \frac{{\hat F}}{f_\pi \sqrt{M_P}}~~.
\label{n19}
\eea

The same relations have been obtained assuming $P^*$ pole dominance in
\cite{isgur}, and in \cite{wolfenstein92}, combining PCAC and heavy quark
spin symmetry.
The formula (\ref{n18}) satisfies the asymptotic scaling (\ref{n4}): in the
chiral limit $v \cdot p_\pi \to 0$, however, the scaling is modified as in
(\ref{n16}), because $\Delta_P \sim 1/M_P$. 
The scaling at $q^2 =0$ is easily derived from (\ref{n13})
\be
\frac{F_1^{B\pi}}{F_1^{D\pi}}\Big|_{(q^2=0)} \simeq 
\sqrt{ \frac{M_D}{M_B}}
\label{n20}
\ee
as in (\ref{n14}). Therefore, if we use the input $D \to K$ to fix
the effective coupling $g {\hat F}$ in (\ref{n18}), we obtain the same results
as in table \ref{bpi1}, as expected. 

Alternatively, one can take the estimates $g\approx 0.38$ (\ref{eq:038}) and
${\hat F} = 0.30 \pm 0.05$ GeV$^{3/2}$ (\ref{eq:eqf}), which give
\bea
F_1^{D \pi} (0) & = &\frac{{\hat F} g}{f_\pi \sqrt{M_D}} \simeq 0.63    
\label{n23}\\
F_1^{B \pi} (0) & = & \frac{{\hat F} g}{f_\pi \sqrt{M_B}}\simeq 0.38 ~~.
\label{n24}
\eea
The uncertainties on these results, as arising from QCD sum rule 
approximations,
can be estimated around $30\%$.
We stress that at 
the charm scale the $1/m_c$ corrections are potentially relevant: nevertheless
the prediction (\ref{n23}) is in rather good agreement 
with data. In table \ref{bpi2} we quote form factors and branching ratios
obtained from (\ref{n18}), (\ref{n19}) and chiral symmetry.
\begin{table}[htb]
\caption{Predictions for semileptonic $D$ and $B$ decays in
a pseudoscalar meson, in the effective lagrangian approach.
We have neglected the $\eta ~- \eta '$ mixing. The branching ratios and
the widths for $B$ must be multiplied for $|V_{ub}/0.0032|^2$. We assume
$\tau_{B_s}=\tau_{B^0} =\tau_{B^+}= 1.55~ps.$}
\label{bpi2}
\centering
\begin{minipage}{10truecm}
\begin{tabular}{l c c c}
Decay & $F_1 (0)$  & BR & Exp. data
 \\ \hline
$D^0 \to \pi^{-}$ & 0.63 &   $2.4 \cdot 10^{-3}$ &
$(3.9^{+ 2.3}_{ - 1.2})\cdot 10^{-3}$ \\
$D^+ \to \eta$ & 0.26 & $4.9 \cdot 10^{-4}$ & \\
$D_s \to \eta$ & 0.51 &  $2.3 \cdot 10^{-2}$ & \\
$D_s \to K^0$ & 0.63 &  $3.0 \cdot 10^{-3}$ & \\
$B^0 \to \pi^- $ & 0.38 &  $1.6 \cdot 10^{-4}$ & 
$(1.63 \pm 0.46 \pm 0.34)\cdot 10^{-4}$
\\
$ B_s \to K $ & 0.38 &  $1.6 \cdot 10^{-4}$ & \\
\end{tabular}
\end{minipage}
\end{table}

The effective lagrangian result (\ref{n18}) shows the dominance of the $P^*$
pole near the kinematic end-point. 
The $1/m_Q$ corrections to the leading results (\ref{n23}), (\ref{n24})
 have been presented in \cite{Boyd-Grin}, where it is shown that 
 at the order $1/m_Q$ one has to introduce four
 new relevant couplings in the
 lagrangian and in the effective current.
   Two of the occurring
 parameters, $\rho_1$ and $\rho_2$, parameterize the $1/M$ corrections to
 the leptonic decay constants $f_P$ and $f_{P^*}$, as shown by (\ref{c11}).
 The others, $g_1$ and $g_2$, are related to the subleading corrections
 to the coupling $g_{P^* P^* \pi}$ and $g_{P^* P \pi}$, as shown 
 in (\ref{g:1/m}).
 In terms of these couplings, the form factor $F_1$ is \cite{Boyd-Grin}:
 \bea
F_1 & = &  \frac{{\hat F}}{2 \sqrt{M_P} f_{\pi}} \left[
 \frac{M_P + v \cdot p_{\pi}}{v \cdot p_{\pi} + \Delta} \left ( 1 + 
 \frac{\rho_1 - 2 \rho_2}{M_P} \right) \left( g + \frac{g_1 - g_2 }{M_P} 
 \right) - \left( 1 + \frac{\rho_1 + 2 \rho_2}{M_P} \right) \right]
 \nn \\
 & = & \frac{1}{2 f_\pi } \left( 
 \frac{M_P + v \cdot p_{\pi}}{M_P ( v \cdot p_{\pi} + \Delta )}
  g_{P^* P \pi} f_{P^*} - f_P \right)~~,
  \label{n25}
  \eea
 where, in the second expression, $g_{P^* P \pi}$, $f_{P^*}$ and $f_P$
 include their own $1/M$ corrections. 
 
 The chiral logarithmic corrections to the process $P \to \pi l \nu_l$ in the
effective theory have been computed in \cite{fle}: evaluating the
one-loop chiral diagrams, it is found that the chiral-corrected form factors
$F_1$ and $F_0$ at the leading order in $1/m_Q$ have the same form as in
(\ref{n18}), (\ref{n19}) but are expressed in terms of the 
chirally renormalized leptonic decay
constants
$f_P^{ren}$ and $f_\pi^{ren}$, the heavy meson coupling $g_{eff}$ to the axial
vector Goldstone current 
(see also section \ref{sec:chirg})
and the hyperfine mass splitting $\Delta^{ren}$, i.e.
\bea
F_1 & = & \frac{g_{eff} f_P^{ren} M_P}{2 f_{\pi}^{ren} 
( v \cdot p_\pi + \Delta_P^{ren})} 
\label{v22}
\\
F_0 & = & \frac{f_P^{ren}}{f_\pi^{ren}}~~. 
\label{v23}
\eea

A more detailed analysis is presented in \cite{fag}, where all the
non-analytic terms arising from chiral loops are kept. The $SU(3)$ violation
in the pole term of the amplitude is of the order of $40 \%$, 
but a rather large value of $g$ is used in 
the numerical estimate, $g \simeq 0.7$.
For smaller values of  $g$, e.g. $g = 0.3$, 
the chiral violation between $P \to \pi$ and
$P \to K$ pole amplitudes reduces to $10 \%$.

\subsection{ $B \to V$ semileptonic decays}
\label{sec-brho}

We now discuss the semileptonic decays of a heavy meson $P$ into a light
vector meson $V = \rho$, $K^*$, $\phi$.

In the following, we will discuss how to relate the $B$ and $D$ form factors,
following as before two different approaches: the scaling approach and the
effective chiral lagrangian approach.
Let us begin with a review of the available experimental data.
In the  $D \to K^* l \nu_l$ decay, the most extensively studied channel,
the quality of the data does not yet allow to
determine the $q^2$ dependence of the form factors. The analysis is 
performed assuming a simple pole formula for 
the form factors $V(q^2), \, A_1(q^2)$, and $A_2(q^2)$,
with pole masses given by the nearest resonance (i.e. $2.1$ GeV for the 
vector form factor and $2.5$ GeV for the two axial form factors). The 
average of three Fermilab experiments gives the results \cite{pdb}:
\be
V(0)\; = \; 1.1 \pm 0.2; \; \; \; A_1(0)\; = \; 0.56 \pm 0.04; \; \; \; 
A_2(0)\; = \; 0.40 \pm 0.08.
\label{v3}
\ee       

Data have also been obtained for the decay $D_s \to \phi l \nu_l$,
but the errors on the form factors are still large, and we shall not use them.

In the case of semileptonic $D$ decays, 
due to the limited  $q^2$ range, the pole assumption does not
sensibly affect the results (\ref{v3}); we have for instance extracted the
form factors assuming $A_1$ constant in $q^2$, but $A_2$ and $V$ 
pole-dominated, finding discrepancies of the order of $10\%$, 
which are within the quoted
uncertainties in (\ref{v3}). The $q^2$ dependence of the form factors is on 
the
contrary extremely important in $B$ decays, as we  discuss below.

For  $B$ mesons the semileptonic rates are strongly Cabibbo-suppressed: 
the CLEO II Collaboration has only recently presented the new measurement 
giving \cite{cleobrux}
\bea
BR(B^0  \to   \rho^- \ell^+ {\nu}_l) & = & (3.88 \pm 0.54 \pm 1.01) \cdot 
10^{-4}
\; \; {\rm WSB} \nonumber \\
BR(B^0  \to   \rho^- \ell^+ {\nu}_l) & = & (2.28 \pm 0.36 \pm 0.59)
 \cdot 10^{-4}\; \; {\rm ISGW}
 \label{v4}
\eea
where the first value is obtained using the WSB model \cite{wsb85}
in the Montecarlo code which evaluates the efficiencies, and the
second one is based on the use of the ISGW model \cite{isgw89}.

Before discussing semileptonic $B$ decays in more detail, let us stress that
another source of information on the weak matrix elements between $B$ and 
$K$ or $K^*$ is represented by non leptonic $B$ decays. As a matter of fact,
the factorization hypothesis allows to relate non-leptonic to semileptonic
rates. The color-suppressed decays $B \to K (K^*) J/\psi$ give, in this
approximation, indications on the form factors 
$B \to K (K^*)$ at $q^2 = M^2_{J/\psi}$.
There are two relevant experimental figures: 
  the ratio of vector and pseudoscalar
widths, measured by Argus \cite{argus94}
and CLEO II \cite{cleo94}, whose
averaged value is
\be
R= \frac{\Gamma (B \to J/\psi K^*)}{\Gamma (B \to J/\psi K)}=1.68 \pm 0.33~~,
\label{v5}
\ee
and the fraction of longitudinal polarization 
\be
\frac{\Gamma_L}{\Gamma} =  
\frac{\Gamma (B \to J/\psi K^*)_L}{\Gamma (B \to J/\psi K^*)}=0.74 \pm 0.07~~, 
\label{v6}
\ee
which corresponds to the average of  the measurements of 
Argus: $\Gamma_L /\Gamma =
0.97 \pm 0.16 \pm 0.15$ \cite{argus94}, CLEO II: $\Gamma_L /\Gamma =
0.80 \pm 0.08 \pm 0.05$ 
\cite{cleo94}, and CDF: $\Gamma_L /\Gamma =
0.65 \pm 0.10 \pm 0.04$ \cite{cdf95}.

Detailed phenomenological
analyses have been performed in \cite{gourdin94,aleksan94,gourdin95}, where it 
has 
been
shown that most of the models fail to explain the previous data, in
particular the fraction of longitudinal polarization (\ref{v6}).
We point out that all the current models use the hypothesis of factorization
\cite{facto}, which in general works satisfactorily in $B$ decays \cite{bdecII} 
but could have corrections in specific channels, like the colour-suppressed
$B \to J/\psi K^*$. Possible non-factorizable contributions are introduced in
\cite{shamali}.

By the definition
\be
x= \frac{A_2^{BK^*}}{A_1^{BK^*}}\Big|_{(M^2_{J/\psi})} ~~~~~~~~
y= \frac{V^{BK^*}}{A_1^{BK^*}}\Big|_{(M^2_{J/\psi})}
\label{v7}
\ee
one gets, assuming factorization \cite{gourdin94},
\be
R = 1.081 \left( \frac{A_1^{BK^*}}{F_1^{BK}}(M^2_{J/\psi}) \right)^2
\left[ (a - b x )^2 + 2 (1 + c y^2) \right]
\label{v8}
\ee
\be
\frac{\Gamma_L}{\Gamma} = \frac{(a - b x )^2}{(a - b x)^2 + 2 (1 + c y^2)}~~.
\label{v9}
\ee
The coefficients $a$, $b$ and $c$ are dimensionless combinations of  masses;
from the data one gets
\be
a= 3.16 ~~~~ b=1.31 ~~~~ c=0.19~~. 
\label{v10}
\ee

\subsubsection{Scaling approach to $B \to V$ form factors}
\label{subsec-brhoscal}

The scaling approach, valid at the leading order in $1/m_Q$, 
is similar to the case $P \to \pi$; the scaling laws for the form factors
are derived from the asymptotic behaviour of the matrix element, i.e.
\be
<V (p')|J_\mu| P(v)> \sim \sqrt{m_Q} \left( 1 + {\cal O} \left(
\frac{\Lambda_{QCD}}{m_Q} \right) + {\cal O} \left(
\frac{v \cdot p'}{m_Q} \right) + {\cal O} \left(
\frac{m_{V}}{m_Q} \right)\right)~~.
\label{v11}
\ee

For  $D \to K^*$ the violation to (\ref{v11}) can be important, namely of
order $m_{K^*}/M_D$. This uncertainty is reflected in different
choices of the scaling laws at $q^2 \simeq q^2_{max}$. 
For example one can follow the approach called ``soft-scaling'' in 
\cite{aleksan94}, and adopted
in \cite{casa93}, i.e.
\bea
V(q^2_{max}) &\approx &   \frac {M_P + M_V} {\sqrt {M_P}}\nn \\
A_1(q^2_{max}) &\approx &  \frac{\sqrt{M_P}}{M_P + M_V}\nn \\
A_2(q^2_{max})  &\approx & \frac {M_P + M_V}{\sqrt{M_P}}~~. \label{v12}
\eea

The second choice  we shall consider is ``hard-scaling'':
\bea
V(q^2_{max}) &\approx &   \sqrt {M_P}\nn \\
A_1(q^2_{max})&\approx &  \frac{1}{\sqrt{M_P}}\nn \\
A_2(q^2_{max}) &\approx & 
\sqrt{M_P}~~. \label{v13}
\eea
The two  scenarios, (\ref{v12}) and (\ref{v13}), differ by subleading terms of
the order
$M_{V}/M_P$, which can be nevertheless numerically important.
Needless to say, some form factors might exhibit  soft-scaling and others 
hard-scaling, in different combinations.

The scaling laws allow to relate the $D$ and $B$ form factors near 
$q^2_{max}$:
as discussed above, the dependence on  $q^2$  is practically
unknown, and, as we stressed already, $B$ transitions depend 
strongly on the extrapolation  to
$q^2=0$.
The vector form factor $V$ is generally believed to be pole-dominated, as
discussed in section \ref{sec-formfact}, while for $A_1$ and $A_2$ the 
theoretical
situation is unclear.
In \cite{aleksan94} the soft scaling laws are justified 
by extending  the heavy-to-heavy scaling
relations down to the light final meson case. In the same limit one finds
that $A_2/A_1$, $V/A_1$ and $F_1/A_1$ should have a polar behaviour in 
$q^2$: assuming $F_1$ as pole dominated, this implies a constant $A_1$ and a
pole  behaviour for $A_2$ and $V$. Nevertheless the extension of the
heavy-to-heavy scaling laws to the heavy-to-light case remains arbitrary, and 
should be considered as an ansatz.
As we will discuss explicitly in the next section, the effective lagrangian
approach leads to the soft-scaling solution (\ref{v12}): this follows  from 
the
factor $(M_P + M_V)$ contained in the definition (\ref{v1}) of the form
factors.

To simplify the discussion, we  assume that $A_2$ is 
dominated by the nearest pole,
while for $A_1$ we  consider two possibilities: the pole-dominance, and 
a flat $A_1$ constant in $q^2$. We have considered four different
possible scenarios:
soft scaling and $A_1$ pole-dependent (called soft-pole),
soft scaling and $A_1$ constant (called soft-constant),
hard scaling and $A_1$ pole-dependent (called hard-pole), and finally
hard scaling and $A_1$ constant (called hard-constant).
For each of them we have computed, using as inputs the
$D \to K^*$ form factors (\ref{v3}), the branching fraction and the 
ratios of decay
widths $\Gamma_+/\Gamma_{-}$ and 
$\Gamma_L/\Gamma_T$ for the process $B^0 \to \rho^- l^+ \nu_l$. 
Here $\Gamma_T$ and $\Gamma_L$ refer to $\rho$ with 
transverse and longitudinal polarization respectively, 
$\Gamma_+$ and $\Gamma_{-}$ to
$\rho$ with positive and negative helicities. We have computed 
the longitudinal fraction $\Gamma_L/\Gamma$,  and the ratio of the
vector to scalar BR's for $B \to J/\psi K (K^*)$.  
The results are presented in table \ref{brho1}, where we have
extrapolated from $D \to K^*$ to $B \to K^*$, and then we have equated, 
by chiral symmetry, the form factors $B \to K^*$ and $B \to \rho$. 
\begin{table}
\caption{Predictions for form factors and widths for $B^0 \to\rho^- l \nu_l$.
$\Gamma_T$ and $\Gamma_L$ refer to $\rho$ with respectively
transverse and longitudinal polarization, $\Gamma_+$ and $\Gamma_{-}$ to
$\rho$ with positive and negative helicities. The branching ratios (BR) and
the widths for $B$ must be multiplied for $|V_{ub}/0.0032|^2$. We assume
$\tau_{B_s}=\tau_{B^0} =\tau_{B^+}= 1.55~ps.$}
\label{brho1}
\centering
\begin{minipage}{12truecm}
\begin{tabular}{l c c c c c}
Extrapolation & soft-pole & hard-pole & soft-cost. & hard-cost. & data
 \\ \hline
$A_1(0)$  & 0.21 & 0.17 & 0.42 & 0.33 & \\
$A_2 (0)$ & 0.27 & 0.34 & 0.27 & 0.34 & \\
$V(0)$    & 0.64 & 0.81 & 0.64 & 0.81 & \\
${\Gamma_+}/{\Gamma_{-}}$ & 0.045 & 0.13 & 0.005 & 0.06 & \\
${\Gamma_L}/{\Gamma_{T}}$ & 0.39 & 0.15 & 1.80 & 0.30 & \\
BR ($10^{-4}$)& 2.8 & 2.8 & 6.7 & 3.5 & $3.88 \pm 0.54\pm 1.01$ \\
${\Gamma_L}/{\Gamma} (J/\psi K^*)$ & 0.27 & 0.02 & 0.50 & 0.18 & $0.74 \pm
0.07$ \\
${\Gamma (K^*)}/{\Gamma (K)}$ & 1.74 & 1.68 & 3.35 & 2.27 & 
$1.68 \pm 0.33$ \\
\end{tabular}
\end{minipage}
\end{table}

Let us comment on table \ref{brho1}.
First of all, all the four scenarios give a rather low value for
the ratio $\Gamma_L/\Gamma (J/\psi K^*)$: the soft-scaling,
$A_1$ constant (third column of table \ref{brho1}),
 is the closest one to the experiment, but it produces a
too high value for the branching ratio $B \to \rho^- l \nu_l$
and for  $\Gamma(K^*)/\Gamma (K)$. The value of the latter ratio depends 
mainly on  the ratio $A_1/F_1$ (at $q^2 = M^2_{J/\psi}$), and could
be smaller for a larger value of $F_1^{B \;K}$: however, a too large value
would disagree with the measured branching ratio for the 
$B \to \pi l \nu_l$ (excluding large $SU(3)$ violation).
This scenario is  preferred  in \cite{aleksan94}, where however
the upper limit for the $B \to \rho$ and the $B \to \pi$ data
are not taken into account
(soft-scaling is also applied to $F_1$, obtaining a larger value for it
and a better agreement for $\Gamma (K^*)/\Gamma (K)$).

The second comment is that
a constant $A_1$ gives a higher value 
than pole behaviour for the $B \to \rho$
branching ratio, when scaling is used. 
The situation improves assuming pole dominance for  $A_1$  for the 
semileptonic
branching ratio $B \to \rho$, and also for the ratio
$\Gamma (K^*)/\Gamma (K)$, as it can be seen in the first column of the table.
But, at the same time, the longitudinally polarized fraction $\Gamma_L/\Gamma
(J/\psi \; K^*)$ decreases, because $A_2/A_1$ at $q^2 = M^2_{J/\psi}$ grows.

Hard scaling decreases the value of
$A_1$, and it raises $A_2$ and $V$: this produces a smaller 
semileptonic $BR (B \to \rho)$, but also lowers $\Gamma_L/\Gamma (J/\psi \;
K^*)$ (because $A_2/A_1$ grows, see (\ref{v9}))  and $\Gamma(K^*)/\Gamma (K)$.
Hard scaling, together with a 
pole-dominated $A_1$, as in the second column, 
leads to a value
for    $\Gamma_L/\Gamma (J/\psi \; K^*)$ in  disagreement with the
data: its combination with a constant $A_1$, as in the fourth column of the
table, improves the agreement with the data, even if  
$\Gamma_L/\Gamma (J/\psi \; K^*)$ remains  still rather small, even smaller 
than in
soft-pole scenario for $A_1$.

Summarizing, the previous analysis indicates that a constant $A_1$ requires 
strong scaling in order to get
agreement with the data
 of the semileptonic $BR (B \to \rho)$.
 In \cite{aleksan94} a different result is obtained, which,
however, does not take into account the results for the semileptonic
$B \to \rho$ transition.

  If $A_1$ is single-pole dominated, 
soft scaling is required in order to get a reasonable (not too small) value 
for 
$\Gamma_L/\Gamma (J/\psi \; K^*)$: these data are however difficult to explain
without spoiling other phenomenological requests.

It should be stressed, however, that the figures of table \ref{brho1} have 
large
uncertainties. Leaving aside theoretical uncertainties that are however
significant, the quoted numbers have an uncertainty due
to the experimental errors of the $D \to K^* l \nu_l$ form factors (\ref{v3})
used as inputs. We notice that, for instance, $A_2 (0)^{D \; K^*}$ is quoted 
with an error of about 20 \%: this error alone implies 30 \% uncertainty in
$\Gamma_L/\Gamma (J/\psi \; K^*)$ and  15 \% in
$\Gamma (K^*) /\Gamma (K)$. Finally we have  used flavour 
$SU(3)$ symmetry to relate $B \to \rho$ and $B \to K^*$ form factors and this 
is another source of theoretical error which in principle should be taken into
account.

\subsubsection{Effective lagrangian approach}
\label{subsec-brhoeff}

Light vector resonances have been discussed in the framework
of the effective heavy meson chiral lagrangian in
\ref{subsec:lvr};
 applications to semileptonic decays have been developed in 
\cite{casa93}.
Chiral loop contributions to $D \to K^* l \nu_l$ have been partially taken 
into
account in \cite{ko93}. We now review this subject. 

In the effective chiral lagrangian framework, five different diagrams
contribute at the leading order in $1/m_Q$
to the matrix element $<V(p',\epsilon)|J_\mu | P (p)>$. They are analogous
to the diagrams of fig. \ref{fig:semilep}.
Four of them are polar diagrams: $V(q^2 )$ takes contribution from the $1^-$
pole diagram, proportional to the coupling $\lambda$ among $P$, $P^*$ and
$\rho$ introduced in (\ref{1.11}).
 The $\zeta$ and $\mu$ couplings in
(\ref{rob:zm})  give
the vertex $P P^{**} \rho$, where $P^{**}$ is a positive-parity states
of the doublet $( 0^+, 1^+ )$ and
the corresponding polar diagrams, with exchange of a
$1^+$ meson, contribute to $A_1$ and $A_2$. 
The $\mu$ and $\lambda$ terms have
dimension higher than the $\zeta$ or $g$
terms; nevertheless they are the lowest-order contributions to 
$V$ and $A_2$.
The fourth polar diagram, with exchange of a $0^-$ meson, is
proportional to the coupling $\beta$ of the vertex $P P \rho$
(see formula (\ref{1.11})) and is relevant only for the form factor
$A_0$. 
Finally there is a direct diagram, which arises from the effective
current term proportional to $\alpha_1$ in the heavy-to-light current
(\ref{c5}). All these vertices can be found in appendix A.

Computing the diagrams 
for $q^2 \simeq q^2_{max}$
and at leading order in $1/m_Q$,
one gets \cite{casa93}:
\be
V(q^2_{\rm max})=\frac {g_V} {\sqrt{2}} \lambda {\hat F} 
\frac {M_P + M_{V}}{\sqrt{M_P}}\frac{1}{M_V + \Delta_{P^*P}}
\label{v14}
\ee
\be        
A_1(q^2_{\rm max})= - \frac{2 g_V}{\sqrt{2}}
\frac{\sqrt{M_P}}{M_P +M_{V}}
 \left[ \alpha_1  - \frac{{\hat F}^+ (\zeta /2 - \mu M_{V})}
{M_V + \Delta_{PP^{**}}} \right]
\label{v15}
\ee
\be
A_2(q^2_{\rm max})=-\frac {\mu g_V {\hat F}^+}{\sqrt{2} (M_V + 
\Delta_{PP^{**}})} \frac {M_P + M_{V}}{\sqrt{M_P}} 
\label{v16}
\ee
\be
A_0 (q^2_{\rm max})=\frac{g_V}{2 \sqrt{2}} \frac{\beta {\hat F} 
\sqrt{M_P}} 
{M_{V} (M_V + \Delta')} + \frac {g_V} {\sqrt{2}} \frac{\alpha_1\sqrt {M_P}} 
{M_{V}}
\label{v17}
\ee
where the $\Delta$ are appropriate mass splittings.

 From (\ref{v14})-(\ref{v17}), we can extract information on the scaling
behaviour of the form factors as well as some indications on their
$q^2$ behaviour. 

As to the scaling, we notice that  
(\ref{v14})-(\ref{v16}) imply the ``soft-scaling'' of eq.
(\ref{v12}), since the coupling of the effective lagrangian is flavour
independent.  Moreover, the form factors $V$ and $A_2$ have only the pole
structure, which at $q^2 \simeq q^2_{\rm max}$ is signalled by the factor
$1/(M_V + \Delta)$, while $A_1$ contains a pole term but also a non-polar
one, proportional to $\alpha_1$. This suggests  a more complex
 $q^2$ behaviour of $A_1$.

The non-polar term could be a general polynomial in $q^2$: the simplest way 
to take into account the indications coming from the effective lagrangian is
to describe $A_1$ as a sum of a constant term and a pole term, i.e. we 
write:
\be
A_1 (q^2) = a + \frac{b}{1- q^2/M^2_{P^{**}}}
\label{v18}
\ee
while keeping for $A_2$ and $V$ the nearest-pole behaviour.
The parameters $a$ and $b$ are flavour dependent, and scale  differently
at $q^2 = q^2_{\rm max}$. Assuming soft scaling, as suggested by
(\ref{v14}) - (\ref{v17}), we find the following scaling laws:
\bea
a(B \to V) & = & \frac{M_D + M_V}{M_B + M_V} \sqrt{\frac{M_B}{M_D}} 
a(D \to V)\label{v18.1}\\  
b(B \to V) & = & \frac{M_D + M_V}{M_B + M_V} \sqrt{\frac{M_D}{M_B}} 
b(D \to V)~~.\label{v18.2}
\eea 

The formulas (\ref{v14})-(\ref{v16}) give the form factors 
 at $q^2_{\rm max}$: at the leading order in $1/m_Q$, the value in $q^2 = 0$
 for pole-dominated terms is:
 \be
 F(0) = \frac{2 (M_V + \Delta )}{M_P} F(q^2_{\rm max})~~.
 \label{v18.3}
 \ee
As for $A_1$, we identify the term proportional to $\alpha_1$ in
(\ref{v15}) with the constant term  $a$ of (\ref{v18}), and the other with
the pole term, proportional to $b$. In this way one gets:
\be
V(0)=g_V\sqrt{2} \lambda {\hat F} 
\frac {M_P + M_{V}}{M_P^{3/2}}
\label{v18.4}
\ee
\be        
a = -  g_V\sqrt{2}
\frac{\sqrt{M_P}}{M_P +M_{V}}\alpha_1 
\label{v18.5}
\ee
\be
b = g_V 2 \sqrt{2}{\hat F}^+ \frac{\zeta /2 - \mu M_{V}}
{(M_P + M_V)\sqrt{M_P}} 
\label{v18.6}
\ee
\be
A_2(0)= -\sqrt{2} \mu g_V {\hat F}^+ 
\frac {M_P + M_{V}}{M_P^{3/2}}~~. 
\label{v18.7}
\ee

 From the experimental data (\ref{v3}) on $D \to K^*$ we can fix the effective 
couplings 
appearing in the previous formulas: from (\ref{v18.4}), using ${\hat F} =
 0.30 \; GeV^{3/2}$
and $g_V = 5.8$ (see (\ref{rob:gv})), we obtain
\be
|\lambda | = 0.41 \; GeV^{-1}~~.
\label{lambda}
\ee
We shall see, in the next section, how the sign of $\lambda$ can be fixed.
It is interesting to observe that this result agrees with
the second, but not with the first determination obtained in \cite{allc}
by a light cone  sum rules calculation (the first determination
gives a higher value; also in \cite{frlbb} a higher value of
$\lambda$ is obtained).

Similarly, from (\ref{v18.7}) and ${\hat F}^+ = 0.46\; GeV^{3/2}$ 
\cite{cola92}
we get
\be
\mu = -0.10 \; GeV^{-1}~~.
\label{mu}
\ee

Having fixed  $\lambda$ and $\mu$, one can compute the form factors $V$ and 
$A_2$
for the $B \to K^*$ and $B \to \rho$ matrix elements, because these couplings
are heavy-flavour independent at the leading order. The result is
\be
|V^{B \to V}(0) | = 0.50 ~~~~~~~~~~ A_2^{B \to V}(0) = 0.19~~.
\label{v18.8}
\ee
The mass difference between $K^*$ and $\rho$ is numerically irrelevant in
 (\ref{v18.8}). 

Concerning $A_1$, from the input $A_1^{D \; K^*} (0)$, we derive
\be
(a + b)^{D \; K^*} = 0.56~~.
\label{v19}
\ee

The knowledge of the $q^2$ behaviour of $A_1^{D \; K^*}$ 
would allow to extract simultaneously $a$ and $b$, 
but at  present we can only introduce an arbitrary 
parameter $r$, defined as
\be
r = \frac{a}{a + b}
\label{v20}
\ee
where $a$ and $b$ are relative to the $A_1^{D \to K^*}$ ($r$ is not heavy
flavour independent). 

When $r$ varies from $r =0$ to $r=1$, we have a smooth transition from a 
pure pole dominance ($r = 0$) to a constant $A_1$ ($r = 1$).
The analysis of section \ref{subsec-brhoscal} has shown that a pure pole
behaviour (soft-pole case) leads to a rather low  value for 
$\Gamma_L/\Gamma (J/ \psi \; K^* )$, while a constant $A_1$ gives 
a semileptonic branching ratio $B \to \rho$ too high: therefore we expect
that the two component form factor (\ref{v18}) can explain better, 
for some intermediate
value of $r$,  the large longitudinal polarization in 
$B \to J/\psi K^*$ and at the same time can agree with the data for $B \to \rho 
l\nu$.

In figs. \ref{longitudo}, \ref{rapo} and \ref{gamma} we plot respectively
$\Gamma_L/\Gamma (J/ \psi \; K^* )$, $\Gamma (K^*)/\Gamma (K)$ and
$BR (B^0 \to \rho^- l^+ \nu_l)$ as a function of $r$. We have assumed
$A_2$ and $V$ pole-dominated, with the values (\ref{v18.8}) at $q^2 = 0$.
We see that all the three observables grow when the $q^2$ dependence 
of $A_1$ becomes flatter, i.e. $r \to 1$. This is a satisfactory feature for
$\Gamma_L/\Gamma (J/\psi \; K^*)$, but  a large $r$ gives  too large
values for the
semileptonic $BR$ and for the $\Gamma (K^* )/\Gamma (K)$ ratio.

\begin{figure}[htb]
\epsfxsize=9cm
\epsfysize=9cm
\epsfbox{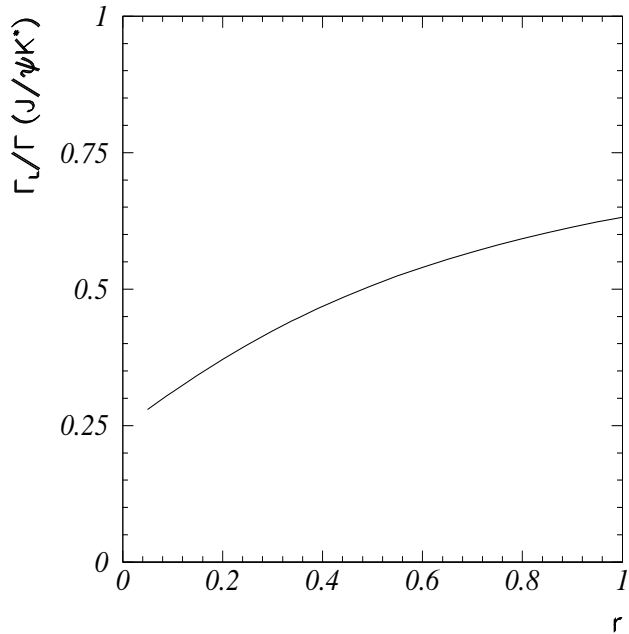}
\vspace{3mm}
\caption{Ratio $\Gamma_L/\Gamma$ for the decay $B \to J/\psi K^*$ as a function 
of the parameter $r$. $\Gamma_L$ is the width for longitudinally polarized 
$K^*$.}
\label{longitudo}
\end{figure} 
\begin{figure}[htb]
\epsfxsize=9cm
\epsfysize=9cm
\epsfbox{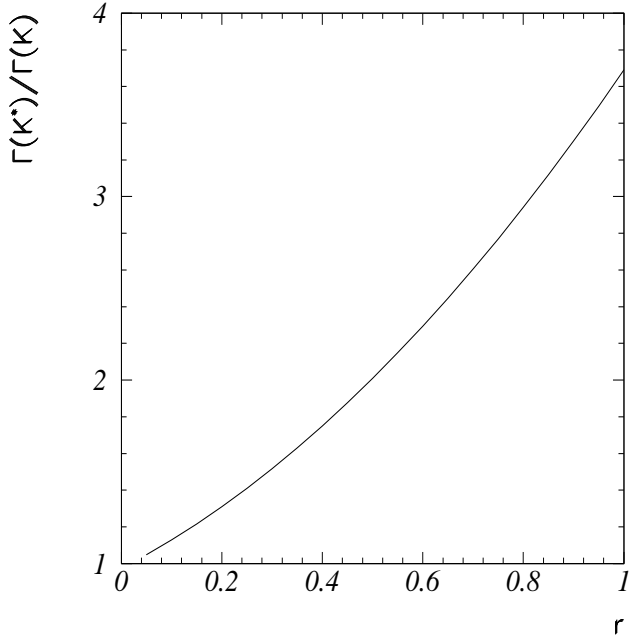}
\vspace{3mm}
\caption{Ratio of the widths $B \to J/\psi K^*$ and $B \to J/\psi K$ as a 
function
of the parameter $r$.}
\label{rapo}
\end{figure} 
\begin{figure}[htb]
\epsfxsize=9cm
\epsfysize=9cm
\epsfbox{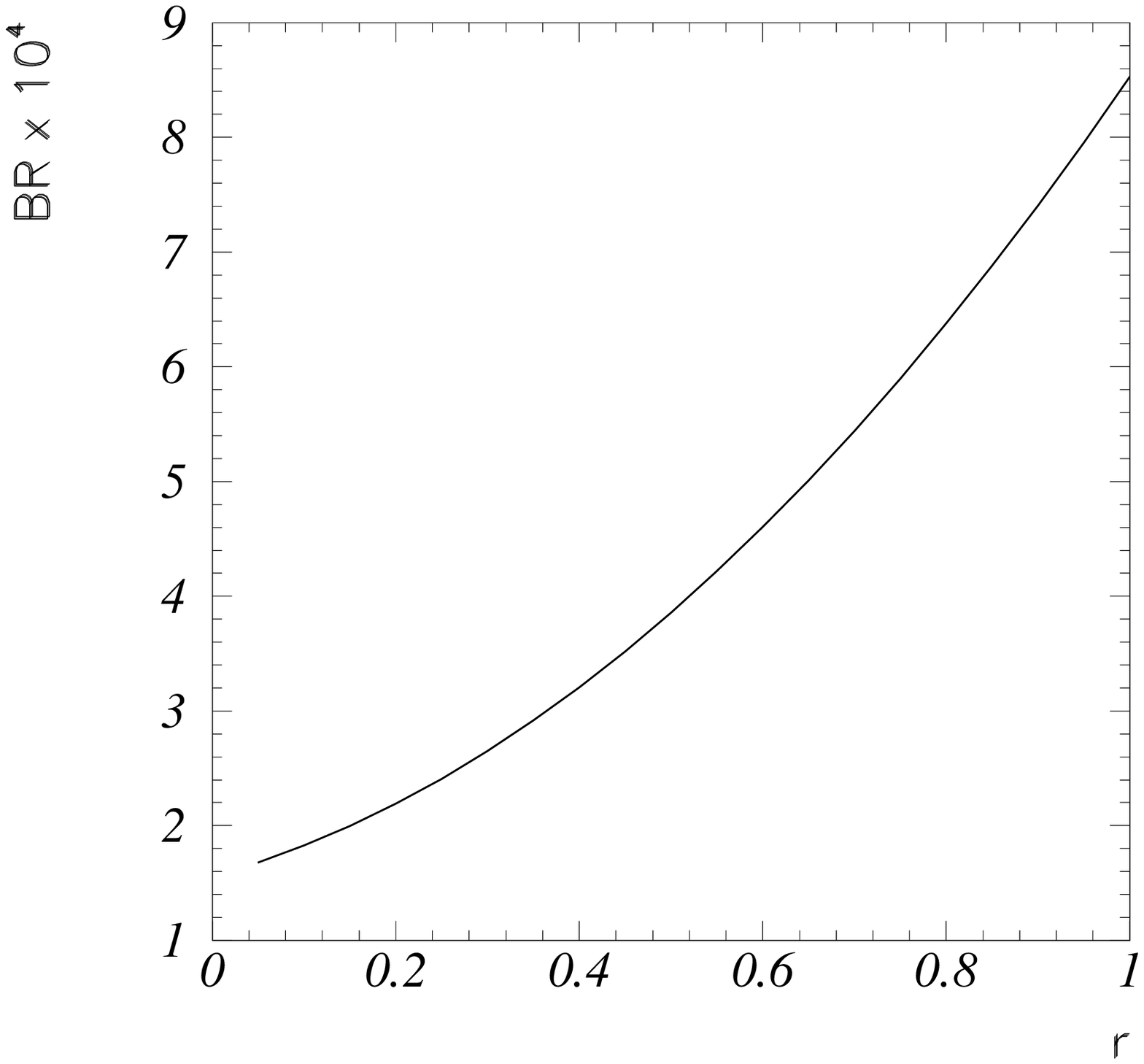}
\vspace{3mm}
\caption{Branching ratio of the decay $B^0 \to \rho^- l \nu_l$ as a function 
of
the parameter $r$. We have taken $V_{ub} = 0.0032$.}
\label{gamma}
\end{figure} 
In table \ref{brho2} we quote
the values of various observables at $r = 0.5$, where a good compromise
 is obtained: 
$\Gamma_L/\Gamma (J/ \psi \; K^* ) = 0.51$. This value is still smaller
than the data, but one should not forget that the factorization assumption
could receive sizeable corrections. 
\begin{table}[htb]
\centering
\caption{Predictions for form factors and widths for $B^0 \to\rho^- l \nu_l$ 
and for
$B \to J/\psi K^{(*)}$, with $r = 0.5$.
$\Gamma_T$ and $\Gamma_L$ refer to $\rho$ with, respectively,
transverse and longitudinal polarization, $\Gamma_+$ and $\Gamma_{-}$ to
$\rho$ with positive and negative helicities. The branching ratios (BR) and
the widths for $B$ must be multiplied for $|V_{ub}/0.0032|^2$. We assume
$\tau_{B_s}=\tau_{B^0} =\tau_{B^+}= 1.55~ps.$}
\label{brho2}
\centering
\begin{minipage}{6truecm}
\begin{tabular}{l c c}
Observable & r = 0.5 & data \\ \hline
$A_1(0)$  & 0.28 & \\
$A_2 (0)$ & 0.19 & \\
$V(0)$    & 0.50 & \\
${\Gamma_+}/{\Gamma_{-}}$ & 0.013 & \\
${\Gamma_L}/{\Gamma_{T}}$ & 1.60 & \\
BR ($10^{-4}$)& 3.8 & $3.88 \pm 0.54 \pm 1.01$ \\
${\Gamma_L}/{\Gamma} (J/\psi K^*)$ & 0.51 & $0.74 \pm
0.07$ \\
${\Gamma (K^*)}/{\Gamma (K)}$ & 2.01 & 
$1.68 \pm 0.33$ 
\end{tabular}
\end{minipage}
\end{table}

 From the value of $r$ one can extract $a$ and $b$ separately: from the scaling
relations (\ref{v18.1}) and (\ref{v18.2}) we have for $r = 0.5$:
\be
a ( B \to V) = 0.21 ~~~~~~~~~~~~~~ b(B \to V) = 0.07
\label{v21}
\ee
i.e. $A_1^{B\rho} (0) = 0.28$, as quoted in table \ref{brho2}. The knowledge 
of
$a$ and $b$,
together with the identifications (\ref{v18.5}) and (\ref{v18.6}), fixes
the couplings $\alpha_1$ and the linear combination $\zeta/2 - \mu M_V$. 
The results depend on the value of $r$; for $r = 0.5$ we get:
\be
\alpha_1 = -0.07 \; GeV^{1/2} \hspace{2cm}  \frac{\zeta}{2} - \mu M_V = 0.14
\label{alphazeta}
\ee
($a$ and $b$ are taken as positive). From (\ref{mu}) we can extract finally 
$\zeta$:
\be
\zeta = 0.10~~.
\label{zeta}
\ee

The previous phenomenological analysis has to be taken cautiously, due to the 
large uncertainties. Subleading corrections, $q^2$ dependence of the form 
factors,
breaking of factorization and chiral violations could easily lead to 
substantial modifications of the chosen scenario.
New experimental data will hopefully clarify the situation, and allow to 
distinguish among different models. We shall adopt in the following 
the effective lagrangian  results of table \ref{brho2} (in particular
$r = 0.5$).   

In table \ref{bucomp} we present the values of the form factors 
of the $b \to u$
transitions in different models.
\begin{table}[htb]
\caption{Form factors at $q^2 = 0$ for $b \to u$  transitions
in different models}
\label{bucomp}
\centering
\begin{minipage}{12truecm}
\begin{tabular}{l c c c c}
Reference & $F_1^{B \to \pi}$ & $A_1^{B \to \rho}$ & 
$A_2^{B \to \rho}$ & $V^{B \to \rho}$ \\ \hline
This paper & $0.38$ & $0.28$ & $0.19$ & $0.50$ \\
\hline
QCD sum rules & & & & \\\hline
DP \cite{dompav88} & $0.4 \pm 0.1$ & & & \\
CZ \cite{chern90} & $0.36$ & & & \\
BBD \cite{bbd91} & $0.24 \pm 0.025$ & & & \\
Narison \cite{narison92} & $0.23 \pm 0.02$ & $0.35 \pm 0.16$ &
$0.42 \pm 0.12$ & $0.47 \pm 0.14$ \\
Ball \cite{ball93} & $0.26 \pm 0.02$ & $0.5 \pm 0.1$ &
$0.4 \pm 0.2$ & $0.6 \pm 0.2$ \\
BKR \cite{belyaev93} & $0.24 - 0.29$ & & & \\
ABS \cite{ali94} & & $0.24 \pm 0.04$ &
 & $0.28 \pm 0.06$ \\ \hline
Quark models & & & & \\ \hline
BSW \cite{wsb85} & $0.33 $ & $0.28$ &
$0.28$ & $0.33$ \\
ISGW \cite{isgw89} & $0.09$ & $0.05$ &
$0.0.02$ & $0.27$ \\
FGM \cite{fgm95} & $0.21 \pm 0.02$ & $0.26 \pm 0.03$ &
$0.30 \pm 0.03$ & $0.29 \pm 0.03$ \\ \hline
Lattice & & & & \\
\hline
APE \cite{ape95} & $0.35 \pm 0.08$ & $0.24 \pm 0.12$ &
$0.27 \pm 0.80$ & $0.53 \pm 0.31$ \\
Abada et al. \cite{abada94} & $0.30 \pm 0.14 \pm 0.05$ & $0.22 \pm 0.05$ &
$0.49 \pm 0.21 \pm 0.05$ & $0.37 \pm 0.11$ \\
UKQCD \cite{ukqcd95} & & $0.27^{+7+3}_{-4-3}$ & & 
\end{tabular}
\end{minipage}
\end{table}
\newpage
\section{Radiative decays}
\label{chap-bep}
\subsection{Flavour conserving radiative decays: $D^* \to D \gamma$ }
\label{sec:bepp1}

In this section we shall consider the decay 

\be D^*_a \to D_a \gamma \hskip 3 pt , \label{5.1.1} \ee
and the related processes for the $B$ case: $B^*_a\to B_a \gamma$.
In (\ref{5.1.1}) $a=1,2,3$ is the light quark index corresponding
to $u,d,s$.
The matrix element for this radiative transition is as follows:

\be \MM(D^*_a \to D_a \gamma)  
=\hskip 3 pt i\,   e \mu_a 
\epsilon^{\mu \nu \alpha \beta} 
\epsilon^*_\mu \eta_\nu p_\alpha p^{\prime}_\beta  \hskip 3 pt . 
\label{5.1.2}  \ee

 In (\ref{5.1.2}) $\epsilon_\mu$ is the photon polarization, whereas
the coupling $\mu_a$ comprises two terms:
\be
\mu_a =  \mu_a^{\ell}   +  \mu_a^h ~~ , \label{5.1.5.bis}
\ee
corresponding to the decomposition:

\be \MM(D^*_a \to D_a \gamma) = 
 e \; \epsilon^{* \mu} \; <D_a(p^{\prime})|J_\mu^{em}|D^*_a (p, \eta)>
=  e \; \epsilon^{* \mu} \;
<D_a(p^{\prime})|J_\mu^{\ell}+ J_\mu^{h}|D^*_a (p, \eta)>~~.
\label{5.1.3} \ee

Here $J_\mu^{\ell}$ and $J_\mu^{h}$ are the light and 
the heavy quark parts of the electromagnetic current:
\be
J_\mu^{\ell}=
\frac{2}{3} \bar{u} \gamma_\mu u -
\frac{1}{3} \bar{d} \gamma_\mu d -
\frac{1}{3} \bar{s} \gamma_\mu s =  \sum_{a=1}^3 e_a 
\bar{q}_a \gamma_\mu q_a \label{5.1.4}
\ee
and
\be
J_\mu^{h}=
\frac{2}{3} \bar{c} \gamma_\mu c -
\frac{1}{3} \bar{b} \gamma_\mu b =  \sum_{Q=c,b} e_Q 
\bar{Q} \gamma_\mu Q  . \label{5.1.5} 
\ee
Correspondingly, eq. (\ref{5.1.5.bis}) becomes
\be
\mu_a  =   \mu_a^{\ell}   +   \mu_a^h =
 \frac{e_a}{\Lambda_a}  +  \frac{e_Q}{\Lambda_Q}~~,   \label{5.1.5.ter}
\ee
where $\Lambda_a$ and $\Lambda_Q$ are mass parameters to be determined.

Let us consider the two currents $J_\mu^{h}$ and 
$J_\mu^{\ell}$ separately. The matrix element of
$J_\mu^h$ 
can be obtained from the Lagrange 
density:

\be \LL^{\prime \prime}=-{e \over 2 m_Q} e_Q {\bar h}_v \sigma^{\mu \nu} h_v 
F_{\mu \nu} \hskip 5 pt , \label{eq : lag} \ee

\noindent which allows the transition $Q \to Q \gamma$
and can be expressed in terms of the Isgur-Wise 
universal form factor $\xi(v \cdot v^{\prime}) $
as follows:

\be <D_a(p^{\prime})| J_\mu^h |D_a^*(p, \eta)>= \, e_c
<D_a(p^{\prime})|\bar{c} \gamma^\mu c|D_a^*(p, \eta)>=
    i \frac{2}{3}  \sqrt{M_{D_a} M_{D_a^*}} \xi(v \cdot v^{\prime}) 
\epsilon_{\mu \nu \alpha \beta} 
\eta^\nu v^\alpha v^{\prime \beta}  \hskip 3 pt , \label{5.1.6} \ee

\noindent where $p^{\prime}=M_D v^{\prime}$, $p=M_{D^*} v$ and $v 
\cdot v^{\prime} \simeq 1$ because:

\be 0=q^2=m^2_D + M_{D^*}^2 -2 M_D M_{D^*} v \cdot v^{\prime} \hskip 3 pt . 
\label{5.1.7} \ee

Taking into account the normalization of $\xi(v \cdot v^{\prime}) $:
$\xi(1)=1 $, one gets,
for the charm case, 
\be
\mu_a^h \; = \; \frac{2}{3 \Lambda_c}  \label{5.1.8}
\ee
($\mu_a^h \; = \; - \; \frac{1}{3 \Lambda_b}$ for the  $b$ case), with
\be
\Lambda_c={\sqrt{ M_{D_a} M_{D^*_a}}} \label{lambdac}
\ee
(resp. $\Lambda_b={\sqrt{ M_{B_a} M_{B^*_a}}}$); in  
the leading order in $1/m_c$ one finds:

\be
\Lambda_c\; =\; m_c \label{5.1.8sec}
\ee
(resp. $\Lambda_b\; =\; m_b$),
independently of the light quark label $a$.

Let us now consider the second term in (\ref{5.1.5.bis}), i. e.
$\mu_a^{\ell}$, which cannot
be computed within HQET since it involves
light quarks; we shall now show that the chiral effective 
theory can be employed to get 
information on this quantity. We shall examine two approaches: the first 
one is based on the calculation of chiral loop corrections \cite{abj};
the second is based on the use of Vector Meson Dominance (VMD), together 
with the effective chiral lagrangian for light and heavy mesons 
\cite{defazio1}. Other approaches used to compute (\ref{5.1.1}) 
are based on quark models \cite{cina1,defazio2,eichten};
bag model \cite{singer} and QCD sum rules  \cite{eletsky,aliev} 
(for a previous review of theoretical results see 
\cite{kamal}).
 
The first approach we consider is based on the chiral loop corrections
to the tree diagram \cite{abj}.
Let us start with the definition of $\mu_a^{\ell}$:
\be
\mu_a^{\ell} \; = \; \frac{e_a}{\Lambda_a} \; ; \label{5.1.8bis}
\ee
in the limit of $SU(3)$ symmetry
the constants ${\Lambda_a}$'s are equal, i. e. 
one gets ${\Lambda_a}^{-1} \; = \; \beta$,
where $\beta $ is an unknown constant which can also contain effects 
suppressed
by powers of $1/m_c$.

The leading $SU(3)$ violations to (\ref{5.1.8bis}) are obtained by considering
the loop diagrams of fig. \ref{figbep1}, with the results \cite{abj}:
\be
\mu_1^{\ell}  = \frac{2}{3} \beta  -  g^2\frac{m_K}{4 \pi f_K^2}
-  g^2\frac{m_\pi}{4 \pi f_\pi^2}
\label{5.1.a}
\ee
\be
\mu_2^{\ell}  = -  \frac{1}{3} \beta +  
g^2\frac{m_\pi}{4 \pi f_\pi^2}
\label{5.1.b}
\ee
\be
\mu_3^{\ell}  = -  \frac{1}{3} \beta  +  g^2\frac{m_K}{4 \pi f_K^2}~~.
\label{5.1.c}
\ee
Here $g$ is the strong coupling constant of the vertex $D^* D \pi$
(in the $m_c \to \infty$ limit)
defined in (\ref{wise}).
If one considers only the leading $SU(3)$ violations,
one should put $f_K =  f_\pi$\footnote{In the 
analysis of Ref. \cite{abj}
$f_K  =  1.22 f_\pi$  is used.}. This has to be the case if one uses 
the value $g = 0.38$, eq. (\ref{eq:038}). As a matter of fact, as discussed in
\cite{hyp}, 
the $SU(3)$ invariant coupling $g$ is obtained assuming a unique value
$f_\pi \approx 130 MeV$ for all the light pseudoscalar meson decay constants
in the sum rule.
\begin{figure}[htb]
\epsfxsize=12cm
\epsfysize=3.5cm
\epsfbox{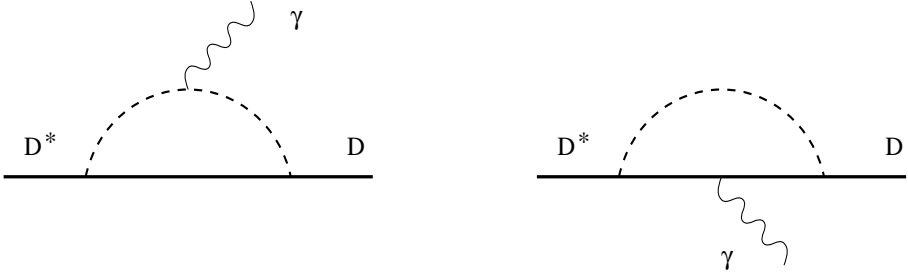}
\vspace{3mm}
\caption{Chiral loops contributing to the radiative decays
$D^{*} \to D \gamma$}
\label{figbep1}
\end{figure} 
The analysis of \cite{abj} provides a pattern for $SU(3)$ violations
in $D^*$ radiative decays, and can be in principle used to
determine $g$ and $\beta$, independently of the theoretical determinations 
based on the the QCD sum rule for $g$.
One can use the
two formulas:
\be
\Gamma(D^*_a \to D_a \gamma ) =  \frac{\alpha}{3} 
\frac{M_{D^*_a}}{M_{D_a}} 
|\mu_\alpha|^2 |{\vec k}|^3 \label{gg}
\ee
($ {\vec k}$ = photon momentum),
\be
\Gamma(D^{*+} \to D^0 \pi^+ )  =  \frac{g^2}{6\pi f_\pi^2} \label{gp}
|{\vec p_\pi}|^3
\ee
and the experimental results contained in table \ref{table:fer}
\cite{cleo,pdb}, together 
with the condition $g<1$ (which is experimentally satisfied \cite{acc}).
Because of the large experimental error (especially in the channel $D^{*+}
\to D^+ \gamma$), one gets, however, a rather broad range of
values for $g$ 
\cite{abj}\footnote{Similar results are obtained in \cite{cho}.}:
\be
0.3  < g  <  0.7 ~~.\label{5.1.g}
\ee
The value of $g$ obtained in this way is an 
effective coupling which takes into account part of the $1/m_c$ corrections,
as it is obvious from the fact that 
$\mu_a^{h}$ has not been neglected in 
comparison with $\mu_a^{\ell}$ ($\mu_a^{h}$ is not
negligible because 
$m_c$ is not sufficiently large: in this analysis one uses 
$\Lambda_c  =  m_c  =  1.7$ GeV).
It is nevertheless 
interesting to observe that the small values for $g$ in (\ref{5.1.g}) are
in broad agreement
with the results of the QCD sum rules quoted in section \ref{subsec:QCD}.

The analysis of \cite{abj} shows 
that smaller values of $g$ favour values of $\beta$
near the non relativistic quark model result \cite{eichten},
 where $\beta  =  m_q^{-1}$, and $m_q \approx 300 - 500 ~  MeV$ is a
typical value of the light quark constituent mass. 
In particular, from $g=0.38$ and $m_c=  1.7  $ GeV one gets the value
$\beta = 1.9  $ GeV$^{-1}$.
The pattern displayed by eq.(\ref{5.1.c}), 
$\mu_3^{\ell}  = -  \frac{1}{3} (\beta  -   3 g^2 {m_K}/(4 \pi 
f_K^2))$ $= -{1}/(3 \Lambda_3)$, 
can be interpreted, in the quark model,
as due to a  constituent strange quark having a 
mass $m_s = {\Lambda_3}$ larger than $m_q=\Lambda_1$, or
$\Lambda_2$,  which is what one would naively expect.

Instead of considering loop effects one can therefore take into account
$SU(3)$ violations by choosing explicitly different $\Lambda_a$'s. Quark model
calculations that assume $SU(3)$ violations are considered in 
\cite{cina1,defazio2}. In particular, the calculation of
ref. \cite{defazio2} in the semirelativistic quark model of ref. 
\cite{pietroni} 
discussed in section \ref{sub:cqm}, as we stated already, makes use of
the Salpeter equation \cite{salpeter}, i.e. a wave equation which
takes into account relativistic kinematics, with
an interquark potential modelled on the Richardson's potential 
\cite{richardson}. As 
we have observed, this model is able to explain the reduction
of the value of the strong ${D^* D \pi}$ coupling constant from the 
(non relativistic) quark model prediction $g =1$ down to $g \approx 0.33$,
as a consequence of the relativistic kinematics relevant to the light quark
in the heavy meson; since this small value is favoured by the
QCD sum rule  analyses \cite{qcdsr,bel}, as well as by the data
(eq.(\ref{5.1.g})) one may take this as an indication that relativistic
kinematics  plays a role also in the case of the radiative decays.
The results of the analysis in \cite{defazio2} for
the constants $\Lambda_a$ are displayed in table \ref{table:bep2}, together 
with the results of the
chiral loop calculation \cite{abj}, i.e. the results based on
eqs. (\ref{5.1.a})-(\ref{5.1.c}). In the same table
we also report the parameters of the model \cite{defazio1} based
on Vector Meson Dominance, to be discussed below.
\begin{table}[htb]
\caption{Theoretical inputs for mass parameters in radiative D decays. 
$\chi$-loop represents the chiral loop calculation, VMD is the model based on 
the effective lagrangian supplemented by the hypothesis of Vector Meson 
Dominance; RQM refers to the relativistic quark model; $\Lambda_Q$
and $\Lambda_a$ are mass parameters (in GeV).}
\label{table:bep2}
\centering
\begin{minipage}{12truecm}
\begin{tabular}{l c c c c c c} 
Decay mode  & \multicolumn{2}{c}{$\chi$-loop} & \multicolumn{2}{c}{VMD} & 
\multicolumn{2}{c}{RQM} \\ \hline
 & $\Lambda_Q$ & $\Lambda_a$ & $\Lambda_Q$ & $\Lambda_a$ & $\Lambda_Q$ & 
$\Lambda_a$ \\ \hline
$ D^{*+} \to D^+ \gamma $ & 1.7 & 0.61 & 1.9 & 0.50 & 1.57 & 0.48  \\ 
\hline
$ D^{*0} \to D^0 \gamma $ & 1.7 & 0.79 & 1.9 & 0.50 & 1.57 & 0.48  \\ 
\hline
$ D^*_s \to D_s \gamma $ & 1.7 & 1.11 & 2.0 & 0.59 & 1.58 & 0.497  \\ 
\hline
$ B^{*+} \to B^+ \gamma$ & 5.0 & 0.79 & 5.3 & 0.51 & 4.93 & 0.59  \\ 
\hline
$ B^{*0} \to B^0 \gamma$ & 5.0 & 0.61 & 5.3 & 0.51 & 4.93 & 0.59  \\ 
\hline
$ B^*_s \to B_s \gamma $ & 5.0 & 1.11 & 5.4 & 0.60 & 4.98 & 0.66  \\ 
\end{tabular}
\end{minipage}
\end{table}

In the case of the chiral loop calculation \cite{abj}, we 
have assumed
as an input
$g=0.38$, which is
the intermediate value among the different QCD sum rules results
\cite{qcdsr,bel,hyp}; 
on the other hand
$\beta = 1.9~{\rm GeV}^{-1}$ is fitted from the experimental CLEO data
of table \ref{table:fer}, using the branching ratio of $D^{*0} \to D^0 \gamma$
as an input. As for $\Lambda_c$, following \cite{abj}, we
take $\Lambda_c= 1.7  $ GeV; on the other hand for
$\Lambda_b$  we take the value $\Lambda_b =  5$ GeV
which, similarly to the $\Lambda_c$ 
case, is slightly larger than the value derived by QCD sum rules.
 
Let us now discuss the
model based on Vector Meson Dominance (VMD) \cite{defazio1}. 
In  this model  the calculation of $\mu^{\ell}_a$ is based on
the results obtained by the effective chiral lagrangian
approach. The idea is
to use VMD to express $<D_a|J_\mu^{\ell}|D^*_a >$ in terms
of $<D_a V|D^*_a>$ ($V=$ light vector meson resonance) and then
to employ information from heavy meson weak decays to compute
$<D_a V|D^*_a>$. In other terms one writes:
\bea
<D_a(p^{\prime})|J_\mu^{\ell}|D^*_a (p, \eta)> &=& \nonumber \\
= e_a \sum_{V, \lambda}  <D_a(p^{\prime}) V(q, {\epsilon_1}(\lambda))|
D^*_a(p, \eta)> & i &  {  <0|\bar{q}_a \gamma_\mu q_a 
|V(q, {\epsilon_1(\lambda)})>
\over  q^2-m_V^2} 
 \label{eq : 25} 
\eea
where $q^2=0$ and the sum is over the vector meson resonances $V=
\omega$, $\rho^0$, $\phi$ and over their helicities. 
The vacuum-to-meson current matrix 
element appearing in (\ref{eq : 25}) 
is given, assuming $SU(3)$ flavour symmetry, by:

\be 
 <0|\bar{q}_a \gamma_\mu q_a 
|V(q, {\epsilon_1})>
={\epsilon_1}^\mu f_V  
Tr(V T^a) \hskip 3 pt , \label{eq : 26} 
\ee
where $(T^a)_{lm}  =  \delta_{al} \delta_{am}$ and,
as usual,   $a=1,2,3$ for $u,d,s$ 
respectively. From $\omega \to e^+ e^-$ and $\rho^0 \to e^+ e^-$ decays 
\cite{pdb} one has $f_\omega =  f_\rho  = 
f_V  = 0.17 \hskip 3 pt {\rm GeV}^2$; from 
$\phi \to e^+ e^-$ one
obtains $f_{\phi}=f_V + \delta f$, with $\delta f=0.08~ 
{\rm GeV}^2$, which implies
a relevant $SU(3)$ violation.
Using (\ref{eq : 26}) and the 
strong lagrangian containing
the vertex $D^* D V$ (see eq.(\ref{1.11})):
\be
{\cal L} = i \lambda < H_b \sigma^{\mu \nu} F_{\mu \nu}(\rho)_{ba}
{\bar H}_a>~,
\ee
one can compute (\ref{eq : 25}).
The results in terms of the mass constants $\Lambda_a$ are as follows:
\be 
\Lambda_a^{-1}  =  - 2
\sqrt{2} g_V \lambda {\sqrt{\frac{M_{D^*}}{M_D}}} \rho_a
\label{27} 
\ee
where $\rho_1=\rho_2=f_V /m^2_{\omega}$, $\rho_3=f_\phi/ 
m^2_{\phi}$, $g_V=5.8$. 
Equation (\ref{lambda}) only gives the absolute value of $\lambda$, 
but eq. (\ref{27}) clearly shows that $\lambda < 0$ if $\Lambda_a$ has
to be interpreted, as in the quark model, as  a mass parameter.
Therefore we take (see eq. (\ref{lambda})):
\be
\lambda \simeq -0.41~  {\rm GeV}^{-1}~~. 
\label{lambdaprimo} 
\ee

The results of this approach are reported in table \ref{table:bep2}, 
together with  the chiral loop ($\chi$-loop) and the 
relativistic quark model predictions.

 From eqs. (\ref{gg}) and (\ref{gp}) and from table \ref{table:bep2} we get the
decay rates and branching ratios (BR) for both $D^*$ and 
$B^*$ decays; they are reported in table \ref{table:bep3} for the three
models examined so far. For the chiral loop calculation and the VMD approach
we use the same value $ g=0.38$ for the strong $BB^* \pi$ coupling constant,
whereas for the third column 
we take $g \simeq 0.39$ as predicted by the relativistic quark model
\cite{defazio2}.
\begin{table}[htb]
\caption{Theoretical predictions for $D^*$ and $B^*$ widths and branching
ratios. The radiative decay widths are computed by the 
parameters of the preceding table.}
\label{table:bep3}
\centering
\begin{minipage}{12truecm}
\begin{tabular}{l c c c c } 
Decay rate/ BR & $\chi$-loop & VMD & RQM &\\ \hline
$\Gamma(D^{*+})$ & $39.5 \hskip 3 pt {\rm KeV}$ & $40.0 \hskip 
3 pt $KeV & $46.2 \hskip 3 pt $KeV & 
\\ \hline
$ BR(D^{*+} \to D^+ \pi^0)$ & $31.5 \%$  & $31.1 \%$  & 
$31.3 \; \%$ & \\ 
\hline
$BR(D^{*+} \to D^0 \pi^+)$ & $68.1  \%$  & $67.3 \%$  & 
$67.7 \; \%$ & \\ 
\hline
$BR(D^{*+} \to D^+ \gamma)$ & $0.4 \%$ &  $1.6 \%$ & 
 $1.0 \; \%$ & \\ 
\hline \\ \hline 
$\Gamma(D^{*0})$ & $28.3 \hskip 3 pt $KeV  & $37.1 
\hskip 3 pt $KeV  & $41.6 \hskip 3 pt $KeV &  \\ 
\hline
$ BR(D^{*0} \to D^0 \pi^0)$ & $63.6 \%$  & 
$51.5 \%$  & $50.0 \;  \%$ & \\ 
\hline
$BR(D^{*0} \to D^0 \gamma)$ & $36.4 \% \; (input) $  & $48.5
\%$  & $50.0 \;  \%$ & \\ 
\hline  \\ \hline
$\Gamma(D^*_s)=
\Gamma(D^*_s \to D_s \gamma)$ & $0.06 \hskip 3  pt $KeV & 
 $0.35 \hskip 3  pt $KeV  & $0.38  \hskip 3 
 pt $KeV &  \\ \hline 
\\ \hline
$\Gamma(B^{*+})=\Gamma(B^{*+} \to B^+ \gamma)$ & $0.14 
\hskip 3  pt $KeV   & $0.37
\hskip 3  pt $KeV   & $0.24
\hskip 3  pt $KeV   &  \\ \hline 
$\Gamma(B^{*0})=\Gamma(B^{*0} \to B^0 \gamma)$ & $0.09
\hskip 3  pt $KeV   & $0.12
\hskip 3  pt $KeV   & $0.092  
\hskip 3  pt $KeV   &  \\ \hline
$\Gamma(B^{*}_s)=\Gamma(B^{*}_s \to B_s \gamma)$ & $0.03
\hskip 3  pt $KeV   & $0.09
\hskip 3  pt $KeV   & $0.08
\hskip 3  pt $KeV   &  \\
\end{tabular}
\end{minipage}
\end{table}

We see that the chiral loop calculation, which uses the data on $D^{*0}
\to D^0 \gamma$ to fix the light mass scale, reproduces quite well the 
$D^{*+}$ decay branching ratios; also the quark model and the VMD  predictions
(that are parameter free) are in reasonable
agreement with the data.

Let us compare these results with other approaches.
The result of the calculation in \cite{cho}, based
on the ideas of HQET, is as follows:
$\Gamma(D^{*0} \to D^0 \gamma)  =  8.8 \pm 17.1$ KeV and
$\Gamma(D^{*+} \to D^+ \gamma)  =  8.3 \pm 8.1$ KeV,
if one uses $m_c =  1.7 $ GeV. For the $B$ case, 
with $m_b =  5.0 ~ $GeV, 
 $\Gamma(B^{*0})  =  0.13 \pm 0.20 $ KeV, and
$\Gamma(B^{*+})  =  0.66 \pm 0.93$ KeV are found. Clearly the results of
this calculation are dominated by
large experimental uncertainties.

The result of  a QCD sum rule calculation \cite{aliev},
which updates previous analyses \cite{eletsky}, is as follows:
$\Gamma(D^{*0} \to D^0 \gamma)  =  2.43 \pm 0.21$ KeV,
$\Gamma(D^{*+} \to D^+ \gamma)  =  0.22 \pm 0.06$ KeV, and 
$\Gamma(D^{*}_s \to D_s \gamma)  =  0.25 \pm 0.08$ KeV.
$D^*$ decays have been also studied in the framework
of the bag model \cite{singer} with the following results (for the value
$\lambda=1$ of the relevant parameter in that paper): $\Gamma (D^{*+}) \simeq
80 $ KeV and $\Gamma (D^{*0}) \simeq 60$ KeV, a factor $1.5 - 2$
larger than the results contained in table \ref{table:bep3}; 
this model also predicts
$\Gamma (D^{*+} \to D^+ \gamma) \simeq 1$ KeV and the correct
ratio $(D^0 \gamma)/(D^0 \pi^0 )$. Correspondingly, the computed $B^*$ 
radiative width is also larger than the entries in table \ref{table:bep3}.

Let us finally observe that similar calculations can
be performed for the radiative decays of positive parity charmed meson 
resonances \cite{defazio1,kilian,luke}. 
For neutral resonances
the computed branching ratios are of the of the order $10^{-4} - 10^{-3}$;
for the  
charged
parity resonances the computed branching ratios are much smaller,
due to an almost complete cancellation between the two contributions in the 
e.m. current \cite{defazio1}.  

\subsection{Weak radiative decay: $B \to \ell \nu  \gamma$ }

Another interesting process where the formalism of
the effective chiral
lagrangian can be applied is
the radiative leptonic decay channel:
\begin{equation}
B^- \to \mu^- {\bar \nu_{\mu}} \gamma \hskip 3 pt .  \label{channel} 
\end{equation}
It has been suggested \cite{atwood,burdmanf,fdfz3}
that this decay channel can be used to extract the $B^*$
decay constant $f_{B^{*}}$; since in the $m_b \to \infty$ limit
$f_{B^{*}} =  f_{B}  = {\hat F}/{\sqrt {M_B}}$, the analysis of
(\ref{channel}) can represent an alternative way
to measure ${\hat F}$ as compared to the purely leptonic decay channel
\begin{equation}
B^- \to \mu^- {\bar \nu_{\mu}} \hskip 3 pt .  \label{leptonic} 
\end{equation}

The branching ratio for 
the purely leptonic channel is given by:
\be
 BR (B^- \to \mu^- {\bar \nu_\mu})  \simeq   2.6  
\lq {V_{ub} \over 0.003} \rq^2
\lq{f_B \over 200  MeV} \rq^2  10^{-7}~~  , \label{bmn}
\ee
where one uses 
$\tau_{B^-}=1.55$ ps. This  
result is two order of magnitudes smaller than
the present experimental upper bound put by CLEO
\cite{CLEOIIa}:  $BR (B^- \to \mu^- {\bar \nu_\mu})< 2.1 \cdot 10^{-5}$.
For the channel 
$B^- \to e^- {\bar \nu_e}$ one expects a  much smaller $BR$
(of the order $10^{-12} -10^{-11}$), 
because of the helicity suppression; on the other hand
in the channel $B \to \tau \nu_\tau$, the helicity suppression 
is absent and
the expected $BR$ is of the order $ 7 \times 10^{-5}$, but
the 
$\tau$ identification represents a serious experimental
problem.

Because of the small value that is expected for the leptonic channel 
(\ref{bmn}), the radiative decay
(\ref{channel}) may be a serious competitor. 
Various estimates \cite{fdfz3,eil2} 
of its branching ratio indicate that 
the  radiative decay rate 
is larger than the purely leptonic
one by almost an order of magnitude, 
mainly because the radiative decay, differently from the leptonic channel,
is not helicity-suppressed due to 
the  photon in the final state.

Let us now describe how one could extract the value $ {\hat F}$
from (\ref{channel}). First of all one should distinguish between the two
classes of diagrams describing the 
radiative process.
The first class contains 
bremsstrahlung diagrams where the photon is emitted from 
the $B^-$ or from the charged lepton leg.
 This
contribution vanishes in the limit $m_\mu \to 0$ and
is negligible  also for finite lepton mass.

The relevant diagrams for this process are of the type depicted
in fig. \ref{fig:iiwcorr} \cite{fdfz3}. Other possible diagrams 
are chiral loop contributions, where, instead of the single particle
intermediate state, one has a chiral loop with the $B$ and a pseudoscalar
particle (these contributions are discussed in 
\cite{burdmanf}). One considers only the resonant pole diagrams, for which
no problem of double counting arises.
\begin{figure}[htb]
\epsfxsize=15cm
\epsfysize=3cm
\epsfbox{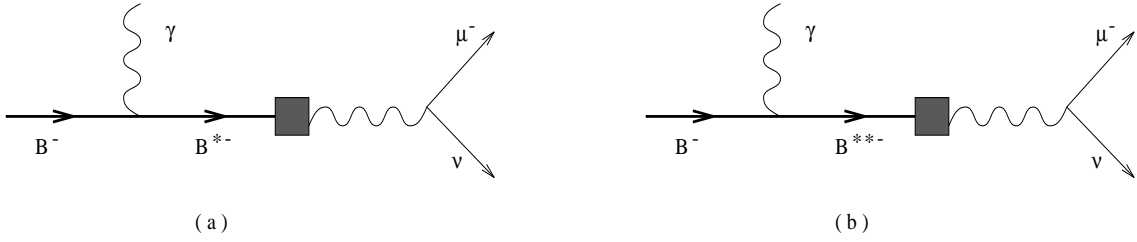}
\vspace{3mm}
\caption{Diagrams dominating the $B^- \to \ell^- {\bar \nu}_\ell \gamma$ 
decay mode in the limit $m_\ell \to 0$. $B^*$ is the vector meson state, 
$B^{**}$ is the $1^+$ axial vector meson state.}
\label{fig:iiwcorr}
\end{figure} 
Let us assume, following \cite{burdmanf} and  \cite{fdfz3}, 
that in the pole diagrams the intermediate state is a 
$J^P=1^- (B^*)$ or a positive parity $B^{**}$ meson.
The amplitude with intermediate $P(=B^*, B^{**})$ state
is written as follows:
\begin{equation}
{\cal M}_{SD}^{(P)}= {G_F \over \sqrt{2}} V_{ub} 
{\cal A}(B \to P \gamma) 
{i \over (p-k)^2-m^2_P} <0|{\bar u} \gamma^{\mu} (1-\gamma_5)b|P>
{\it l}_{\mu} \hskip 3 pt , 
\label{2new} 
\end {equation}
\noindent where $p$ and $k$
are the $B$ and photon momenta respectively,
${\it l}_{\mu}={\bar \ell}(p_{\it l}) \gamma_{\mu}(1-\gamma_5) 
\nu(p_{\nu})$ is the lepton current, ${\cal A}(B \to P \gamma)$ is the 
amplitude of the process $B \to P \gamma$, and $P$ indicates the pole. 
When in (\ref{2new}) one takes  $P=B^*$, the matrix element
becomes proportional to $f_{B^*}$; therefore, if the contribution
of the higher mass resonances  is negligible
(and we shall see that this is indeed the case), and
for light leptons in the final 
state, the radiative flavour changing $B$ decay can be used to measure
the decay constant $f_{B^*}$, provided  the amplitude  
${\cal A}(B^* \to B \gamma)$ is known. A direct
measurement of
the $B^*$ width would be extremely difficult
(as mentioned in section \ref{sec:bepp1}, it is less than 1 KeV     
in all the models); however
this amplitude can be indirectly obtained 
by using the heavy flavour symmetry already employed
in section \ref{sec:bepp1} to relate the radiative $D^*$ and $B^*$ decays.

If  the partial
width $\Gamma(D^{*0} \to D^0 \gamma)$ 
is measured (only the branching ratio is available
so far, see table \ref{table:fer}, then one could extract, from the
amplitude
\begin{eqnarray}
{\cal A}(D_a^*(v,\eta) \to D_a(v^{\prime}) \gamma(q, \epsilon))= 
&i&  e  \left[{e_c \over \Lambda_c}  + {e_a \over \Lambda_a} 
\right] \nonumber 
\\
&& M_{D^*} \sqrt{M_D M_{D^*}} \epsilon_{\mu \nu \alpha \beta} 
\epsilon^{* \mu} \eta^{\nu}
v^{\alpha} v^{\prime \beta}  \hskip 3 pt , 
\label{3new} 
\end{eqnarray}
the mass constant $\Lambda_a$ to be
used in the formula giving
${\cal A}(B^* \to B \gamma)$:

\bea
{\cal A}(B_a^*(v,\eta) \to B_a(v^{\prime}) \gamma(q, \epsilon))= 
&i&  e  \left[{e_b \over \Lambda_b}  + {e_a \over \Lambda_a} 
\right] \nonumber 
\\
&& M_{B^*} \sqrt{M_B M_{B^*}} \epsilon_{\mu \nu \alpha \beta} 
\epsilon^{* \mu} \eta^{\nu}
v^{\alpha} v^{\prime \beta}  \hskip 3 pt , \label{33} 
\eea

The expression for the amplitude ${\cal M}_{SD}^{(B^*)}$ 
giving the contribution of the
$B^*$ pole to the decay $ B^- \to \mu^- {\bar \nu_\mu} \gamma$ is  
therefore as follows:
\begin{equation}
{\cal M}_{SD}^{(B^*)}= {C_1 f_{B^*} \over (v \cdot k + \Delta)} 
\epsilon_{\mu \sigma \alpha \beta}  \l^\mu  \epsilon^{* \sigma}  
v^{\alpha}  k^{\beta}  , 
\label{5due}   
\end{equation}
where $\Delta =M_{B^*}-M_B$, and $C_1$ is given by:
\begin{equation}
C_1={G_F \over \sqrt{2}} V_{ub}  {M_{B^*} \over 2 M_B} {\sqrt {M_{B^*} M_B}}
\; e \; \Big[{e_b \over \Lambda_b} +
{2  \over { 3\Lambda_1}} \Big] \hskip 3 pt . \label{10} \end{equation}
 From (\ref{5due}) one can compute the contribution of the $B^*$ pole to 
$BR(B^- \to \mu^- {\bar \nu}_\mu \gamma)$ as a function of the parameter 
${\hat F}$. 
Before considering some numerical
predictions, let us study the effect of the
 $B^{**}$ pole, i.e. the positive parity
heavy meson having  $s_\ell=1/2$ (only the axial vector states $1^+$ 
can contribute as poles to the 
decay; moreover, the state $B_1$, having 
$s_\ell=3/2$, has vanishing coupling to the weak current in the limit
$m_b \to \infty$ \cite{casa93} (see the discussion after 
(\ref{c6}); 
therefore only the state 
$B_1$ (with $s_\ell=1/2$) gives a contribution in the same limit.

The contribution
of this state is:
\begin{equation}
{\cal M}_{SD}^{({\tilde B}_1)}= i {C_2 f_{B_1} \over
(v \cdot k + \Delta^{\prime}) } 
(\epsilon \cdot v  k_{\mu}  -v \cdot k  \epsilon_{\mu}) 
{\it l}^{\mu}\hskip 3 pt  
\label{7new} 
\end{equation}
\noindent where $\Delta^{\prime}=M_{B_1}-M_B \simeq 500 \, MeV$,
\begin{equation}
<0|{\bar b} \gamma_\mu \gamma_5 q|B_1(p,\eta)>=f_{B_1}
M_{B_1}\eta_\mu ~~ , \label{fb1} \end{equation}
\noindent  and 
\begin{equation}
C_2={G_F \over \sqrt{2}} V_{ub} { M_{B_1}\over 2 M_B} 
\sqrt{M_B M_{B_1}}  e 
\Big[{2 e_b \tau_{1/2}(1) \over M_B} + 
{e_a  \over \Lambda_a^{\prime}} \Big] \hskip 3 pt . \label{11} 
\end{equation}
\noindent 
The function  $\tau_{1/2}(v \cdot v^{\prime})$ 
is the universal form factor, analogous to the Isgur-Wise function,
that appears in
the matrix element of the heavy quark current
$J_\mu^h$ between a positive $s_\ell=1/2$ 
and a negative parity heavy meson state.
$\tau_{1/2}(v \cdot v^{\prime}) $
has been introduced in \cite{IWtau} and computed 
in \cite{CNP} by QCD sum rules, with the
result $\tau_{1/2}(1)\simeq 0.24$.

Using this input and 
the ratio 
$(f_{B_1} / f_{B^*}) (\Lambda_a  \Lambda^{\prime}_a)$ 
estimated  in \cite{fdfz3}, one obtains
\be
{\Gamma^{(B_1)} \over \Gamma^{(B^*)}} \simeq 0.1 \; ,
\ee
\noindent 
which confirms the previous hypothesis, i.e.
that  the $B_1$ pole represents only a small
contribution to the final result.

The results are sensitive to  the experimental input 
$\Gamma(D^{*0} \to D^0 \gamma)$. For reasonable values
of this quantity, which has not been measured yet, one
finds for
$BR(B^- \to \mu^- {\bar \nu_\mu} \gamma)$ 
a result in the range $10^{-7}  -  10^{-6}$, i.e. 
a radiative branching ratio 
larger 
than the leptonic $BR(B^- \to \mu^- \nu)$;
the enhancement is obviously still higher if the electron 
leptonic decay channel is considered, which contributes by a factor of 2.
Therefore, in principle, the decay channel
$B^- \to \mu^- {\bar \nu}_\mu \gamma$ can be used as a way 
to measure the 
leptonic  $B^*$ decay constant.

\subsection{Weak radiative decays: $B \to V \gamma$ }
\label{sec:bepp3}

This section is devoted to the analysis
of the
exclusive flavour changing radiative $B$ decays with a vector
meson in the final state. These channels, at short 
distances, are described by processes
\be
b \to s \gamma   \label{inclusive}
\ee
or
\be
b \to d \gamma \; , \label{inclusived}
\ee
which are dominated
by a penguin diagram with the top as intermediate quark; these
decays
have been intensively studied in the past \cite{bertolini} and experimental
data have been collected both for the inclusive decay
\cite{cleobsg}:
\be
BR (B \to X_s \gamma)  =  (2.32 \pm 0.57 \pm 0.35 ) \times
10^{-4} \; , \label{cleoincl}
\ee
and for the exclusive decay process
\be
B \to K^*\gamma \label{bkg}
\ee
for which the following result has been obtained \cite{cleobkg}:
\be
BR (B \to K^{*}\gamma) =  
(4.3^{{\displaystyle +1.1}}_{{\displaystyle -1.0}}
\pm 0.6) \times 10^{-5} \; . \label{result}
\ee
The process (\ref{bkg}) has been studied within the framework of the heavy
quark effective theory \cite{isgurgamma,santorelli,gatto,donnell,griffin}
(for  reviews see \cite{ali}),
by QCD sum rules \cite{paver0,paver,ball2,braunbkg}, 
\cite{narison} or Lattice QCD 
\cite{latticeqcd}. In this section
we shall review the application of the heavy meson effective chiral lagrangian
\cite{Casalbuoni1}
to the decay process (\ref{bkg}) as well as to the other exclusive
decay channels
\bea
&B_s & \to  \phi \gamma \label{bfg} \\
&B & \to  \rho \gamma \label{brg} \\
&B & \to  \omega \gamma~~. \label{bog} 
\eea

Before doing this, let us consider however the simpler approach
already considered in section \ref{subsec-brhoscal}, consisting in
the application
of the light and heavy flavour symmetries \cite{gatto}.

Let us begin with the $ b \to s \gamma$
transition. All the exclusive processes arising from 
this elementary decay are dominated
at the quark level by the short
distance $b\to s\gamma$ hamiltonian \cite{bertolini} given by
\be
H_\gamma=Cm_b\bar s\sigma^{\mu\nu}(1+\gamma_5)b F_{\mu\nu}~~~+~~~{\rm h.c.}
\label{Hg}
\ee
(neglecting terms of order $m_s/m_b$). $F_{\mu\nu}$ is the
electromagnetic tensor, $\sigma_{\mu\nu}=\frac{i}{2}[\gamma_\mu,\gamma_\nu]$
and $C$ is given by
\be
C=\frac{G_F}{\sqrt{2}}\frac{e}{16\pi^2}V_{tb}V_{ts}^* F_2\left(\frac{m_t^2}
{m_W^2}\right)
\ee
where $ F_2$ is a factor including perturbative QCD corrections, and
slightly dependent
on the top quark mass. For $m_t=175$ GeV, it has the value: $F_2=0.63$.
Together with the short distance hamiltonian, one should
take into account also long distance effects, such as transitions
mediated by four quark operators; they will be discussed below.

The short distance hadronic matrix element relevant to the transition
$\bar B\to K^*\gamma$ ($\bar B= B^-$ or ${\bar B}^0$) can be expressed as
follows:
\bea
\langle K^*(p^\prime,\epsilon)|\bar s\sigma^{\mu\nu}(1+\gamma_5)b|{\bar B}
(p)\rangle  &=& i\Big\{A(q^2)\left[p^\mu\epsilon^{*\nu}-p^\nu\epsilon^{*\mu}-
i\epsilon^{\mu\nu\lambda\sigma}p_\lambda\epsilon^*_\sigma\right] \nn\\
&+& B(q^2)\left[p^{\prime\mu}\epsilon^{*\nu}-p^{\prime\nu}\epsilon^{*\mu}-
i\epsilon^{\mu\nu\lambda\sigma}p^\prime_\lambda\epsilon^*_\sigma\right] \nn\\
&+& H(q^2)(\epsilon^*\cdot p)\left[p^\mu p^{\prime\nu}-p^\nu p^{\prime\mu}-
i\epsilon^{\mu\nu\lambda\sigma}p_\lambda p_\sigma^\prime\right]\Big\}
\label{ffrad}
\eea
where we have used the property
$\frac{i}{2}\epsilon^{\mu\nu\lambda\sigma}\sigma_{\lambda\sigma}\gamma_5=
-\sigma^{\mu\nu}$.

Now, as first noted in \cite{isgurgamma}, as a
consequence of the equations of motion of the heavy quark
\be 
\frac{1+\slash{v}}{2}b=b \; ,
\ee
in the $b$ rest frame one has
\be
\gamma^0 b=b \; ,
\ee
which means that
\be
{\bar q}^a\sigma_{0i}(1+\gamma_5)Q=-i{\bar q}^a\gamma_i(1-\gamma_5)Q \; .
\label{relat}
\ee
Therefore the form factors (\ref{ffrad}) can be related to those describing
the weak semileptonic transition $B \to K^*$ (or $B \to \rho $,
using $SU(3)$ symmetry). 
In computing the width for the decay $B\to K^*\gamma$ 
only the combination of form factors $A+B$ 
is relevant, whereas
$H(q^2)$ contributes to amplitudes with virtual photons.
Using (\ref{relat}), the form factors
$A$, $B$ and $H$ are related to the form factors $V$, $A_1$
and $A_2$  as follows:
\bea
A(q^2) &=& i\left\{\frac{q^2-M_B^2-m_{K^*}^2}{M_B}
\frac{V(q^2)}{M_B+m_{K^*}}-\frac{M_B+m_{K^*}}{M_B}A_1(q^2)\right\}
\label{aq2} \\
B(q^2) &=& i\frac{2M_B}{M_B+m_{K^*}} V(q^2)  \label{bq2} \\
H(q^2) &=& \frac{2 i}{M_B}\left\{
\frac{V(q^2)}{M_B+m_{K^*}}+\frac{1}{2q^2}\frac{q^2+M_B^2-m_{K^*}^2}
{M_B+m_{K^*}}  A_2(q^2)\right\}~~. \label{hq2} 
\eea

These relations are strictly valid for $q^2\approx q^2_{max}$. However,
following \cite{isgurgamma,burdman2,donnell}
 (see however 
\cite{griffin} ),
one can assume their
validity down to the value $q^2=0$ which is the kinematical point relevant
for decays with a real photon in the final state.

In order to compute $A(0)+B(0)$, one needs the values of the form factors
for the transition $B \to K^*$; they have been computed 
by $ D\to K^*$ semileptonic decays using heavy flavour symmetry 
(see table \ref{brho1} of section \ref{subsec-brhoscal}, neglecting
$SU(3)$ flavour breaking). Using the soft pole column result, we
have $V^{B \to K^*}(0)=0.64$ and $A_1^{B \to K^*}(0)=0.21$.
We observe, however, that, on the basis of the scaling relations given in 
the previous section, eq. (\ref{v12}), since in this case we
assume the same pole behaviour for $V(q^2)$ and
 $A_1(q^2)$, the term in $A_1$ in eq.(\ref{aq2}) is subleading
(in $1/m_Q$) for any value of $q^2>0$ and should be neglected
in comparison to the term proportional to $V(q^2)$, if one works
consistently at a given order in the $1/m_Q$ expansion. In this way
one gets the result:
\be
|A(0)+B(0)|=0.53 \; . \label{resab}
\ee
Following the discussion in section \ref{subsec-brhoscal}, we may estimate 
for
this result a theoretical 
uncertainty of at least $\pm 30 \%$.
We  consider now the radiative width, which is
given by
\be
\Gamma(B\to K^* \gamma)=\left(\frac{M_B^2-m^2_{K^*}}{2M_B}\right)^3
\frac{2|C|^2m_b^2}{\pi}|A(0)+B(0)|^2 \; .
\ee
One gets
\be
BR(B\to K^{*}\gamma)=\left[ 2.4\times \left( |V_{ts}|/0.039 \right)^2
\right]\times 10^{-5} ~. \label{thbkg1}
\ee
In the
previous formula we have used $|V_{tb}|\simeq 1$, $m_b=4.7~$GeV and
$\tau_{B^+}\simeq\tau_{B^0}\simeq\tau_{B_s}\simeq 1.55~ps$.
The  result based on (\ref{resab}) agrees within
the errors with the experimental finding (\ref{result}). 
It should be noted that eq.(\ref{thbkg1}) does not include
long distance effects due the $c \bar c$ quark loop \cite{milana} that 
raise the branching ratio by $\approx 20\%$, thus improving
the agreement; we shall discuss them in more detail below.
A similar analysis, with obvious changes, applies to the decay
$B_s\to\phi\gamma$ and one obtains 
\be
BR(B_s\to \phi\gamma) \approx BR(B\to K^* \gamma)~,
\ee
due to approximate $SU(3)$ light flavour symmetry.

Now we discuss how the $b \to s \gamma$ exclusive decays
can be described by the effective chiral lagrangian approach.

At the lowest order in the derivatives of the pseudoscalar field, the weak
tensor current between light pseudoscalar and negative parity heavy mesons
is as follows:
\be
L_{\mu\nu}^a=i\frac{ {\hat F}}{2} <\sigma_{\mu\nu}(1+\gamma_5)H_b
\xi^\dagger_{ba}> ~,\label{lmunu}
\ee
and it has the same transformation properties of the quark current
${\bar q}^a\sigma^{\mu\nu}(1+\gamma_5)Q$. Together with (\ref{lmunu}) 
we also consider
the weak effective current (\ref{c1})
corresponding to the quark $V-A$ current 
${\bar q}^a\gamma^\mu(1-\gamma_5)Q$:
\be
L_\mu^a=i\frac{ {\hat F}}{2}\langle\gamma_\mu(1-\gamma_5)H_b\xi^\dagger_{ba}
\rangle~~. \label{lmu}
\ee
We put the same coefficient $i {\hat F}/2$ in both (\ref{lmunu}) and 
(\ref{lmu})  because, due to (\ref{relat}), the relation
\be
L_{0i}=-iL_i \label{l0i}
\ee
must be satisfied.

We also introduce the weak tensor current containing the light vector meson
$\rho^\alpha$ and reproducing the bilinear ${\bar 
q}^a\sigma^{\mu\nu}(1+\gamma_5)Q$
\be
L_{1a}^{\mu\nu}=i\alpha_1\left\{g^{\mu\alpha}g^{\nu\beta}-\frac{i}{2}
\epsilon^{\mu\nu\alpha\beta}\right\}\langle\gamma_5H_b\left[
\gamma_\alpha(\rho_\beta-\VV_\beta)_{bc}-\gamma_\beta(\rho_\alpha-
\VV_\alpha)_{bc}\right]\xi^\dagger_{ca}\rangle~~.
\ee
$L_{1a}^{\mu\nu}$ is related to the vector current $L_{1a}^\mu$, 
eq. (\ref{c5}),
introduced
 to describe at the meson level the quark current operator
${\bar q}^a\gamma^\mu(1-\gamma_5)Q$  
taken  between light vector particles and heavy mesons:
\be
L_{1a}^\mu=\alpha_1\langle\gamma_5 H_b(\rho^\mu-\VV^\mu)_{bc}\xi^\dagger_{ca}
\rangle
\ee
(we keep only the leading term in $1/m_Q$).
We notice that in order to construct the tensor current we have imposed
(\ref{l0i}).

To compute $B\to K^*\gamma$ we consider a pole diagram 
having as intermediate state  between the current and the $B K^*$ 
system either a $1^+$ or a $1^-$  heavy meson; moreover we  add 
a direct term.
The effective lagrangian and the effective tensor
currents can  reliably describe
the process only for large values of $q^2$, i.e. $q^2\approx q^2_{max}=(M_B-
m_{K^*})^2$. This is a general feature of the chiral lagrangian
approach. Again, in order to extend our
results to small values of $q^2$, we shall assume a polar dependence in $q^2$ 
(with pole mass suggested
by dispersion relations). By following a procedure
similar to section \ref{subsec-brhoeff}, 
we obtain the results of table \ref{table:bep4} that are valid for any $q^2$
and in the limit $m_Q\to\infty$. We notice that in writing the various
contributions in table \ref{table:bep4}
we have left the dependence of $p\cdot p^\prime$
on $q^2$, $p\cdot p^\prime= (M_B^2+m_{K^*}^2-q^2)/2$, in the
term arising from the $1^-$ pole and we have assumed that the direct term
has a pole dependence with  mass given by the $1^+$ pole. These
choices can be justified as follows. The results in table \ref{table:bep4}
satisfy, for $q^2\approx q^2_{max}$, the  relations 
(\ref{aq2} - \ref{hq2}) between form
factors of vector and tensor currents.
Eqs. (\ref{aq2}) and (\ref{bq2}) coincide with the relations found in ref. 
\cite{isgurgamma};
as for (\ref{hq2}), the result of \cite{isgurgamma}:
\bea
H(q^2) = \frac{2 i}{M_B}\Big\{
\frac{V(q^2)}{M_B+m_{K^*}} &+&\frac{1}{2q^2}\Big( 
\frac{q^2+M_B^2-m_{K^*}^2}
{M_B+m_{K^*}}  A_2(q^2) \nn \\
&+& 2 m_{K^*} A_0 (q^2)-(M_{B}+m_{K^*}) 
A_1 (q^2) \Big) \Big\}
\eea
differs from (\ref{bq2})  
for terms that are subleading in the limit $m_Q\to\infty$ 
and can be neglected. 
\begin{table}
\caption{Terms contributing to the various form factors of the transition
$B\to K^*\gamma$. $m_P$ is the pole mass ($m_P=5.9~$GeV for the direct
and $1^+$ term; and $5.43~$GeV for the $1^-$ contribution). $p\cdot p^\prime=
(M_B^2+m_{K^*}^2-q^2)/2$.}
\label{table:bep4}
\centering
\begin{minipage}{12truecm}
\begin{tabular}{l c c c } 
${\rm Form~Factor}$ & ${\rm Direct}$ & $1^-$ & $1^+$  \\ \hline\\
$A(q^2)$ &$\displaystyle{
\frac{i\sqrt{2}g_V\alpha_1}{\sqrt{M_B}}\frac{q^2_{max}-m_P^2}
{q^2-m_P^2}}$ &$\displaystyle{
\frac{i2\sqrt{2}{\hat F}\lambda g_V(p\cdot p^\prime)}
{(m_P^2-q^2)\sqrt{M_B}}}$ &$\displaystyle{
\frac{-i\sqrt{2M_B}{\hat F}^+  g_V(\zeta-2\mu
m_{K^*})}{m_P^2-q^2}}$ \\\\ \hline\\
$B(q^2)$ &0 &$\displaystyle{
\frac{-i2\sqrt{2}{\hat F}\lambda g_V M_B^{3/2}}{m_P^2-q^2}}$
 & 0 \\\\ \hline\\
$H(q^2)$ & 0&$\displaystyle{
\frac{-i2\sqrt{2} {\hat F} \lambda g_V}{(m_P^2-q^2)\sqrt{M_B}}}$ &
$\displaystyle{
\frac{-i2\sqrt{2M_B}{\hat F}^+ g_V \mu}{(m_P^2-q^2)M_B}}$ \\
\end{tabular}
\end{minipage}
\end{table}

Following \cite{isgurgamma} and \cite{burdman2} and our previous 
discussion, we assume that the results (\ref{aq2} - \ref{hq2}) hold
also for small values of $q^2$, which justifies the above mentioned choices
in table \ref{table:bep4}.

Before computing the entries of table \ref{table:bep4} let us observe that
the results in the columns "Direct" and $1^+$ are subleading
as compared to those in the column $1^-$. In other terms 
\be
\frac{(A+B)_{Direct+1^+}}{(A+B)_{1^-}}= O \left( \frac{1}{m_b} \right )
\ee
for any value of (positive) $q^2$. Therefore, consistently with
the neglect of the $O({1}/{m_b})$ contributions, we do not keep them, 
which means that only the term arising from the exchange of the
$1^-$ particle is taken into account. In this way one obtains:
\be
A(q^2)+B(q^2)= -i \sqrt{\frac{2}{M_B}}\frac{{\hat F} \lambda g_V}{M^2_{B^*_s}}
(q^2+M_B-m^2_{K^*}) \; , \label{abq2}
\ee
which gives, for
${\hat F}= 0.30 ~{\rm GeV}^{3/2}$, $g_V=5.8$ (see eq. (\ref{rob:gv}))
and $\lambda=-0.41 ~ {\rm GeV}^{-1}$
(eq.(\ref{lambdaprimo})), the result:
\be
|A(0)+B(0)|=0.41,  \label{ab0}
\ee
and, therefore, $
BR(B\to K^{\star }\gamma)=\left[ 1.4\times \left( V_{ts}/0.039 \right)^2
\right]\times 10^{-5}$.
A similar analysis, with obvious changes, applies to the decay 
$B_s\to\phi\gamma$. In this case one obtains $|A(0)+B(0)|=0.42$ and 
$BR(B_s\to \phi\gamma)=\left[1.6 \times \left( V_{ts}/0.039 \right)^2
\right]\times 10^{-5}$.
In table \ref{table:bep5} we compare the analyses based on the scaling approach
and on effective chiral lagrangian to the results
of  QCD sum rule calculations \cite{paver,braunbkg}; 
other QCD sum rules
analyses \cite{ball2,narison} agree with \cite{paver}
and \cite{braunbkg}. We note
that the results based on the use of the heavy flavour symmetry
(first and second row in the table) are generally smaller than
the QCD sum rules outcome.
\begin{table}
\caption{Theoretical values of the $B \to K^* \gamma$
coupling $|A(0)+B(0)|$ in different approaches: soft pole, 
chiral lagrangian, QCD sum rules calculations based on the 
evaluation of three point function and light cone sum rules respectively.}
\label{table:bep5}
\centering
\begin{minipage}{6truecm}
\begin{tabular}{l c c c c } 
Model & & $|A(0)+B(0)|$   & &\\ \hline
soft pole (sec.\ref{subsec-brhoscal})& & 0.53 && 
\\ 
$\chi$-lagrangian eq.(\ref{abq2})& & $0.41$ && 
\\ 
QCD sum rules \cite{paver}& &$ 0.70\pm 0.10$ && 
\\ 
QCD sum rules \cite{braunbkg}& &$ 0.64\pm 0.10$ && 
\\ 
\end{tabular}
\end{minipage}
\end{table}

Also lattice QCD \cite{latticeqcd} has been used to compute
the transition $B \to K^* \gamma$; however in this approach 
the couplings for this decay
are computed  near the zero recoil point and for
a heavy quark mass smaller than its physical value. Therefore
a double extrapolation is needed to compute them and it is hard 
to compare these outcomes, that should be considered as still
preliminary, with QCD sum rules or chiral lagrangian approaches.
As for the comparison with the experimental data, as we have already 
mentioned, 
one should take into account also the so called long 
distance (LD) effects, that we now discuss.

Let us begin with the decays (\ref{brg}) and (\ref{bog}), where these effects
are larger. 
The decays $B \to \rho \gamma$  and $B \to \omega \gamma$ take
contributions both from the short distance and the long distance mechanisms.
The former is generated by a hamiltonian similar to eq. (\ref{Hg}),
with obvious modifications ($s \to d$ and $V_{ts} \to V_{td}$). For it 
an analysis similar to the one employed in $B \to K^* \gamma$ applies,
but it is obvious that this contribution is Cabibbo suppressed as
compared to $B \to K^* \gamma$, which can explain why the
$\rho \gamma$ final state is more difficult to measure (and 
indeed it has not been observed yet). Because of the smallness of
the short distance contribution, LD effects are more important in
these decays than in $B \to K^* \gamma$ or $B_s \to \phi \gamma$ .
For $B^+ \to \rho^+ \gamma$ decay, their contribution has been
estimated by QCD sum rules \cite{a,k}. The ratio
of the long distance to the short distance amplitudes,
as expressed by
\be
\frac{A_{LD}(B^+ \to \rho^+ \gamma)}{A_{SD}(B^+ \to \rho^+ \gamma)}  = R 
\left|
\frac{V_{ub}V_{ud}}{V_{td}V_{tb}}\right|~,
\ee
is estimated to be \cite{a}: $R= -0.30 \pm 0.07$, i.e. a 
significant contribution. LD effects mainly contribute to the 
weak annihilation diagrams and are therefore relevant for 
$B^+ \to \rho^+ \gamma$, but less important for $B \to \rho , \omega \gamma$.
For $B \to K^* \gamma$ one does not expect significant contributions
from the weak annihilation diagrams because of the CKM suppression
(CKM non suppressed terms can contribute by non factorizable diagrams, whose 
role, however, has been found to be very small \cite{colriaz}). Other LD 
contributions come from the four quark operator $O_2 \approx \bar 
c_L \gamma_\mu c_L \bar s_L \gamma^\mu b_L$ (or $O_2 \approx \bar 
c_L \gamma_\mu c_L \bar d_L \gamma^\mu b_L$ for $ B \to \rho \gamma$).
It contributes via a 
 charm quark loop (the up quark loop gives a negligible contribution),
with the photon emitted by the charm quark line, and adds about $20\%$
to  the $B \to K^* \gamma$ width \cite{milana}; by duality this 
contribution  may be seen as the result of a mechanism where $K^* J/\psi$
are produced {\it via} $O_2$, with
photon conversion of $J/\psi$ \cite{soares,golowich1,deshpande,golowich2};
even though
less reliable, the estimates of LD contributions based on this mechanism seem
to agree \cite{soares} with  the result of the charm loop calculation. 
Similar results  hold also for the $B\ \to \rho \gamma$ decays
\cite{milana,at,cng}.

\subsection{Weak radiative decays: $B \to K e^+e^-,\; B \to K^* e^+e^-$ }

In this section we discuss the decays
\bea
&B & \to K e^+ e^- \label{K}\\
&B& \to K^* e^+ e^- ~.  \label{ee}
\eea

They occur dominantly via a quark process $b\to s\gamma^*\to
s e^+ e^-$ ($\gamma^*=$ virtual photon). In the effective lagrangian
for $b\to s e^+ e^-$ we have to include also the already
mentioned long-distance
contributions; they produce
$\psi-\gamma$ or $\psi^\prime-\gamma$
conversion, and are seen
as peaks in the lepton pair invariant mass distribution.
The effective lagrangian has been derived in \cite{inami}, \cite{nn}
and we shall not report it here for the sake of simplicity.

Let us begin with $B\to K e^+ e^-$. 
The transition $B\to K\gamma$ can occur only
by virtual photons and is described at
short distance by the hamiltonian (\ref{Hg}) and,
therefore, by the hadronic matrix element
\be 
\langle K(p^\prime)|\bar s\sigma^{\mu\nu}(1+\gamma_5)b|{\bar B}(p)\rangle  
=iS(q^2)\left[p^\mu p^{\prime\nu}-p^\nu p^{\prime\mu}-
i\epsilon^{\mu\nu\lambda\sigma}p_\lambda p_\sigma^\prime\right]~.
\ee
The form factor $S(q^2)$
can be related, by using the heavy quark equation of
motion and eq. (\ref{relat}), \cite{isgurgamma}, to
the form factors $F_1(q^2)$ and $F_0(q^2)$ for the weak transition $B \to K$
{\it via} the vector current. In this way one finds \cite{isgurgamma}:
\be
S(q^2)=\frac{1}{M_B}\left[-F_1(q^2)+\frac{m^2_K-M^2_B}{q^2}\left(F_1(q^2)-
F_0(q^2)\right)\right] ~.\label{iwg}
\ee

Let us now consider the calculation of the transition $B \to K$
by tensor current in the effective chiral lagrangian approach
\cite{Casalbuoni1}. 
At the tree level the relevant hamiltonian is given by (\ref{lmunu})
together with the strong hamiltonian $BB^*\pi$ of eq. (\ref{wise}).
The
short distance diagrams are similar to those of $B \to K^* \gamma$, 
that we have described in some detail
in the preceeding section. In this case, however, at  tree level,
there is no direct coupling, and the only surviving term, in the limit
$m_b\to\infty$, and for $q^2\approx q^2_{max}$, is the pole contribution.

Assuming 
a $q^2$ dependence of a simple pole, with $M_P=M_{B^*_s}$
(as discussed before,
this seems a reasonable assumption for $F_1$  related form factors),
 one gets the result:
\be
S(q^2)=\frac{S(0)}{1-q^2/M^2_{B_s^*}}
\ee
with
\be
S(0)=\frac{{\hat F} g }{f_\pi M^2_{B^*_s}\sqrt{M_B}}(M_{B^*_s}+
M_B-m_K)~~. \label{pole}
\ee   
When expressed in terms of the form factors
of the $B\to K$ transition by vector current,  this result is:
\be
S(q^2)=-\frac{2F_1(q^2)}{M_B} \label{sf}
\ee
which coincides with the Isgur and Wise relation (\ref{iwg})
only at $q^2\approx q^2_{max}$ and $M_B\to\infty$,
namely in the range of validity of
the effective hamiltonian. Exactly as in the case of
$B\to K^* \gamma$
transition,
we find that some form factors (in this case $F_0$) are subleading
when $m_Q\to\infty$, which is expected because the $0^+$ state, 
contributing to $F_0$, cannot couple to the antisymmetric tensor
current $\bar s\sigma_{\mu\nu}(1+\gamma_5) b$. 

The numerical result 
of this analysis is as follows:
using $g \simeq 0.38$ and ${\hat F} = 0.30$ GeV$^{3/2}$, 
one obtains
\be
S(0) \simeq -0.13 ~{\rm GeV}^{-1}~~.
\ee

This result already contains some SU(3)  corrections,
namely in the meson masses,
but the bulk of the chiral corrections should come 
from loops containing pseudoscalar bosons ($\pi$, $K$ and $\eta$).
They have been computed in \cite{fag}. Three classes of corrections
are found. First one has correction to the pole amplitude (\ref{pole});
second there are corrections to the direct term (called 
{\it point} contribution in \cite{fag}); last there is the renormalization
of the $B$ meson wavefunction. Taking into account only the
nonanalytic corrections that arise from the loop
diagrams, and having no uncertainty related to unknown analytic
higher order terms in the phenomenological lagrangian, one finds
\cite{fag} a correction of $-51\%$ to the dominant pole contribution
(the correction to the direct term is much smaller).

This analysis, as stressed
in \cite{fag}, is not conclusive,
since the analytic corrections could be significant;
nevertheless it is interesting  to observe that, with
the nonanalytic correction alone, the outcome of the effective chiral
lagrangian becomes $S(0)\approx -0.06$ GeV$^{-1}$, which
is compatible with the result of a QCD sum rules
analysis based on  three point functions \cite{scrimieri}:
$S(0)= -0.05\pm 0.01$ GeV$^{-1}$ .

The results we have reported, 
together with  the  long distance contributions \cite{inami,nn}, 
can be used to obtain
the distribution $d\Gamma(B\to K e^+ e^-)/dq^2$ in the invariant mass
squared of the lepton pair $Q^2$. This distribution is dominated
by the contribution of the resonances $J/\psi$ and $\psi^\prime$;
however, as discussed for instance in \cite{scrimieri}, for
$Q^2$ far from the resonance masses, one could still obtain 
from the data, when available, useful information on the
short distance dynamics.

The same analysis can be performed for
$d\Gamma(B\to K^* e^+ e^-)/dQ^2$.
The long distance contribution can be evaluated starting from
experimental data on the nonleptonic decay modes
$B\to K^{\star} J/\psi$ and $B\to K^{\star} \psi^\prime$.
For the short distance part one needs the form factors $A(q^2)$,
$B(q^2)$ and $H(q^2)$ we have defined in eq. (\ref{ffrad}).
In the effective
lagrangian approach one finds, from table \ref{table:bep4}
\bea
A(q^2) &=& \frac{i\sqrt{2}{\hat F}\lambda g_V(m^2_B+m^2_{K^*}-q^2)}
{(M_{B_s^*}^2-q^2)\sqrt{M_B}} \\
B(q^2) &=&\frac{-i2\sqrt{2}{\hat F}\lambda g_V M_B^{3/2}}
{M_{B_s^*}^2-q^2}
\\
H(q^2) &=& \frac{-i2\sqrt{2} g_V}{\sqrt{M_B}} \left[ \frac{ {\hat F} 
\lambda}{M_{B_s^*}^2-q^2 } + \frac{ {\hat F}^+ \mu}
{M_{B_s^{**}}-q^2 } \right ]
\eea
where $B_s^{**}$ is the $1^+$ $\bar s b$ resonance.
$A(0)$ and $B(0)$ have been discussed in section \ref{sec:bepp3}; as for 
$H(0)$, using (\ref{lambdaprimo})
and the result of section \ref{subsec-brhoeff} (obtained with $r=0.5$):
$\mu = -0.10$ GeV$^{-1}$, together with ${\hat F} =0.30$ GeV$^{3/2}$ and
${\hat F}^+ =0.46$ GeV$^{3/2}$, eq. (\ref{c4bis}), one gets:
\be
|H(0)|=0.04 ~{\rm GeV}^{-2} \label{h0}
\ee
to be compared with the QCD sum rule result \cite{scrimieri}
$|H(0)|\approx 0.10~ {\rm GeV}^{-2}$. Analogously to the $A(0)+B(0)$
case (see table \ref{table:bep5}),
we see that the chiral lagrangian approach at  tree level
gives significantly smaller results than  QCD sum rules,
which might indicate either a relevant $1/m_Q$ correction or,
most probably, a relevant contribution from chiral loop,
similarly to the case of $S(0)$ discussed above.

The distribution in the invariant mass of the lepton pair is 
largely dominated by the long distance contributions
\cite{riaz,deshpande}, exactly as for the $Ke^+ e^-$ final state;
nevertheless an accurate measurement of the lepton pair spectrum below
$c\bar c$ resonances would display the effects of the short distance 
hamiltonian.
This measurement would therefore
complement the analysis of the $B\to K^*\gamma$ decay process,
providing further information on the fundamental parameters appearing
in the short distance hamiltonian as well as on the validity of the
effective lagrangian approach.

\newpage
\section{Symmetries for heavy quarkonium states}    
\label{chap-al}
Quarkonium, a heavy quark-antiquark bound state, is one of the most 
interesting systems for the study of quantum chromodynamics (QCD). The 
physical 
idea is that quarks with mass larger than the QCD scale $\Lambda_{QCD}$
 would form 
bound states resembling positronium \cite{appo}. Many properties of
quarkonium can be predicted by the use of non-relativistic potential models. 
The overall description one obtains in this way for the charmonium 
and bottomonium spectroscopies is quite satisfactory provided corrections 
originated by leading relativistic terms are included and the possible 
multichannel structure of the phenomenon is taken into account for certain 
expected effects \cite{appo}.

The heavy quark and anti-quark are bound in these models by an 
instantaneous potential, meaning that gluons have typical interaction times 
much shorter than the time scale associated with the motion of the heavy
quarks. We indicate with $k$ the relative momentum and with $v_r=k/m_Q$ the 
relative velocity between the two heavy quarks of mass $m_Q$. It is 
interesting  
to examine the dependence of these quantities on the quark mass. 

For instance, Buchm\"uller and Tye \cite{BTye} have studied a QCD-motivated 
potential reproducing the behaviour $1/r$ for small $r$, and behaving as $r$ 
at large distances (this model is similar to the model in
\cite{richardson}). Analogous results can be obtained using other models, 
such as Quigg and Rosner \cite{Quigg} or Grant and Rosner \cite{Grant}, 
indicating that, by increasing the quark mass, the kinetic energy and the 
residual momentum increase, whereas the relative velocity decreases. 
Going further up with the mass $m_Q$, the
heavy quarks separation becomes smaller, and eventually the Coulomb 
term of the potential energy, proportional to $\alpha_s/r$ will dominate 
($\alpha_s$ is the strong coupling constant evaluated at the momentum scale 
$1/r$). Taking $r$ of the order of the size of the bound state $1/(m_Q v_r)$
in the potential energy and equating its average value to the kinetic term
we get $<v_r> \approx \alpha_s$, with $\alpha_s$ evaluated at the momentum 
scale
$m_Q v_r$, going to zero in the limit $m_Q \to \infty$. We notice that in such 
a coulombic regime the relative velocity decreases logarithmically. 

Concerning the spin symmetry, the coupling of the gluon to the spin of the
heavy quark is expected to be of the order $p_g/m_Q$, with the gluon
momentum $p_g \approx k$. Therefore the quantity $k/m_Q=v_r$ gives
information on
the degree of spin decoupling, and strictly in the limit $m_Q \to \infty$ one 
has an exact spin symmetry.

The heavy quark flavour symmetry, on the contrary, is badly broken in
quarkonium systems. In general the gluon radiation exchanged between static 
quarks gives rise to infrared divergences. In a bound state, potential and 
kinetic energy play a delicate balance against each other \cite{BW}. 
The regularization of the infrared divergences then
implies a large breaking of flavour symmetry because of the explicit
appearance of the heavy quark mass in the kinetic energy.

For charmonium, potential models give $<v_r> \approx 0.5$, for bottomonium 
$<v_r> \approx 0.25$; one expects corrections, even important, to the 
``leading order" velocity and spin symmetry description, especially for
charmonium.
 
\subsection{Non-relativistic QCD description}

An effective approach to quarkonium is given by the non-relativistic heavy 
quark QCD 
description, which provides a general factorization formula for annihilation
decay rates of heavy quarkonium \cite{bbl}. It consists in exploiting 
the fact that in quarkonium the heavy quark moves with a small relative 
velocity and  nonrelativistic quantum chromodynamics (NRQCD) is a good 
approximation. 
The lowest order dynamics is given by the Schr\"odinger equation 
for the heavy quarks. The resulting effective theory  \cite{nrqcd} 
consists in fact of a nonrelativistic Schr\"odinger field theory for the heavy
quarks coupled to the relativistic theory for gluons and light quarks.
Relativistic corrections can be included systematically into this picture at 
any given order in the heavy quark velocity $v$. 

In this framework the scales entering the problem are written in terms 
of the heavy quark velocity $v$ and mass $m_Q$. As shown before, the typical
velocity of the heavy quark decreases as the mass increases. When $m_Q$ is
sufficiently large, the heavy quark and antiquark are non-relativistic ($v <<
1$) and the scales $m_Q$, $m_Q v$ (typical three-momentum of the heavy quark in 
the meson rest frame), and $m_Q v^2$ (typical kinetic energy) are well separated:
\be
\Lambda^2_{QCD} \ll (m_Q v^2)^2 \ll (m_Q v)^2 \ll m_Q^2
\ee 
In NRQCD the effects at the scale $m_Q$ are taken into account through the 
coupling constants of 4-fermion operators, while the effects of the lower 
momentum scales $m_Qv$, $m_Qv^2$, and $\Lambda_{QCD}$ are included into matrix 
elements organized in terms of their dependence on $v^2$. The lagrangian is
obtained from QCD by introducing an ultraviolet cut-off of the order
$m_Q$, which excludes relativistic heavy quarks from the theory. It also 
excludes
light quarks and gluons with momenta of order $m_Q$. Then the heavy-quark and
heavy-antiquark degrees of freedom are decoupled by a Foldy-Wouthuysen
transformation. The full NRQCD lagrangian consists of the part describing the
heavy quarks and anti-quarks in terms of a non-relativistic Schr\"odinger
theory with separate two-component fields for the quarks and anti-quarks 
${\cal{L}}_{heavy}$, plus a fully relativistic part for the light quarks and 
gluons ${\cal{L}}_{light}$, plus a correction term $\delta {\cal{L}}$ 
reproducing the relativistic effects of full QCD in terms of new local 
interactions:
\be
{\cal{L}}_{NRQCD}={\cal{L}}_{light}+{\cal{L}}_{heavy}+\delta {\cal{L}}
\ee
with 
\be
{\cal{L}}_{heavy}=\psi^{\dag}\left( iD_o+\frac{{\bf D}}{2m_Q} \right) \psi +
\chi^{\dag}\left( iD_0+\frac{{\bf D}}{2m_Q} \right) \chi~~,
\ee
where $\psi$ and $\chi$ are the two component fields for quarks and 
anti-quarks 
and $D_0$ and ${\bf D}$ are the time and space components of the covariant 
derivative. The term $\delta {\cal{L}}$ contains all possible gauge invariant 
counterterms, whose coefficients must be matched with QCD in order to 
avoid ultraviolet divergences in the calculation of long distance 
quantities and to reproduce the results of full QCD. In principle the NRQCD 
lagrangian consists of infinitely many terms. However they can be classified 
in 
powers of the heavy quark velocity $v$ and their relative importance can be 
established. 

The annihilation of quarkonium can be reproduced in this framework only 
indirectly, through its effects on $Q\bar Q$ scattering amplitudes. At long 
distance, these amplitudes can be described adding to the lagrangian four 
fermion operators that annihilate and create a heavy quark-antiquark pair. Due 
to the optical theorem the imaginary parts of the coefficients of the four 
fermion operators are related to the annihilation of heavy quarkonium. It 
should be noted that the annihilation decay rates for heavy quarkonium are 
small perturbations of the energy levels. In this approach the contributions 
to annihilation widths from the dimension-6 four fermion operators contain
extra suppression factors, due to the coefficients of the operators. The 
widths are of order $\alpha_s^2(m_Q) v$ or smaller, while the splitting 
between 
radial excitations is of order $m_Qv^2$.

\subsection{Heavy quarkonium effective theory}

Quarkonium, in the heavy quarkonium effective theory 
\cite{hqqet1,hqqet2,hqqet3}
is described as a bound state in the 
particle-antiparticle sector of the HQET 
\cite{hq1,hq2,hq3}. In quarkonium systems the internal motion of the heavy 
quarks 
cannot be neglected due to the delicate balance between the potential and 
kinetic energy in the bound state. This suggests to go beyond the static limit 
to describe quarkonia states. One must therefore keep the kinetic energy 
operator even when working at the lowest order. The kinetic energy operator is 
spin symmetric, but it includes a factor of $1/m_Q$.
Therefore heavy flavour symmetry is lost while spin symmetry is still present.
The leading order lagrangian 
is 
\cite{hqqet2}:
\be
{\cal L}_0= {\bar h}_v^{(+)}(ivD)h_v^{(+)}+{\bar h}_v^{(+)}\frac{(iD)^2}{2m_Q} 
h_v^{(+)}-{\bar h}_w^{(-)}(iwD)h_w^{(-)}+{\bar h}_w^{(-)}\frac{(iD)^2}{2m_Q} 
h_w^{(-)}~~,
\label{lhqqet}
\ee
where as usual a sum over heavy quark velocities is understood.
The heavy field is obtained from the field $Q$ of QCD by removing the 
dominant part of the heavy quark momentum:
\be
Q=\exp (-im_Qvx) \left(1+\frac{1+\slash v}{2} \frac{\slash D}{2m_Q}+ {\cal O} 
(1/m_Q^2) 
\right) h_v^{(+)}(x)
\ee 
and $h_v^{(+)}$ describes a quark with velocity $v$. In a similar way 
$h_w^{(-)}$ describes an anti-quark moving with velocity $w$. The lagrangian 
in (\ref{lhqqet}) is the starting point of an effective lagrangian description 
of heavy quarkonium decays. The two velocities of the heavy quarks differ only 
by a quantity of the order of $\Lambda_{QCD}/m_Q$ so that it is convenient to 
work in the limit in which the two heavy quark velocities become equal. 
This limit can be taken consistently starting from the effective 
lagrangian (\ref{lhqqet}). In any case a mass dependence in the lowest order 
dynamics is unavoidable so that heavy flavour symmetry is destroyed. In this 
picture spin symmetry, as we anticipated,  still holds  
since the kinetic energy operator is 
spin symmetric. 

As to relativistic corrections (proportional to the relative 
velocity of the heavy quarks) and non-perturbative corrections, they turn out 
to have the same origin in this approach, 
namely they come from higher order terms of the $1/m_Q$
expansion. Short and long distance contributions for the inclusive 
annihilation decays are separated by means of the operator product expansion. 
The distance scale is 
given by the Compton wavelength of the heavy quark. The annihilation rates are 
written in an expansion in $({\tilde \Lambda}/m_Q)$ where $\tilde \Lambda$
is the inverse Bohr radius of the system. The coefficients of this expansion 
can be calculated perturbatively.

This approach is similar to the one of non-relativistic 
QCD and in fact in that case the lowest order dynamics is basically the one
obtained by adding the kinetic energy operator to the static HQET part. The
two approaches are not completely equivalent though: for example if we 
consider
two operators like the gluon field strength $[iD,iD]$ and $(iD)^2$, they have
the same dimension, but they are not equivalent if an expansion in relative
velocity $v/c$ is considered, as in the case of NRQCD (for a detailed
comparison of the two approaches see \cite{hqqet3}). In any event if the same 
set of
assumptions is applied in the two cases, both the approaches yield the same
results up to the order $({\tilde \Lambda}/m_Q)^2$.

\subsection{Heavy-meson effective theory}

This approach consists in constructing a heavy meson multiplet field and 
writing a lagrangian including the exact and approximate symmetries of the 
problem \cite{noi0}. The procedure is analogous to the one introduced in the 
preceeding chapters of this review treating of heavy-light mesons. Symmetry 
breaking terms can be easily 
added to the formalism as we shall show in the following. The velocity 
description and spin symmetry are still useful, but flavour symmetry is 
broken. 
As in the single heavy quark case \cite{wise,yan,burdman92}, an effective 
lagrangian describing the low-momentum interactions of heavy quarkonia with 
light mesons can be written down. The heavy quarkonium multiplets are 
described 
by a simple trace formalism \cite{hq5}, which can be also applied to the 
description of the $B_c$ system.

\subsubsection{Heavy quarkonium states}

A heavy quark-antiquark bound state, characterized by the radial number $m$, 
the orbital angular momentum $l$, the spin $s$, and the total angular momentum 
$J$, is denoted by:
\be
m \,^{2s+1}l_{J}
\label{q2.1}
\ee
Parity $P$ and charge conjugation $C$, which determine selection rules for 
electromagnetic and hadronic transitions, are given by:
\bea
P=(-1)^{l+1}  \label{q2.2} \\  
C=(-1)^{l+s} \label{q2.3}
\eea
and are exactly conserved quantum numbers for quarkonium, together with $J$.
If spin dependent interactions are neglected, it is natural to describe the 
spin singlet $m\,^{1}l_{J}$ and the spin triplet $m\,^{3}l_{J}$ by means of a
single multiplet $J(m,l)$. For the case $l=0$, when the triplet $s=1$ 
collapses into a single state with total angular momentum $J=1$, 
this is readily realized: 
\be
J= \frac{(1+\slash v)}{2}[H_{\mu}\gamma^\mu-\eta\gamma_5]
\frac{(1-\slash v)}{2}~~.
\label{q2.4}
\ee
Here $v^{\mu}$ denotes the four velocity associated to the multiplet $J$;
$H_{\mu}$ and $\eta$ are the spin 1 and spin 0 components respectively; the
radial quantum number has been omitted. 
The expressions for the general wave $J^{\mu_1 \ldots \mu_l}$
are given in the appendix C. In the sequel $K_{l}^{\mu_1 \ldots \mu_{l}}$ 
represents the spin singlet component $~^1l_J$, $H_{l+1,l,l-1}^{\mu_1 \ldots 
\mu_{l+1}}$ the spin triplet $~^3l_J$ in the wave $J^{\mu_1 \ldots \mu_l}$.

 From eqs. (\ref{q2.2}) and (\ref{q2.3}) one has the following transformation 
properties of $H_l$ and $K_l$ under parity and charge conjugation:
\bea
H_{l+1}^{\mu_1 \ldots \mu_{l+1}} & \stackrel{P}{\rightarrow} &
 H^{l+1}_{\mu_1 \ldots \mu_{l+1}}  \nn \\
H_{l}^{\mu_1 \ldots \mu_{l}} & \stackrel{P}{\rightarrow} &
 -H^{l}_{\mu_1 \ldots \mu_{l}}  \nn \\
H_{l-1}^{\mu_1 \ldots \mu_{l-1}} & \stackrel{P}{\rightarrow} &
 H^{l-1}_{\mu_1 \ldots \mu_{l-1}}  \nn \\
K_{l}^{\mu_1 \ldots \mu_{l}} & \stackrel{P}{\rightarrow} &
-K^{l}_{\mu_1 \ldots \mu_{l}}  
\label{q2.10}
\eea
\bea
H_{l+1,l,l-1}^{\mu_1 \ldots \mu_{l+1,l,l-1}} & \stackrel{C}{\rightarrow} &
(-1)^{l+1} H_{l+1,l,l-1}^{\mu_1 \ldots \mu_{l+1,l,l-1}}  \nn \\
K_{l}^{\mu_1 \ldots \mu_{l}} & \stackrel{C}{\rightarrow} &
(-1)^{l} K_{l}^{\mu_1 \ldots \mu_{l}}  
\label{q2.11}
\eea
As one can easily verify, the previous transformations laws are reproduced 
by assuming that the multiplet $J^{\mu_1 \ldots \mu_l}$ transforms as
follows:
\bea
J^{\mu_1 \ldots \mu_{l}} & \stackrel{P}{\rightarrow} &
\gamma^0 J_{\mu_1 \ldots \mu_{l}} \gamma^0 \nn \\
v^\mu & \stackrel{P}{\rightarrow} &   v_\mu
\label{q2.12}
\eea
\bea
J^{\mu_1 \ldots \mu_{l}} & \stackrel{C}{\rightarrow} &
(-1)^{l+1} C J^{\mu_1 \ldots \mu_{l}T} C 
\label{q2.13}
\eea
where $C=i\gamma^2\gamma^0$ is the usual charge conjugation matrix.
\par
Under heavy quark spin transformation one has
\be
J^{\mu_1 \ldots \mu_l} \rightarrow S J^{\mu_1 \ldots \mu_l} S'^\dagger  
\label{q2.19}
\ee
with $S,\,S' \in SU(2)$ and $[S,\slash v]= [S',\slash v]=0$. As long as one 
can 
neglect spin dependent effects, one will require invariance of the allowed 
interaction terms under the transformation (\ref{q2.19}).

Finally under a Lorentz transformation $\Lambda$ we have:
\be
J^{\mu_1 \ldots \mu_{l}} \rightarrow \Lambda^{\mu_1}_{~\nu_1} \ldots
\Lambda^{\mu_l}_{~\nu_l} D(\Lambda) J^{\nu_1 \ldots \nu_l} D(\Lambda )^{-1}
\label{lorentz}
\ee
where $D(\Lambda)$ is the usual spinor representation of $\Lambda$.

\subsubsection{$B_c$ meson states}

The study of $B_c$ meson decays gives important information about the QCD 
dynamics and the weak interactions; moreover the $B_c$ system allows to use 
theoretical insight and phenomenological information obtained from charmonium 
and bottomonium. One important difference is that the total widths of excited 
$B_c$ levels are about two orders of magnitude smaller than the total widths 
of 
charmonium and bottomonium excited levels, as the excited $B_c$ system does 
not 
have strong or electromagnetic annihilation decay channels and can only decay 
weakly. For the $B_c$ there is a large probability for the decay modes with a 
heavy meson in the final states as it can be seen in a simple constituent quark 
picture. 
The $B_c$ mesons spectra and their decay modes have been studied using 
potential 
models (see for example \cite{potbc1,potbc2,delubc,potbc3,potbc4}), 
lattice calculations \cite{lattbc} and QCD sum rules \cite{sumrulbc}. 
QCD perturbative calculations \cite{probc} and fragmentation 
functions \cite{fragbc} were used to study its production. The approximate 
spin 
symmetry independence of the system can be implemented in an effective meson 
lagrangian and the corresponding symmetry relations impose restrictions on the 
form factors of the exclusive weak semileptonic decays of $B_c$ \cite{hqetbc}. 
In the following we shall consider only this effective theory 
approach, as it is related to the material of the previous section. The 
consequences of spin symmetry for hadronic matrix elements may be derived 
using 
a trace formalism \cite{hq5,hqetbc} analogous to the one used for quarkonium:
\be
H^{(c\bar b )} = {{(1+\slash v)}\over 2}\left[
B_c^{*\mu}\gamma_\mu-B_c\gamma_5\right] {{(1-\slash v) }\over 2}
\ee
where $H^{(c\bar b )}$ is the $4\times 4$ matrix representing the lowest-lying 
pseudoscalar and vector meson $c\bar b$ bound states. Under spin symmetries on 
the heavy quark and antiquark, the heavy meson field transforms as
\be
H^{(c\bar b )}\rightarrow S_c\ H^{(c\bar b )}\ S_{ b}^\dagger
\ee
The definitions are analogous to the ones given for quarkonium. 
We shall not examine these systems any longer and we shall refer the
interested reader to the existing literature, since no experimental data
are available yet, and, therefore, the analysis would be only
speculative. To give only an example, it has been suggested that the
above formalism
can be applied to the study of the semileptonic decay $B_c \to D \ell \nu$, 
which 
could provide a way of extracting the mixing angle $|V_{ub}|$ \cite{hqetbc}.

\section{Heavy quarkonium decays}
\label{chap-al1}

The heavy quark spin symmetry leads to general relations for the differential 
decay rates in hadronic transitions among quarkonium states that 
essentially reproduce the results of a QCD double multipole expansion 
\cite{Yann} 
for gluonic emission. Further use of chiral symmetry leads to differential 
pion decay distributions valid in the soft regime \cite{noi0}, \cite{noi1}
(see also \cite{dobook}). At the lowest order in the 
chiral expansion for the emitted pseudoscalars we find a selection
rule allowing only for even (odd) number of emitted pseudoscalars for 
transitions between quarkonium states of orbital angular momenta different by 
even (odd) units. Such a rule can be violated by higher chiral terms, by 
chiral
breaking, and by terms breaking the heavy quark spin symmetry. Specialization
to a number of hadronic transitions reproduces by elementary tensor 
construction the known results from the expansion in gluon multipoles,
giving a simple explanation for the vanishing of certain coefficients
which would otherwise be allowed in the chiral expansion. In certain cases,
such as for instance $~^3P_0 \to ~^3P_2 \pi\pi$, $~^3P_1 \to ~^3P_2 \pi\pi$, 
or $D-S$ transitions via $2\pi$, the final angular and mass distributions are
uniquely predicted from heavy quark spin and lowest order chiral expansion.

The effective heavy-meson description of quarkonium does not seem
to present special advantages to
describe heavy quarkonium annihilation. 

At the heavy quark level such annihilations can be described introducing 
four-fermion operators. The optical theorem then relates heavy quarkonium 
annihilation rates to the imaginary part of $Q{\bar Q} \to Q{\bar Q}$ 
scattering 
amplitudes. 

Heavy quarkonium annihilations were among the first tests of perturbative QCD 
on 
the assumption that one could factor out the non-perturbative bound state 
features and use asymptotic freedom to calculate the quark-antiquark short 
distance annihilation process. Such an approach has met a general 
phenomenological success and indeed it has provided  a basic support to the 
quarkonium picture and to the asymptotic freedom. Calculations for $P$ state 
annihilation were however disturbed by infrared divergences \cite{barga} 
appearing within the simplest perturbative-nonperturbative separation scheme 
and 
requiring a suitable prescription to obtain physical predictions. The 
separation between short distance and long distance effects in $P$-states has 
been 
recently reexamined \cite{bbl1p1}, leading to a factorization prescription 
which 
introduces an additional long-distance parameter setting the problem of 
infrared 
regularization.

Very recently the theoretical and experimental analysis of quarkonium
production has given interesting results. In the following we shall
concentrate on quarkonium decays. It is worth mentioning that a
new set of data \cite{proda} has encouraged to deepen the theoretical 
understanding 
of quarkonium production from suitable models \cite{promo} to more 
sophisticated 
calculations in the framework of QCD \cite{proth}.

\subsection{Radiative decays}

Here we discuss radiative transitions $Q\bar{Q} \to Q\bar{Q} \gamma$, where 
the
recoiling system $Q\bar{Q}$ has a mass close to that of the radiating system.
In such a case radiative transitions are usually studied in the framework of
multipole radiation. Radiation can occur through electric or magnetic 
multipole 
transitions, when allowed by the conservation rules of spin, parity and charge 
conjugation. Radiative decays provide a simple test of the formalism. We 
expect that
this approach should reproduce the well established results of QCD motivated 
potentials \cite{radpo}. The application demonstrates the power of the 
formalism 
in the evaluation of radiative decay amplitudes between the $S-$ and $P-$wave 
states, both for charmonium and bottomonium. If absolute predictions are to be 
made, the formalism requires data to fix the unknown parameters in the 
effective lagrangian.

The analysis of radiative decays in quarkonium
can also be carried out directly in terms of reduced matrix elements of the
appropriate interaction Hamiltonian, using the usual angular momentum
procedures \cite{chwi}. The two procedures are equivalent, as in this approach
spin and angular momentum are described directly within the multiplet field.

We write the lagrangian for radiative decays as follows:
\be
{\cal L}=\sum_{m,n} \delta (m,n) <{\overline{J}}(m)\; J_{\mu} (n)> v_{\nu} 
F^{\mu\nu} + h.c.
\label{q3.1}
\ee
where a sum over velocities is understood, $F^{\mu\nu}$ is the electromagnetic 
tensor, the indices $m$ and $n$ represent the radial quantum numbers, $J(m)$ 
stands for the multiplet with radial number $m$ and $\delta (m,n)$ is a 
dimensional parameter (the inverse of a mass), to be fixed from 
experimental data and which also depends on the heavy flavour.
The lagrangian (\ref{q3.1}) conserves parity and charge conjugation 
and is invariant under the spin transformation of eq. (\ref{q2.19}). It 
reproduces 
the electric dipole selection rules $\Delta \ell=\pm 1$ and $\Delta s=0$.
It is straightforward to obtain the corresponding radiative widths:
\be
\Gamma (^3P_J \to \;^3S_1 \gamma)= {\frac {\delta^2} {3\pi}} p^3 {\frac 
{M_{S_1}} {M_{P_J}}}
\label{q3.2}
\ee
\be
\Gamma (^3S_1 \to \;^3P_J \gamma)= {\frac {(2J+1)} {9\pi}} \delta^2 p^3 
{\frac {M_{S_1}} {M_{P_J}}}
\label{q3.3}
\ee
\be
\Gamma (^1P_1 \to \;^1S_0 \gamma)= {\frac {\delta^2} {3\pi}} p^3 {\frac 
{M_S} {M_P}}
\label{q3.4}
\ee
where $p$ is the photon momentum. Once the radial numbers $n$ and $m$ have 
been
fixed, the lagrangian (\ref{q3.1}) describes four no spin-flip transitions 
with a
single parameter; this allows three independent predictions.
For the triplet states they are reported in table \ref{table:aldo}, where we 
give
the ratio of the width for the state with $J=1$ to the state with 
$J=0$ and that for the state with $J=2$ to the state with $J=0$, within a 
given 
multiplet. The theoretical numbers refer to eqs. (\ref{q3.2}-\ref{q3.4}) and
to the results obtained by Cho and Wise \cite{chwi}. 

For the state $h_c(1P)$ ($\;^1P_1$ state) no data on radiative widths are 
available yet, but extracting  the value of $\delta$ from the data of
the corresponding triplet states, we can predict 
the width of $h_c(1P) \to \eta_c \gamma$. This transition is an electric 
dipole
(E1) transition, which is expected to be the dominant decay mode of the 
$h_c(1P)$ state (with a branching ratio of order 80\% \cite{kuatua}). 
Using for the $h_c(1P)$ mass the value $3526.2\pm 0.2 \, GeV$ of E760 
experiment \cite{E760}, we obtain:
\be
\Gamma (h_c(1P) \to \eta_c \gamma)= 0.45 \pm 0.02 \, MeV
\ee
We note that QCD-motivated potentials using first order relativistically 
corrected wave 
functions give a prediction of the width $\Gamma (h_c(1P) \to \eta_c \gamma)$ 
of 0.39 MeV \cite{1p1}, while QCD predictions based on the factorization 
formulas of \cite{bbl1p1} give $0.45\pm 0.05 (\pm 20\%)$ MeV for the partial 
radiative width and $0.98\pm 0.09 (\pm 22\%)$ MeV for the total decay rate of 
$h_c$, to be compared with the experimental upper bound on the total width of 
$\Gamma_T(h_c(1P)) < 1.1$ MeV at 90\% C.L. \cite{E760}. A similar prediction 
for the $\chi_b (\;^1 P_1 )$ state can be easily extracted from eq. 
(\ref{q3.4}), once the mass of this state and the $\delta$ parameter of the 
corresponding multiplet are known. 
\begin{table}[htb]
\caption {Results for $\Gamma_1/\Gamma_0$ and $\Gamma_2/\Gamma_0$, where
$\Gamma_J$ stays for the radiative width of the process involving
$\;^3P_J$. The number in parentheses have been obtained 
by Cho and Wise (see text)}
\label{table:aldo}
\centering
\begin{minipage}{12truecm}
\begin{tabular}{l c c c c} 
${\rm Process}$ & $\Gamma_1/\Gamma_0 (th.)$ 
& $\Gamma_1/\Gamma_0 (exp.)$ & 
$\Gamma_2/\Gamma_0 (th.) $ & $\Gamma_2/\Gamma_0 (exp.)$\\ \hline
$\psi (2S) \to \chi_{c}(1^3P_J) \gamma$ & $0.82  (0.85)$ & $0.94\pm 0.12$ 
& $0.55  (0.58)$ & $0.84\pm 0.11$ \\ \hline
$\chi_{c}(1^3P_J) \to J/\psi(1S) \gamma$ & $2.05  (2.11)$ & $2.61\pm 1.24$ 
& $2.74  (2.84)$ & $2.93\pm 1.34$ \\ \hline
$\Upsilon (2S) \to \chi_{b}(1^3P_J) \gamma$ & $1.58  (1.56)$ & $1.56\pm 0.41$ 
& $1.56  (1.54)$ & $1.53\pm 0.41$ \\ \hline
$\Upsilon (3S) \to \chi_{b}(2^3P_J) \gamma$ & $1.61  (1.61)$ & $2.09\pm 0.26$ 
& $1.76  (1.76)$ & $2.11\pm 0.28$ \\ \hline
$\chi_{b}(1^3P_J)\to \Upsilon(1S) \gamma$ & $1.25~ ~ ~~~~~$ & - 
& $1.45~ ~ ~~~~~$ & - \\ \hline
$\chi_{b}(2^3P_J)\to \Upsilon(1S) \gamma$ & $1.07~ ~ ~~~~~$ & - 
& $1.12~ ~ ~~~~~$ & - \\ \hline
$\chi_{b}(2^3P_J)\to \Upsilon(2S) \gamma$ & $1.29~ ~ ~~~~~$ & - 
& $1.52~ ~ ~~~~~$ & - \\
\end{tabular}
\end{minipage}
\end{table}

\subsection{Hadronic transitions in heavy quarkonia}

An important class of hadronic transitions between heavy-quarkonium states 
is provided by the decays with emission of two pions, for example:
\be
\psi' \rightarrow \psi \, \pi \pi
\label{q4.1}
\ee

To describe these processes we use the chiral symmetry for the pions and the
heavy-quark spin symmetry for the heavy states. The first one is expected to
hold when the pions have small energies. We notice that
the velocity superselection rule applies
at $q^2=q^2_{max}$, when the energy transfer to the pion
is maximal. Therefore we expect these approximations to be valid in the whole
energy range only if $q^2_{max}$ is small.

Nonetheless a number of interesting properties of these transitions can 
be derived  on the basis of the heavy quark symmetry alone. Therefore,
before specializing the pion couplings by means of chiral symmetry,
we discuss the implications of the heavy quark spin symmetry in 
hadronic transitions.

As an example, we consider transitions of the type $~^3S_1 \to ~^3S_1+h$ 
and $~^1S_0 \to ~^1S_0+h$, where $h$ can be light hadrons, photons, etc.
By imposing the heavy quark spin symmetry, one is lead to describe
these processes by an interaction lagrangian:
\be
{\cal L}_{SS'}=<J'\bar{J}>\Pi_{SS'}+h.c.~~,
\label{q4.2}
\ee
where the dependence upon the pion field is contained in the yet
unspecified operator $\Pi_{SS'}$. It is immediate to derive from
${\cal L}_{SS'}$ the averaged modulus square matrix elements for 
the transitions $^3S_1\to^3S_1+h$ and $^1S_0\to^1S_0+h$ with an
arbitrary fixed number of pions in the light final state $h$.
We obtain:
\bea
|{\cal M} (^3S_1\to^3S_1+h)|^2_{av.}&=&
|{\cal M} (^1S_0\to^1S_0+h)|^2_{av.}\nn\\
&=&4 M_S M_{S'}|\Pi_{SS',h}|^2
\label{q4.3}
\eea
where $M_S$ and $M_S'$ are the average masses of the two
$S$-wave multiplets; $\Pi_{SS',h}$ is the appropriate
tensor for the emission of the light particles $h$, to be calculated from the 
operator $\Pi_{SS'}$. By denoting with $d\Gamma$ the generic differential
decay rate, we have:
\be
d\Gamma(^3S_1\to^3S_1+h)= 
d\Gamma(^1S_0\to^1S_0+h)~~.
\label{q4.4}
\ee

This is the prototype of a series of relations which can be derived
for hadronic transitions as a consequence of the spin independence of the 
interaction terms. In all the known cases they coincide with those calculated 
in the context of a QCD double multipole expansion. We notice however that 
we do not even need to specify the nature of the operator $\Pi$,
which may depend on light fields different from the pseudoscalar
mesons (e.g. the photon,
or a light hadron, etc), provided that the interaction term we are building
is invariant under parity, charge conjugation, and the other 
symmetries relevant to the transition considered.
Indeed the label $h$ in eq. (\ref{q4.3}) stands for an arbitrary combination
of light final state particles. In this sense, this approach provides a 
generalization of the results obtained in the context of the QCD multipole
expansion.

By assuming a spin independent interaction, we can easily extend the 
previous considerations to other transitions \cite{noi1}. In general,
as a consequence of the heavy quark spin symmetry, the allowed
transitions between two multiplets $l$ and $l'$ will be related by a set
of equations, independently of the nature of the light final state $h$.

\subsubsection{Chiral invariant hadronic transitions}

A useful symmetry that can be used in processes involving light quarks
is the chiral symmetry. It is possible to build up an 
effective lagrangian which allows to study transitions among quarkonium states
with emissions of soft light pseudoscalars, 
considered as the Goldstone bosons of the spontaneously broken chiral 
symmetry. 
The chiral symmetry is explicitly broken through light quark mass terms,
which allow for rarer processes that could be, in some circumstances,
kinematically favored.

The relations (\ref{q4.4}) among the differential decay rates are direct 
consequences of the 
assumed dominance of spin independent terms for the operators describing
the corresponding transitions. More detailed predictions
can be obtained by specifying the form of the operators $\Pi$'s appearing
in the expressions of the interaction terms (\ref{q4.2}). We restrict here to 
hadronic transitions with emission of light pseudoscalar mesons.

The light mesons are described as pseudo-Goldstone bosons, included in the 
matrix $\Sigma=\xi^2$ (see formulas (\ref{1.4}), (\ref{1.5}) and the 
discussion 
in the section \ref{subsec:cs}). Frequently occurring quantities are the 
functions of $\xi$ and its derivatives ${\cal A}_\mu$ and ${\cal V}_\mu$
given
in eqs. (\ref{1.7}) and (\ref{1.6}). 

For the subsequent analyses we are interested in the transformations under 
parity and
charge conjugation:
\bea
\Sigma & \stackrel{P}{\rightarrow} & \Sigma^\dagger \nn \\
{\cal A}_\mu &\stackrel{P}{\rightarrow} & -{\cal A}^\mu \nn \\
{\cal V}_\mu &\stackrel{P}{\rightarrow} & {\cal V}^\mu
\label{q5.4}
\eea
\bea
\Sigma & \stackrel{C}{\rightarrow} & \Sigma^T \nn \\
{\cal A}_\mu &\stackrel{C}{\rightarrow} & {\cal A}^T_\mu \nn \\
{\cal V}_\mu &\stackrel{C}{\rightarrow} & -{\cal V}^T_\mu~~.
\label{q5.5}
\eea    
Finally we recall that light vector mesons can be introduced 
as gauge particles  as discussed in section \ref{subsec:lvr}.
Under parity and charge conjugation, one has:
\bea
\rho_\mu &\stackrel{P}{\rightarrow} & \rho^\mu \nn \\
\rho_\mu &\stackrel{C}{\rightarrow} & -\rho^T_\mu ~~.
\label{q5.10}
\eea

By imposing the heavy quark spin symmetry, parity and charge
conjugation invariance, and by assuming that the pseudoscalar meson
coupling are described by the lowest order (at most two derivatives)
chiral invariant operators, we can establish the following selection
rules for hadronic transitions:
\bea
{\rm even~ number~ of~ emitted~ pseudoscalars} &\leftrightarrow & 
\Delta l=0,2,4,...\nn\\
{\rm odd~ number~ of~ emitted~ pseudoscalars} &\leftrightarrow & 
\Delta l=1,3,5,...\nn\\
\label{q5.11}
\eea

In fact the spin independent operator describing $\Delta l=0,2,4,...$
transitions has charge conjugation $C=+1$ (see eq. (\ref {q2.13})). 
On the other hand, the lowest order, chiral invariant terms with
positive charge conjugation are:
\bea
&<{\cal A}_\mu {\cal A}_\nu>\nn\\
&<({\cal V}_\mu-\rho_\mu) ({\cal V}_\nu-\rho_\nu)>
\label{q5.12}
\eea
whose expansion contains an even number of pseudoscalar mesons.
Spin independence of the interaction, on the other hand, requires that the 
$\Delta l=1,3,5,...$ transitions are described by $C=-1$ operators.
At the lowest order we can form just one chiral invariant term with $C=-1$:
\be
<{\cal A}_\mu ({\cal V}_\nu-\rho_\nu)>
\label{q5.13}
\ee
whose expansion contains an odd number $(\ge 3)$of pseudoscalar mesons.

This selection rule is violated at higher orders of the chiral expansion
or by allowing for terms which explicitly break the heavy quark or 
the chiral symmetries.

To further characterize the hadronic transitions respecting chiral
symmetry, we consider below explicit expressions for the most general 
operators $\Pi_{ll'}$. For simplicity, we limit ourselves to those 
contributing
to two or three pion emissions:
\bea
\Pi_{SS'}&=&A_{SS'} <{\cal A}_\rho {\cal A}^\rho>
+B_{SS'} <(v\cdot {\cal A})^2>\nn\\
\Pi_{PS}^\mu&=&D_{PS}~\epsilon^{\mu\nu\rho\sigma} v_\nu <{\cal A}_\rho
({\cal V}_\sigma-\rho_\sigma)>\nn\\
\Pi_{PP'}^{\mu\nu}&=&A_{PP'} <{\cal A}_\rho {\cal A}^\rho>g^{\mu\nu}
+B_{PP'} <(v\cdot {\cal A})^2>g^{\mu\nu}
+C_{PP'}<{\cal A}^\mu {\cal A}^\nu>\nn\\
\Pi_{DS}^{\mu\nu}&=&C_{DS}<{\cal A}^\mu {\cal A}^\nu>
\label{q5.14}
\eea
The constants $A_{ll'}$, $B_{ll'}$, $C_{ll'}$ and $D_{ll'}$ are arbitrary 
parameters of dimension $(mass)^{-1}$, to be fixed from experiments. One can 
easily derive amplitudes, decay rates and distributions for the corresponding 
hadronic transitions.

For instance, the amplitude for the decay (\ref{q4.1}) is given by:
\be
{\cal M}(^3S_1\to^3S_1+\pi\pi)=\frac{4i\sqrt{M_S M_{S'}}}{f_\pi^2}
\epsilon' \cdot \epsilon^* \left( A_{SS'} p_1 \cdot
p_2 +B_{SS'} v\cdot p_1 v\cdot p_2 \right)
\label{q5.15}
\ee
where $\epsilon$ and $\epsilon '$ are the polarization vectors of 
quarkonium states; $p_1$, $p_2$ are the momenta of the two pions. 
It is well known that the use of chiral symmetry arguments \cite{BC}
leads to a general amplitude for the process in question which
contains a third independent term given by:
\be
C_{SS'} \frac{4 i \sqrt{M_S M_{S'}}}{f_\pi^2}
\left(\epsilon'\cdot p_1 \epsilon^*\cdot p_2+
\epsilon'\cdot p_2 \epsilon^*\cdot p_1 \right)~~.
\label{q5.16}
\ee
By combining the soft pion technique with a QCD double multipole expansion,
Yan \cite{Yann} finds $C_{SS'}=0$. It is interesting
to note that, within the present formalism, this result is an 
immediate consequence of the chiral and heavy quark spin symmetries.

Experimentally the amplitude (\ref{q5.15}) describes well the observed pion
spectra in the transitions $\psi' \to \psi\pi\pi$ \cite{Abrams} and 
$\Upsilon(2s) \to \Upsilon(1s)\pi\pi$ \cite{Argus}. The 
spectrum for the transition $\Upsilon(3s) \to \Upsilon(1s)\pi\pi$ seems to 
exhibit an unusual double-peaked shape \cite{Bow} and cannot be fitted using 
(\ref{q5.15}). We observe that in this case, due to the large available 
phase space, probably the soft-pion approximation is however not reliable. 

\subsubsection{Chiral breaking hadronic transitions}

In this section we discuss possible chiral breaking but spin
conserving terms \cite{noi1},
 which are important for transitions forbidden in the $SU(3)
\times SU(3)$ symmetry limit. Examples of such kind of transitions are
\be
^3P_{J'} \to \;^3P_J \pi^0, \; \;^3P_J \eta~~.
\label{q6.1}
\ee
The transitions
\be
\psi' \to J/\psi \pi^0, \, J/\psi \eta
\label{q6.3}
\ee
require terms which  violate also the spin symmetry and will be discussed
in the next section.

We first discuss the masses and mixings of the octet and singlet $\eta'$
pseudoscalar light meson states. The term which gives mass to the pseudoscalar
octet, massless in the chiral limit, is
\be
{\cal L}_{m}=\lambda_0 <\hat m (\Sigma + \Sigma^{\dagger})>
\label{q6.4}
\ee
Here $\hat m$ is the current mass matrix:
\be
{\hat m}=
\left (\begin{array}{ccc}
m_u & 0 & 0 \nn\\
0 & m_d & 0  \\
0 & 0 & m_s
\end{array}\right )~~.
\label{q6.5}
\ee
The lagrangian
(\ref{q6.4}) gives, in addition, a mixing $\pi^0 - \eta$: the physical states
${\tilde \pi}^0, \; {\tilde \eta}$ turn out to be:
\bea
{\tilde \pi}^0 & = & \pi^0 +\epsilon \eta \nn \\
{\tilde \eta} & = & \eta- \epsilon \pi^0
\label{q6.6}
\eea
where the mixing angle $\epsilon$ is
\be
\epsilon=\frac{(m_d -m_u) \sqrt{3}}{4 (m_s-\dd\frac{m_u+m_d}{2})}
\label{q6.7}
\ee
The $\eta'$, which is a chiral singlet, mixes with
$\pi^0, \; \eta$.
Such a mixing can be described by the term
\be
{\cal L}_{\eta\eta'}=\frac{i f_\pi}{4} \tilde\lambda 
<\hat m(\Sigma - \Sigma^{\dagger})>
\eta'
\label{q6.8}
\ee
where $\hat\lambda$ is a parameter with dimension of a mass. At first order in 
the
mixing angles  the physical states are:
\bea
{\tilde \pi}^0 & = & \pi^0 +\epsilon \eta +\epsilon' \eta' \nn \\
{\tilde \eta} & =  & \eta -\epsilon \pi^0 +\theta \eta' \nn \\
{\tilde \eta}' & = & \eta' -\theta \eta -\epsilon' \pi^0
\label{q6.9}
\eea
where
\bea
\epsilon' & = & \frac{\tilde\lambda (m_d -m_u)}{\sqrt{2} (m^2_{\eta'}- 
m^2_{\pi^0})}
\nn \\
\theta  & = & \sqrt{\frac{2}{3}}~~~\frac{\tilde\lambda \dd\left(
m_s-\frac{m_u+m_d}{2}
\right)}{m^2_{\eta'}- m^2_{\eta}}
\label{q6.10}
\eea
and $\epsilon$ as given in (\ref{q6.7}).

We consider chiral violating, spin-conserving hadronic transitions 
between charmonium states at first order in the chiral breaking mass matrix.
We therefore consider the quantities:
\bea
& &<\hat m(\Sigma +\Sigma^{\dagger})>   \nn \\
& &<\hat m(\Sigma - \Sigma^{\dagger})>
\label{q6.11}
\eea
The first one is even under parity, the second is odd, and both
have $C=+1$.

The only term spin-conserving and of leading order in the current quark masses
contributing to the transition (\ref{q6.1}) is
\be
<J_\mu {\bar J}_\nu > v_\rho \epsilon^{\mu\nu\rho\sigma} \partial_\sigma
\left[ \alpha \frac{i f_\pi}{4} <\hat m(\Sigma - \Sigma^{\dagger})> + \beta
f_\pi\eta'
\right]
\label{q6.12}
\ee
where $\alpha$ and $\beta$ are coupling constants of dimensions
$(mass)^{-2}$.
The direct coupling to $\eta'$ contributes through the mixing (\ref{q6.9}).
The spin symmetry of the heavy sector gives relations among the modulus square
matrix elements of the transitions between the two $p$-wave states. In
particular we find  that
\be
|{\cal M}|^2 (^3P_0 \to ^3P_0\pi)
=|{\cal M}|^2 (^3P_2 \to ^3P_0\pi)=0
\label{q6.13}
\ee
and that all non-vanishing matrix elements can be expressed in terms of
$^3P_0 \to ^3P_1\pi$:
\bea
|{\cal M}|^2 (^3P_1 \to ^3P_1\pi)& =&\frac{1}{4}
|{\cal M}|^2 (^3P_0 \to ^3P_1\pi)   \nn \\
|{\cal M}|^2 (^3P_1 \to ^3P_2\pi)& = &\frac{5}{12}
|{\cal M}|^2 (^3P_0 \to ^3P_1\pi)   \nn \\
|{\cal M}|^2 (^3P_2 \to ^3P_2\pi)& = &\frac{3}{4}
|{\cal M}|^2 (^3P_0 \to ^3P_1\pi)   \nn \\
|{\cal M}|^2 (^1P_1 \to ^1P_1\pi)& =&
|{\cal M}|^2 (^3P_0 \to ^3P_1\pi)
\label{q6.14}
\eea
where $\pi$ stays for $\pi^0$ or $\eta$. The relations (\ref{q6.14}) can be
generalized for any spin conserving transition between $l=1$ multiplets,
leading to the same results of a QCD double multipole expansion \cite{Yann}.
Predictions for widths can be easily obtained from (\ref{q6.12}).
\def\qpiu{Q_v^{(+)}}
\def\qmeno{Q_v^{(-)}}
\def\propiu{\frac {1+\slash v}{2}}
\def\promeno{\frac {1-\slash v}{2}}
\def\opiu{O_1^{(+)}}
\def\omeno{O_1^{(-)}}

\subsubsection{Spin breaking hadronic transitions}

We study here transitions which violate spin symmetry \cite{noi1}.
For heavy mesons there are only two types of operators that can break
spin symmetry. The  reason is that on the quark (antiquark) indices
of the quarkonium wave function act
projection operators $(1+\slash v)/2$ and $(1-\slash v)/2$ which
reduce the original $4\times 4$-dimensional space to a $2\times 2$-dimensional
one. Obviously, in the rest frame, the most general
spin symmetry breaking term is of the form $\vec a\cdot\vec\sigma$,
where $\vec\sigma$ are the Pauli matrices.
In an arbitrary frame one observes that any $\Gamma$-matrix
sandwiched between two projectors $(1+\slash v)/2$, or $(1-\slash v)/2$,
can be reexpressed in terms of $\sigma_{\mu\nu}$ sandwiched between the
same projectors:
\bea
\frac{1+\slash v}{2} 1\frac{1+\slash v}{2}&=&\frac{1+\slash v}{2}\\
\frac{1+\slash v}{2}\gamma_5\frac{1+\slash v}{2}&=&0\\
\frac{1+\slash v}{2}\gamma_\mu\frac{1+\slash v}{2}&=&v_\mu\frac{1+\slash
v}{2}\\
\frac{1+\slash v}{2}\gamma_\mu\gamma_5\frac{1+\slash v}{2} &=&
\frac{1}{2}\epsilon_{\mu\nu\alpha\beta}v^\nu\frac{1+\slash v}{2}
\sigma^{\alpha\beta}\frac{1+\slash v}{2}\\
\frac{1+\slash v}{2}\gamma_5\sigma_{\mu\nu}\frac{1+\slash v}{2} &=&
-\frac{i}{2}\epsilon_{\mu\nu\alpha\beta}\frac{1+\slash v}{2}
\sigma^{\alpha\beta}\frac{1+\slash v}{2}
\eea
and analogous relations with $(1+\slash v)/2\to(1-\slash v)/2$.
We use here $\epsilon_{0123}=+1$. Let us define
\be
\sigma_{\mu\nu}^{(\pm)}=\frac{1 \pm \slash v}{2}\sigma_{\mu\nu}
\frac{1 \pm \slash v}{2}~~.
\ee
In the rest frame, $\sigma_{\mu\nu}^{(\pm)}$ reduce to Pauli matrices. From 
the previous identities it follows that the most general spin symmetry
breaking terms in the quarkonium space are of the form
$G_1^{\mu\nu}\sigma_{\mu\nu}^{(+)}$, or $G_2^{\mu\nu}\sigma_{\mu\nu}^{(-)}$,
with $G_i^{\mu\nu}$ two arbitrary antisymmetric tensors. 
One expects that any insertion of the operator
$\sigma_{\mu\nu}^{(\pm)}$ gives a suppression factor $1/m_Q$.

A relevant example of spin breaking is the splittings of the levels in a
multiplet; one can easily write down the  spin-spin,
spin-orbit and tensor terms \cite{noi1}. 

We apply as an example the formalism to the transitions $\psi' \to J/\psi 
\pi^0$ and $\psi' \to J/\psi \eta$. Of particular interest is the ratio
\be
R=\frac {\Gamma (\psi' \to J/\psi \pi^0)}{\Gamma(\psi' \to J/\psi \eta )}
\label{q7.1}
\ee
which provides for a measure of the light-quark mass ratio
\be
r=\frac {m_d - m_u}{m_s -\dd\frac{m_u+m_d}{2}}~~.
\label{q7.2}
\ee
Using partial conservation of axial-vector current, Ioffe and Shifman
\cite{Ioffe} give the prediction
\be
R=\frac{27}{16} \left[\frac{\vec{p_{\pi}}}{\vec{p_{\eta}}}\right]^3 r^2
\label{q7.3}
\ee
The calculation of $R$ is straightforward with the heavy quark formalism. Eq.
(\ref{q7.3}) will be recovered when neglecting the mixings $\pi^0 - \eta$ and
$\eta -\eta'$ (or a possible direct coupling of $\eta'$).

The most general spin breaking lagrangian for the processes $\psi' \to J/\psi 
\pi^0 , \eta$ is
\bea
{\cal L}=&i\epsilon_{\mu\nu\rho\lambda} \left[ <J'\sigma^{\mu\nu}\bar{J}> -
<\bar{J}\sigma^{\mu\nu}J'>\right] v^{\rho} \times \nn\\
&\partial^{\lambda} \left[\dd\frac{i A}{4} <\hat m (\Sigma -\Sigma^{\dagger})>
+B\eta ' \right] +h.c.
\label{q7.4}
\eea
The couplings $A$ and $B$ have dimension $(mass)^{-1}$; the $B$ term
contributes to the ratio (\ref{q7.1})
 via the mixing $\pi^0 -\eta '$ and $\eta -\eta '$,
in the same way as the $\beta$ coupling in (\ref{q6.12}). There are no terms
with the insertion of two $\sigma$; the two P and C conserving  candidates
\bea
&\epsilon_{\mu\nu\rho\lambda} \left[
<J'\sigma^{\mu\tau}\bar{J}\sigma_{\tau}^{~\nu}> +
<\bar{J}\sigma^{\mu\tau}J'\sigma_{\tau}^{~\nu}>\right] v^{\rho}
\partial^{\lambda} <\hat m (\Sigma -\Sigma^{\dagger})>;\nn\\
&\epsilon_{\mu\nu\rho\lambda} \left[
<J'\sigma^{\mu\nu}\bar{J}\sigma^{\rho\lambda}> +
<\bar{J}\sigma^{\mu\nu}J'\sigma^{\rho\lambda}>\right]
<\hat m (\Sigma -\Sigma^{\dagger})>
\label{q7.5}
\eea
are both vanishing. Using the lagrangian (\ref{q7.4}) and taking into account 
the mixings (\ref{q6.9}) we can calculate the ratio (\ref{q7.1})
\be
R=\frac{27}{16} \left[\frac{\vec{p_{\pi}}}{\vec{p_{\eta}}}\right]^3
\left[ \frac {m_d - m_u}{m_s -1/2 (m_u+m_d)}\right]^2 \left[\frac{
1+\dd\frac{2 B}{3 A}\frac{\hat\lambda f_{\pi}}{m^2_{\eta '}-m^2_{\pi0}}}
{1+\dd\frac{B}{A}\frac{\hat\lambda f_{\pi}}{m^2_{\eta 
'}-m^2_{\eta}}}\right]^2~~.
\label{q7.6}
\ee
If we neglect the mixings $\pi^0 -\eta '$ and $\eta -\eta '$ ($\hat\lambda 
=0$) or
the direct coupling of $\eta '$ ($B$=0) (\ref{q7.6}) reduces to (\ref{q7.3}).

Eq. (\ref{q7.6}) can receive corrections from electromagnetic
contributions to the transition $\psi ' \to J/\psi \pi^0$. It has been
shown that such corrections are suppressed \cite{Don,Malt}.
A second type of corrections is associated with higher order terms in the 
light-quark mass expansion (the lagrangian (\ref{q7.4}) is the first order of 
such an expansion); a discussion can be found in ref. \cite{DW}.

\section{Appendix A}

We list here the Feynman rules for the vertices appearing in the heavy meson
chiral lagrangian and used in the text. Dashed lines refer to light mesons,
solid lines to heavy mesons of fixed masses ($M_P$ or $M_+$) and $J^P$.
The heavy meson propagators, for a state with velocity $v$ and residual momentum
$k$,  are 
\be
\frac i{2(v\cdot k+\frac 3 4\Delta)} ~~~~~ J^P = 0^-
\ee
and
\be 
-\frac {i(g^{\mu\nu}-v^\mu v^\nu)}{2(v\cdot k-\frac 1 4\Delta)} ~~~~~ J^P = 1^-
\ee
where $\Delta = M_{P^*} - M_P$. For the $0^+$ and $1^+$ states one has similar
formulas with the appropriate mass difference $\Delta$.

\begin{eqnarray}
&&{{\lower50pt\hbox{\epsfxsize=2.2truein \epsfbox{v1.eps}}}\hspace{1.5truecm}
\frac{2 M_P}{f_\pi } g \; (\epsilon (1^- ) \cdot q )
}\nonumber
\end{eqnarray}

\vspace{1truecm}
\begin{eqnarray}
&&{{\lower50pt\hbox{\epsfxsize=2.2truein \epsfbox{v2.eps}}} \hspace{1.5truecm}
-\frac{2 M_P}{f_\pi } g \; \epsilon_{\mu\nu\alpha\beta} \;
\epsilon^\mu (1^- )\epsilon^{*\nu} (1^- ) q^\alpha v^\beta 
}\nonumber
\end{eqnarray}

\vspace{1truecm}
\begin{eqnarray}
&&{{\lower50pt\hbox{\epsfxsize=2.2truein \epsfbox{v3.eps}}} \hspace{1.5truecm}
\frac{2 \sqrt{M_P M_{+}}}{f_\pi } h \; (v \cdot q )
}\nonumber
\end{eqnarray}

\vspace{1truecm}
\begin{eqnarray}
&&{{\lower50pt\hbox{\epsfxsize=2.2truein \epsfbox{v4.eps}}} \hspace{1.5truecm}
-\frac{2 \sqrt{M_P M_{+}}}{f_\pi } h \; (v \cdot q ) (\epsilon (1^+ ) \cdot
\epsilon^{*} (1^- ))
}\nonumber
\end{eqnarray}

\vspace{1truecm}
\begin{eqnarray}
&&{{\lower50pt\hbox{\epsfxsize=2.2truein \epsfbox{v5.eps}}} \hspace{1.5truecm}
\frac{4 g_V M_P}{\sqrt{2}} \; \lambda \; \epsilon_{\alpha \beta \mu \nu}\;
\epsilon^{\alpha} (\rho ) \epsilon^{\beta} (1^- ) v^\mu q^\nu
}\nonumber
\end{eqnarray}

\vspace{1truecm}
\begin{eqnarray}
&&{{\lower50pt\hbox{\epsfxsize=2.2truein \epsfbox{v6.eps}}} \hspace{1truecm}
-i \sqrt{2} g_V \sqrt{M_P M_+} \epsilon^{\alpha} (\rho )
\epsilon^{*\beta} (1^+ )  
\left[  \zeta g_{\alpha \beta} + 2 \; \mu ( v \cdot
q \; g_{\alpha\beta} - v_{\alpha} q_{\beta} ) \right]
}\nonumber
\end{eqnarray}

\vspace{1truecm}
\begin{eqnarray}
&&{{\lower50pt\hbox{\epsfxsize=2.2truein \epsfbox{v7.eps}}} \hspace{1.5truecm}
-i \sqrt{2} g_V M_P \; \beta ( v \cdot \epsilon (\rho ) )
}\nonumber
\end{eqnarray}

\section{Appendix B}

In this appendix we list some integrals that are encountered in computing 
loop corrections in the effective chiral theory for heavy mesons 
\cite{Boyd-Grin,fag,goity92,goity95}. In the sequel we put
${\bar \Delta}=2/\epsilon -\gamma+ln(4 \pi) +1$.

\be
i\int \frac{d^{4- \epsilon} 
q}{(2\pi)^{4-\epsilon}} \frac{1}{q \cdot v -\Delta}=0
\ee

\be
i\int \frac{d^{4- \epsilon} q}{(2\pi)^{4-\epsilon}} \frac{1}{q^2 -m^2}=
\frac{1}{16 \pi^2} I_1(m)
\ee

\be
i\int \frac{d^{4- \epsilon} q}{(2\pi)^{4-\epsilon}} 
\frac{1}{(q^2 -m^2)(q\cdot v - \Delta)}=
\frac{1}{16  \pi^2}\frac{1}{\Delta}  I_2(m, \Delta)
\ee

\bea
J^{\mu \nu}(m, \Delta) &=&
i\int \frac{d^{4- \epsilon} q}{(2\pi)^{4-\epsilon}} 
\frac{q^\mu q^\nu}{(q^2 -m^2)(q\cdot v - \Delta)}=
  \nn \\
&=& \frac{1}{16 \pi^2} \Delta [ J_1(m,\Delta) g^{\mu \nu} +J_2(m,\Delta)
v^\mu v^\nu]
\eea
where
\bea
I_1(m)&=& m^2 {\rm ln}\frac{m^2}{\mu^2}- m^2 {\bar \Delta}\nn \\
I_2(m, \Delta)
&=& - 2 \Delta^2 {\rm ln}\frac{m^2}{\mu^2}-4 \Delta^2F(\frac{m}{\Delta})
+2 \Delta^2 (1 +{\bar \Delta})
\eea
and
\bea
F(x)&=& \sqrt{1-x^2}~{\rm tanh}^{-1}\sqrt{1-x^2}~~~~~~|x|\leq 1 \nn \\ 
&=&- \sqrt{x^2-1}~{\rm tan}^{-1}\sqrt{x^2-1}~~~~~|x|\geq 1 
\eea
\bea
J_1(m,\Delta)&=& (-m^2+\frac{2}{3} \Delta^2) {\rm ln}\frac{m^2}{\mu^2}
+\frac{4}{3}( \Delta^2 - m^2)F(\frac{m}{\Delta})\nn \\
&-& \frac{2}{3} \Delta^2 (1 +{\bar \Delta}) +\frac{1}{3}m^2(2 +
3 {\bar \Delta}) \nn \\
J_2(m,\Delta)&=& (2m^2-\frac{8}{3} \Delta^2) {\rm ln}\frac{m^2}{\mu^2}
-\frac{4}{3}(4 \Delta^2 - m^2)F(\frac{m}{\Delta})\nn \\
&+& \frac{8}{3} \Delta^2 (1 +{\bar \Delta}) -\frac{2}{3}m^2(1 +
3 {\bar \Delta})~~.
\eea
Moreover, if we put:
\be
\int \frac{d^{4 - \epsilon} q}{(2\pi)^{4 - \epsilon}} 
\frac{q_\alpha q_\beta}{(q^2 -m^2)(v \cdot q -\Delta)
(v \cdot q - \Delta')}= \frac{i}{16\pi^2} \left( C_1(\Delta,\Delta',m) 
g_{\alpha\beta}
+ C_2(\Delta,\Delta',m)v_\alpha v_\beta \right)~,
\ee
then it follows that
\bea
C_1(\Delta,\Delta',m) & = & \frac{1}{ \Delta - \Delta'} \left[
\Delta J_1 (m, \Delta ) -\Delta' J_1(m, \Delta') \right]\\
C_2(\Delta,\Delta',m) & = & \frac{1}{ \Delta - \Delta'} \left[
\Delta J_2 (m,\Delta) -\Delta' J_2(m, \Delta') \right]~.
\eea
It can be useful to write down explicitly 
\be
C(\Delta,\Delta',m) = C_1(\Delta,\Delta',m) + C_2 (\Delta,\Delta',m) \;,
\ee
which is given by
\be
C(\Delta,\Delta',m)  =  \frac{2 m^3}{9(\Delta - \Delta')} \left[
H (\frac{\Delta}{m},m ) - H(\frac{\Delta'}{m},m) \right]
\ee
with
\bea
H (x,m) & =&  - 9 x^3(1+ {\bar \Delta})
+\frac{9}{2} x {\bar \Delta} + (9 x^3 - \frac{9}{2} x ) 
\log(\frac{m^2}{\mu^2})+
\nn \\
& + &  18 x^3 F(\frac{1}{x}) \;.
\eea

For $\Delta =\Delta'$ the previous formulas reduce to
\bea
C_1(\Delta,\Delta,m)= J_1(m, \Delta) + \Delta \frac{\partial J_1(m, \Delta)}
{\partial \Delta}\\
C(\Delta,\Delta,m)= \frac{2 m^2}{9} H' (\frac{\Delta}{m},m)
\eea
where $H'(x,m) = \frac{d H(x,m)}{d x}$.

\section{Appendix C}

In the $l\neq 0$ case,  the multiplet $J$ for quarkonium is generalized to 
$J^{\mu_1 \ldots \mu_l}$, with a decomposition 
\bea
J^{\mu_1 \ldots \mu_l} & = & \frac{(1+\slash v)}{2} \left[
 H_{l+1}^{\mu_1 \ldots \mu_l\alpha} \gamma_{\alpha}
+ \frac{1}{\sqrt{l(l+1)}} \sum_{i=1}^{l} \epsilon^{\mu_i\alpha\beta\gamma}
v_{\alpha} \gamma_{\beta}H_{l\gamma}^{\mu_1 \ldots \mu_{i-1} \mu_{i+1}\ldots
\mu_l} \right. \nn \\
&+& \frac{1}{l}\sqrt{\frac{2l-1}{2l+1}}\sum_{i=1}^{l} (\gamma^{\mu_i}-
v^{\mu_i}) H_{l-1}^{\mu_1 \ldots \mu_{i-1} \mu_{i+1}\ldots \mu_l} \nn \\
 &-&  \frac{2}{l\sqrt{(2l-1)(2l+1)}} \sum_{i<j} (g^{\mu_i\mu_j}-v^{\mu_i}
v^{\mu_j})\gamma_{\alpha} H_{l-1}^{\alpha\mu_1 \ldots \mu_{i-1} 
\mu_{i+1}\ldots
\mu_{j-1}\mu_{j+1}\ldots \mu_l} \nn \\
&+&  K_l^{\mu_1\ldots\mu_l} \gamma_5 \left] \frac{(1-\slash v)}{2}
 \right. ~~~.
\label{aq2.5}
\eea
In the above equation, $K_l^{\mu_1\ldots\mu_l}$ represents the spin singlet 
$\,^1 l_J$. Since $J=l$, $K_l^{\mu_1\ldots\mu_l}$ is a completely symmetric,
traceless tensor, satisfying the transversality condition:
\be
v_{\mu_1}K_l^{\mu_1\ldots\mu_l}=0~~.
\label{aq2.6}
\ee
The spin triplet $\,^3 l_J$ is represented by 
$H_{l+1}^{\mu_1 \ldots \mu_{l+1}}$ for $J=l+1$, 
$H_{l}^{\mu_1 \ldots \mu_l}$ for $J=l$ and 
$H_{l-1}^{\mu_1 \ldots \mu_{l-1}}$ for $J=l-1$. These three tensors are 
completely symmetric, traceless and satisfy transversality conditions 
analogous to eq. (\ref{aq2.6}). Moreover, in order to avoid orbital momenta
other than $l$, we require that $J^{\mu_1 \ldots \mu_l}$ itself is 
completely symmetric, traceless and orthogonal to the velocity:
\be
v_{\mu_1}J^{\mu_1 \ldots \mu_l}=0~~.
\label{aq2.7}
\ee
This allows to identify the states in (\ref{aq2.5}) with the physical states.
The normalisation for $J^{\mu_1 \ldots \mu_l}$ has been chosen so that:
\bea
< J^{\mu_1 \ldots \mu_l}{\bar J}_{\mu_1 \ldots \mu_l} > =
& 2 & \left( H_{l+1}^{\mu_1 \ldots \mu_{l+1}}
H^{\dagger l+1}_{\mu_1 \ldots \mu_{l+1}}
-H_{l}^{\mu_1 \ldots \mu_{l}}H^{\dagger l}_{\mu_1 \ldots \mu_{l}}\right. \nn 
\\
& + & \left. H_{l-1}^{\mu_1 \ldots \mu_{l-1}}H^{\dagger l-1}_{\mu_1 \ldots 
\mu_{l-1}}                                
-K_{l}^{\mu_1 \ldots \mu_{l}}K^{\dagger l}_{\mu_1 \ldots \mu_{l}} \right)
\label{aq2.8}
\eea
where ${\bar J}=\gamma^0 J^\dagger \gamma^0$ and $< \ldots >$ means the trace
over the Dirac matrices.

For example the expression for the $P$-wave multiplet $J^\mu$ that can be 
obtained from eq. (\ref{aq2.5}) is:
\bea
J^\mu = \left. \frac{1+\slash v}{2} \right [
   & H_2^{\mu\alpha}\gamma_{\alpha} &  +
      \frac{1}{\sqrt{2}}\epsilon^{\mu\alpha\beta\gamma}v_\alpha\gamma_\beta
H_{1\gamma}  \nn \\
&+& \left. \frac{1}{\sqrt{3}} (\gamma^\mu-v^\mu)H_0 +K_1^\mu \gamma_5 \right]
\frac{(1-\slash v)}{2}~~.
\label{aq2.9}
\eea

\vspace{1truecm}
\begin{large}
\begin{bf}
Acknowledgements
\end{bf}
\end{large}
We thank P. Colangelo, N. Paver and S. Stone for most useful comments.


\vspace{1truecm}

\begin{thebibliography}{300}

\bibitem{ltw}For a recent review see H. Leutwyler, 
talk given at Workshop on Chiral Dynamics, Theory
and Experiments, Cambridge, MA, 25-29 July 1994 (hep-ph/9409423), 
in {\it Chiral Dynamics Workshop} (1994) p.14
and references therein.

\bibitem{hq1}
N.Isgur and M.B.Wise, Phys. Lett. {\bf B232} (1989) 113; ibidem
{\bf B237} (1990) 527.

\bibitem{hq2} 
H.Georgi, Phys. Lett. {\bf B240} (1990) 447.

\bibitem{hq3}
B.Grinstein, Nucl. Phys. {\bf B339}
(1990) 253.

\bibitem{hq4}
M.B.Voloshin and M.A.Shifman, Sov. J. Nucl. Phys. {\bf 45} (1987) 292;
ibidem {\bf 47} (1988) 511;
H. Politzer and M.B. Wise, Phys. Lett. {\bf B206} (1988) 681; ibidem {\bf 
B208} (1988) 504;
E.Eichten and B.Hill, Phys. Lett.
{\bf B234} (1990) 511;
J.D. Bjorken in {\it Results and Perspectives in Particle Physics}, proc. of
the 4th Rencontres de Physique de la Valle d'Aoste, La Thuile, Italy 1990,
M. Greco ed. (ed. Frontieres, Gif-sur-Yvette, France, 1990) p. 583; 
N.Isgur and M.B. Wise, Phys. Rev. {\bf D43} (1991) 819.

\bibitem{hq5} 
A. Falk, H. Georgi, B. Grinstein and M.B. Wise, Nucl. Phys. {\bf B343} (1990)
1.

\bibitem{hqrev}
H. Georgi, contribution to the {\it Proceedings of TASI 91}, R.K. Ellis ed.,
World Scientific, Singapore,1991 ;
B. Grinstein, contribution to {\it High Energy Phenomenology}, R. Huerta and
M.A. Peres eds., World Scientific, Singapore, 1991;
N. Isgur and M. Wise, contribution to {\it Heavy Flavours}, A. Buras and M.
Lindner eds., World Scientific, Singapore,1992; 
T. Mannel, contribution to the Workshop {\it QCD 94}, Montpellier, 7-13 July
1994, preprint CERN-TH.7499/94.

\bibitem{neubert}
M. Neubert, Phys. Rep. {\bf 245} (1994) 259; M. Neubert, preprint Cern-TH/96-55,
hep-ph/9604412.

\bibitem{luke90}
M.E. Luke, Phys. Lett. {\bf B252} (1990) 447;
C.G. Boyd and D.E. Brahm, Phys. Lett. {\bf B257} (1991) 393;
R.F. Lebed and M. Suzuki, Phys. Rev. {\bf D44} (1991) 829.

\bibitem{wise}
M.B.Wise, Phys. Rev. {\bf D45} (1992) R2188.

\bibitem{yan}
T.-M. Yan, H.-Y. Cheng, C.-Y. Cheung, G.-L. Lin, Y.C. Lin and H.-L. Yu,
Phys. Rev. {\bf D46} (1992) 1148.

\bibitem{burdman92}
G. Burdman and J.F.Donoghue, Phys. Lett. {\bf B280} (1992) 287; 
Phys. Rev. Lett. {\bf 68} (1992) 2887.

\bibitem{noilr}
R. Casalbuoni, A. Deandrea, N. Di Bartolomeo, R. Gatto, F. Feruglio and G.
Nardulli, Phys. Lett. {\bf B292} (1992) 371.

\bibitem{casa92}
R. Casalbuoni, A. Deandrea, N. Di Bartolomeo, F. Feruglio, R. Gatto and 
G. Nardulli, Phys. Lett. {\bf B294} (1992) 106.

\bibitem{casa93}
R. Casalbuoni, A. Deandrea, N. Di Bartolomeo, F. Feruglio, R. Gatto and 
G. Nardulli, Phys. Lett. {\bf B299} (1993) 139.

\bibitem{schechter}
J. Schechter and A. Subbaraman, Phys. Rev. {\bf D48} (1993) 332.

\bibitem{falkluke}
A.F. Falk and M. Luke, Phys. Lett {\bf B292} (1992) 119.

\bibitem{kilian}
U. Kilian, J.C. K\"orner 
and D. Pirjol, Phys. Lett. {\bf B288} (1992) 360.

\bibitem{ebert}
D. Ebert, T. Feldmann, R. Friedrich and H. Reinhardt, Nucl. Phys. {\bf B434}
(1995) 619.

\bibitem{barwise}
N. Isgur and M. Wise, Nucl. Phys. {\bf B348} (1991) 276.

\bibitem{bargeorgi}
H. Georgi, Nucl. Phys. {\bf B348} (1991) 293.

\bibitem{barman}
T. Mannel, W. Roberts and Z. Ryzak, Nucl. Phys. {\bf B355} (1991) 38.

\bibitem{barhussain}
F. Hussain, J.G. Korner, R. Migneron, Phys. Lett. {\bf B248} 
(1990) 406, ERRATUM-ibid. {\bf B252} (1990) 723.

\bibitem{HUSSAIN}F. Hussain, Dong-Sheng Liu, M. Kramer, J.G. Korner,
S. Tawfiq, Nucl. Phys. {\bf B370} (1992) 259.

\bibitem{qcdsr}
P. Colangelo, G. Nardulli, A. Deandrea, N. Di Bartolomeo, R. Gatto and 
F. Feruglio, Phys. Lett. {\bf B339} (1994) 151.

\bibitem{M.Shifman}M.A. Shifman ed.,{\it Vacuum structure and QCD sum rules},
North-Holland, 1992.

\bibitem{defazio2}
P. Colangelo, F. De Fazio and G. Nardulli, Phys. Lett. {\bf B334} (1994) 175.

\bibitem{rivnc}
G. Nardulli, Riv. Nuovo Cim. {\bf 15} (1992) no.10, 1. 

\bibitem{MARTINELLI}
G. Martinelli, Proc. 6th Rencontres De Blois, Blois, France, 20 - 25
Jun 1994, Ed. Frontieres, Gif-Sur-Yvette.

\bibitem{mannel}
T. Mannel, W. Roberts and Z. Ryzak, Nucl. Phys. {\bf B368} (1992) 204.

\bibitem{ripinv}
M. Luke and A.V. Manohar, Phys. Lett. {\bf B286} (1992) 348.

\bibitem{falk}
A.F. Falk, Nucl. Phys. {\bf B378} (1992) 79.

\bibitem{GL}
J. Gasser and H. Leutwyler, Ann. Phys. {\bf 158} (1984) 142;
Nucl. Phys. {\bf 250} (1985) 465.

\bibitem{CCWZ}
S. Coleman, J. Wess and B. Zumino, Phys. Rev. {\bf 177} (1969) 2239;
C.G. Callan Jr., S. Coleman, J. Wess and B. Zumino, Phys. Rev. {\bf 177} 
(1969)
2247.

\bibitem{Boyd-Grin} 
C.G. Boyd and B. Grinstein, Nucl. Phys. {\bf B442} (1995) 205.

\bibitem{cheng}
H.-Y. Cheng, C.-Y. Cheung, G.-L. Lin, Y.C. Lin, T.-M. Yan and H.-L. Yu,
Phys. Rev. {\bf D49} (1994) 2490.

\bibitem{bando}
M.Bando, T.Kugo and K.Yamawaki, Nucl. Phys. {\bf B259} (1985) 493; and
Phys. Rep. {\bf 164} (1988) 217.

\bibitem{ko93}
P. Ko, Phys. Rev. {\bf D47} (1993) 1964.

\bibitem{bando1}
M.Bando, T.Kugo, S. Uehara, K.Yamawaki and T. Yanagida, Phys. Rev. Lett. {\bf
54} (1985) 1215.

\bibitem{bando2}
M.Bando, T.Kugo and K.Yamawaki, Nucl. Phys. {\bf B259} (1985) 493.

\bibitem{KSRF}
K. Kawarabayashi and M. Suzuki, Phys. Rev. Lett. {\bf 54} (1966) 255;
Riazuddin and Fayyazuddin, Phys. Rev. {\bf 147} (1966) 1071.

\bibitem{gvect}
H. Georgi, Phys. Rev. Lett. {\bf 63} (1989) 1917; Nucl. Phys. {\bf B331} 
(1990)
311.

\bibitem{manvec}
T. Feldmann and T. Mannel, Phys. Lett. {\bf B344} (1995) 334.

\bibitem{cleo}
CLEO Collab., F.Butler et al., Phys Rev. Lett. {\bf 69} (1992) 2041 and
talk given at {\it XXI Rencontres de Moriond}, Les Arcs, March 16-23 1996.

\bibitem{pdb}
Particle Data Group, Review of Particle Properties, Phys. Rev. {\bf D50} 
(1994).

\bibitem{acc}
ACCMOR Collab., S.Barlag et al., Phys. Lett. {\bf B278} (1992) 480.

\bibitem{abj}
J.F. Amundson, C.G Boyd, E. Jenkins, M. Luke, A.V. Manohar,
J.L. Rosner, M.J. Savage and M.B. Wise, Phys. Lett. {\bf B296} (1992) 415.

\bibitem{cho}
P. Cho, H. Georgi, Phys. Lett. {\bf B296} (1992) 408.

\bibitem{cina1}
H.-Y. Cheng, C.-Y. Cheung, G. L. Lin, Y. C. Lin, T.-M. Yan and H.-L. Yu,
Phys. Rev. {\bf D47} (1993) 1030.

\bibitem{suzuki}
M. Suzuki, Phys. Rev. {\bf D37} (1988) 239.

\bibitem{isgur}
N. Isgur and M. B. Wise, Phys. Rev. {\bf D41} (1990) 151.

\bibitem{pham}
T. N. Pham, Phys. Rev. {\bf D25} (1982) 2955.

\bibitem{nussi}
S. Nussinov and W. Wetzel, Phys. Rev. {\bf D36} (1987) 139.

\bibitem{dompav88}
C.A.Dominguez and N. Paver, Z. Phys. {\bf C41} (1988) 217.

\bibitem{richardson} 
J. L. Richardson, Phys. Lett. {\bf B82} (1979) 272.

\bibitem{salpeter}
E. E. Salpeter, Phys. Rev. {\bf 87} (1952) 328.

\bibitem{pietroni}
P. Colangelo, G. Nardulli and M. Pietroni, Phys. Rev. {\bf D43} (1991) 3002.

\bibitem{donnell} P.J. O'Donnell and Q.P. Xu, Phys. Lett. {\bf B336} (1994) 
113.

\bibitem{ovc}
A.A. Ovchinnikov, Sov. J. of Nucl. Phys. {\bf 50} (1989) 519.


\bibitem{bel}
V.M. Belyaev, V.M. Braun, A. Khodzhamirian, R. R\"uckl, Phys.Rev. {\bf D51}
(1995) 6177. 

\bibitem{alievs}
T.M. Aliev, D.A. Demir, E. Iltan, N.K. Pak, Phys. Lett. {\bf B351} (1995) 339.

\bibitem{neubert1}
M.Neubert, Phys. Rev. {\bf D46} (1992) 1076.

\bibitem{APE}
C. Allton, Int. Symposium on Lattice Gauge Theories, Melbourne, Australia
11-15 July 1995.

\bibitem{UKQCD}
D.G. Richards, Int. Symposium on Lattice Gauge Theories, Melbourne, Australia
11-15 July 1995;
UKQCD collab., A.K. Ewan et al., hep-lat/9508030.

\bibitem{MILC}
C. Bernard, Int. Symposium on Lattice Gauge Theories, Melbourne, Australia
11-15 July 1995.

\bibitem{maiani}
J. Bijnens, G. Ecker and J. Gasser, in {\it The Daphne Physics Handbook}, 
L. Maiani, G. Pancheri and N. Paver eds. (1992) p. 115.
 
\bibitem{fle}
R. Fleischer, Phys. Lett. {\bf B303} (1993) 147.

\bibitem{fag}
A. F. Falk and B. Grinstein, Nucl. Phys. {\bf B416} (1994) 771.

 
\bibitem{ros}
J.L. Rosner and M.B. Wise, Phys. Rev. {\bf D47} (1993) 343.

\bibitem{ran}
L. Randall and E. Sather, Phys. Lett. {\bf B303} (1993) 345.

\bibitem{jen}
E. Jenkins, Nucl. Phys. {\bf B412} (1994) 181.

\bibitem{hyp}
N. Di Bartolomeo, F. Feruglio, R. Gatto and G. Nardulli, Phys. Lett. {\bf B347}
(1995) 405.

\bibitem{Ball}
P. Ball, Nucl. Phys. {\bf B421} (1994) 593.

\bibitem{delphi} 
Delphi Collaboration, P.Abreu et al., Phys. Lett. {\bf B345} (1995) 598.

\bibitem{opal} 
Opal Collaboration, R.Akers et al., Z.Phys. {\bf C66} (1995) 19.

\bibitem{isgur2}
S.Godfrey and N.Isgur, Phys. Rev. {\bf D32} (1985) 189.

\bibitem{veseli}
A. Wambach, Nucl. Phys. {\bf B434} (1995) 647; S. Veseli and M.G. Olsson, 
preprint MAPH-96-924, hep-ph/9601307.
 
\bibitem{noi}
P.Colangelo, G.Nardulli, A.A.Ovchinnikov and N.Paver,
Phys. Lett. {\bf B269} (1991) 204.

\bibitem{cola92}
P.Colangelo, G.Nardulli and N.Paver,
Phys. Lett. {\bf B293} (1992) 207.

\bibitem{faz}
P. Colangelo, F. De Fazio, N. Di Bartolomeo, R. Gatto and G.Nardulli,
Phys. Rev. {\bf D52} (1995) 6422.

\bibitem{chern1}
V.L.Chernyak and A.R.Zhitnitsky, JETP Lett. {\bf 25} (1977) 510; Yad. Fiz. 
{\bf 31} (1980) 1053;\\
A.V.Efremov and A.V.Radyushkin, Phys. Lett. {\bf B94} (1980) 245; Teor. 
Mat. Fiz. {\bf 42} (1980) 147;\\
G.P.Lepage and S.J.Brodsky, Phys. Lett. {\bf B87} 
(1979) 359; Phys. Rev. {\bf D22} (1980) 2157.

\bibitem{chern2}
V.L.Chernyak and A.R.Zhitnitsky, Phys. Rep. {\bf 112} (1984) 173.

\bibitem{fil} 
V.M.Braun and I.B.Filyanov, Z.Phys. {\bf C48} (1990) 239.

\bibitem{iswi}
N.Isgur and M.B.Wise, Phys. Rev. Lett. {\bf 66} (1991) 1130.

\bibitem{iwsumrul}
A.V.Radyushkin, Phys. Lett. {\bf B271} (1991) 218;
M.Neubert, Phys. Rev. {\bf D45} (1992) 2451;
E. Bagan, P. Ball, V.M. Braun, and H.G. Dosch, Phys.Lett. {\bf B278} 
(1992) 457;
M.Neubert, Phys. Rev. {\bf D47} (1993) 4063.

\bibitem{latiw}
C.W. Bernard, Y. Shen and A.Soni, Phys. Lett. {\bf B317} (1993)164;
UKQCD Collaboration, Phys. Rev. Lett. {\bf 72} (1994) 462.

\bibitem{2corr}
A. Falk and M. Neubert, Phys. Rev. {\bf D47} (1993) 2965 and 2982;
T. Mannel, Phys. Rev. {\bf D50} (1994) 428;
M. Shifman, N. Uraltsev and A. Vainshtein, Phys. Rev. {\bf D51} (1995) 2217.

\bibitem{ag}
M. Ademollo and R. Gatto, Phys. Rev. Lett. {\bf 13} (1964) 264.

\bibitem{neub94}
M. Neubert, Phys. Lett. {\bf B338} (1994) 84.

\bibitem{manwin94}
T.Mannel and W.Roberts, Z.Phys. {\bf C61} (1994) 293.

\bibitem{goity92}
J.L. Goity, Phys. Rev. {\bf D46} (1992) 3929.

\bibitem{randall93}
L.Randall and M.B. Wise, Phys. Lett. {\bf B303} (1993) 135.

\bibitem{boyd95}
C.G. Boyd and B. Grinstein, Nucl. Phys. {\bf B451} (1995) 177.

\bibitem{jenkins92}
E. Jenkins and M.J.Savage, Phys. Lett. {\bf B281} (1992) 331.

\bibitem{lepratio}
B.Grinstein, E.Jenkins, A.V.Manohar, M.J.Savage and M.B.Wise, Nucl. Phys. 
{\bf B380} (1992) 369.

\bibitem{lee92}
C.L.Y. Lee, M. Lu and M.B. Wise, Phys. Rev. {\bf D46} (1992) 5040.

\bibitem{cheng93}
H.-Y. Cheng, C.-Y. Cheung, G. L. Lin, Y. C. Lin, T.-M. Yan and H.-L. Yu, 
Phys. Rev. {\bf D48} (1993) 3204.

\bibitem{kramer93}
G. Kramer and W.F. Palmer, Phys. Lett. {\bf B298} (1993) 437.

\bibitem{lee93}
C.L.Y. Lee, Phys. Rev. {\bf D48} (1993) 2121.

\bibitem{goity95}
J.L. Goity and W. Roberts, Phys. Rev. {\bf D51} (1995) 3459.

\bibitem{latfp}
A.Abada et al., Nucl. Phys. {\bf B376} (1992) 172;
UKQCD Collab., R.M.Baxter et al., Phys. Rev. {\bf D49} (1994) 1549;
C.Alexandrou et al., Z. Phys. {\bf C62} (1994) 659;
C.W.Bernard, J.N.Labrenz and A.Soni, Phys. Rev. {\bf D49} (1994) 2536;
MILC Collab., C. Bernard et al., FSU-SCRI-95C-28, talk given at LAFEX 
International School on High Energy Physics (LISHEP 95).

\bibitem{falk93}
A.F.Falk, Phys. Lett. {\bf B305} (1993) 268.

\bibitem{cleobrux}
E.H. Thorndike, talk given at the 1995 International Europhysics Conference
on High Energy Physics, Bruxelles, 27 July- 2 August 1995.

\bibitem{grinstein94}
B.Grinstein and P.F.Mende, Nucl. Phys. {\bf B425} (1994) 451.

\bibitem{wsb85}
M. Wirbel, B.Stech and M. Bauer, Z. Phys. {\bf C29} (1985) 637.

\bibitem{isgw89}
N. Isgur et al., Phys. Rev. {\bf D39} (1989) 799; 
N. Isgur and D. Scora, Phys. Rev. {\bf D40} (1989) 1491.

\bibitem{bbd91}
P.Ball, V.M. Braun and H.G. Dosch, Phys. Lett. {\bf B273} (1991) 316.

\bibitem{ball93}
P. Ball, Phys. Rev. {\bf D48} (1993) 3190.

\bibitem{belyaev93}
V.M. Belyaev, A. Khodjamirian and R. Rueckl, Z.Phys. {\bf C60} (1993) 349.

\bibitem{ali94}
A.Ali, V.M. Braun and H. Simma, Z.Phys. {\bf C63} (1994) 437.

\bibitem{santo94}
P.Colangelo and P.Santorelli, Phys. Lett. {\bf B327} (1994) 123. 

\bibitem{narison95}
S. Narison, Phys. Lett. {\bf B345} (1995) 166.

\bibitem{ball91}
P. Ball, V.M. Braun and H.G. Dosch, Phys. Rev. {\bf D44} (1991) 3567.

\bibitem{abada94}
A.Abada et al., Nucl. Phys. {\bf B416} (1994) 675.

\bibitem{uk95}
UKQCD Collaboration, K.C.Bowler et al., Phys. Rev. {\bf D51} (1995) 4905.

\bibitem{ape95}
APE Collab., C.R.Allton et al., Phys. Lett. {\bf B345} (1995) 513.

\bibitem{ukqcd95a}
UKQCD Collab., D.R.Burford et al., Nucl.Phys. {\bf B447} (1995) 425.

\bibitem{ukqcd95}
UKQCD Collaboration, J.M. Flynn et al., Nucl.Phys. {\bf B461} (1996) 327.

\bibitem{gourdin94}
M. Gourdin, A.N. Kamal and X.Y. Pham, Phys. Rev. Lett. {\bf 73} (1994) 3355.

\bibitem{aleksan94}
R. Aleksan, A. Le Yaouanc, L. Oliver, O. Pene and J.C. Raynal, 
Phys. Rev. {\bf D51} (1995) 6235.

\bibitem{gourdin95}
M. Gourdin, Y.Y. Keum and X.Y. Pham, Phys. Rev. {\bf D52} (1995) 1597.

\bibitem{sumrul}
M.A. Shifman, A.I. Vainshtein and V.I. Zakharov, Nucl. Phys. {\bf B147} 
(1979) 385; ibidem 448.

\bibitem{xu93}
Q.P. Xu, Phys. Lett. {\bf B306} (1993) 363.

\bibitem{wolfenstein92}
L.Wolfenstein, Phys. Lett. {\bf B291} (1992) 177.

\bibitem{cleo94}
CLEO Collaboration, M.S.Alam et al., Phys. Rev. {\bf D50} (1994) 43.

\bibitem{argus94}
ARGUS Collaboration, H.Albrecht et al., Phys. Lett. {\bf B340} (1994) 217.

\bibitem{cdf95}
CDF Collaboration, F.Abe et al., Phys. Rev. Lett. {\bf 75} (1995) 3068.

\bibitem{facto}
J.Schwinger, Phys. Rev. Lett. {\bf 12} (1964) 630; R.P.Feynman, 
in {\it Symmetries 
in Particle Physics}, ed. by A.Zichichi, Academic Press 1965, p.167.

\bibitem{bdecII}
For a  review on $B$ decays see: T.E.Browder, K.Honscheid and S.Playfer,
in {\it $B$ decays II}, ed. S.Stone, World Scientific, 1994.

\bibitem{shamali}
A.N. Kamal and F.M. Al-Shamali, preprint Alberta Thy-12-96, hep-ph/9605293.

\bibitem{allc}
T. M. Aliev, D. A. Demir, E. Iltan and N. K. Pak, Phys. Rev. {\bf D53} (1996)
355.

\bibitem{frlbb}
Riazuddin and Fayyazuddin, Phys. Lett. {\bf B337} (1994) 189.

\bibitem{chern90}
V.L. Chernyak and I.R. Zhitnitsky, Nucl. Phys. {\bf B345} (1990) 137.

\bibitem{narison92}
S.Narison, Phys. Lett. {\bf B283} (1992) 384.

\bibitem{fgm95}
R.N. Faustov, V.O. Galkin and A. Yu. Mishurov, Phys. Lett. {\bf B356} (1995) 
516.


\bibitem{defazio1} P. Colangelo, F. De Fazio and G. Nardulli,
Phys. Lett. {\bf B316} (1993) 555.

\bibitem{eichten}
E. Eichten, K. Gottfried, T. Kinoshita, K. D. Lane and T. M. Yan, Phys. Rev. 
{\bf D21} (1980) 203.

\bibitem{singer} G. A. Miller and P. Singer, Phys. Rev. {\bf D37} (1988)
2564; P. Singer and  G. A. Miller, Phys. Rev. {\bf D39} (1988)
825.

\bibitem{eletsky}
V. L. Eletsky and Ya. I. Kogan, Zeit. fur Phys. {\bf C28} (1985) 155.

\bibitem{aliev} 
T. M. Aliev, E. Iltan and N. K. Pak, Phys. Lett. {\bf B334}
(1994) 169.

\bibitem{kamal} A. N. Kamal and Q. P. Xu, Phys. Lett. {\bf B284} (1992) 421.


\bibitem{luke} A. F. Falk and M. Luke, Phys. Lett. {\bf B292} (1992) 119.

\bibitem{atwood}
D.Atwood, G.Eilam and A.Soni, SLAC-PUB-6716, hep-ph/9411367.

\bibitem{burdmanf}
G.Burdman, T.Goldman and D.Wyler, Phys. Rev. {\bf D51} (1995) 111.

\bibitem{fdfz3} P. Colangelo, F. De Fazio and G. Nardulli, Phys. Lett. 
{\bf B372} (1996) 331.

\bibitem{CLEOIIa}
M.Artuso et al., CLEO collaboration, Phys.Rev.Lett. {\bf 75} (1995) 785.

\bibitem{eil2}
G.Eilam, I.Halperin and R.R. Mendel, Phys.Lett. {\bf B361} (1995) 137.

\bibitem{IWtau}
N.Isgur and M.B.Wise, Phys. Rev. {\bf D43} (1991) 819; S. Balk,
J.G. K\"orner, G. Thompson and F. Hussain, Z. Phys. {\bf C59} (1993) 283.

\bibitem{CNP} 
P.Colangelo, G.Nardulli and N.Paver,  Phys. Lett. {\bf B293} (1992) 207.  

\bibitem{bertolini}
S.Bertolini, F.Borzumati and A.Masiero, Phys. Rev. Lett. {\bf 59} (1987)
180; N.G.Deshpande, P.Lo, J.Trampetic, G.Eilam and P.Singer, Phys. Rev.
Lett. {\bf 59} (1987) 183; B.Grinstein, R.Springer and M.B.Wise,
Nucl. Phys. {\bf B339} (1990) 269; R.Grigjanis, P.J.O'Donnell,
M.Sutherland and H.Navelet, Phys. Lett. {\bf B237} (1990) 355;
G.Cella, G.Curci, G.Ricciardi and A.Vicer\'e, Phys. Lett. {\bf B248}
(1990) 181; M.Misiak, Phys. Lett. {\bf B269} (1991) 161.

\bibitem{cleobsg} M. S. Alam et al. (CLEO Coll.), Phys. Rev. Lett.
{\bf 74} (1995) 2885.

\bibitem{cleobkg}R. Ammar et al. (CLEO Coll.),
 Phys. Rev. Lett. {\bf 71} (1993) 674; CLEO Coll. paper EPS0160, submitted
to the EPS Conference on High Energy Physics, Bruxelles, July 1995.

\bibitem{isgurgamma}
N.Isgur and M.B.Wise, Phys. Rev. {\bf D42} (1990) 2388.

\bibitem{santorelli}
P.Santorelli, Z. Phys. {\bf C61} (1994) 449.

\bibitem{gatto}
R. Gatto and G. Nardulli, Nuovo Cim. {\bf 108A} (1995) 1391.

\bibitem{griffin}
P. A. Griffin, M. Masip and M. Mc Guigan, 
Phys. Rev. {\bf D50} (1994) 5751.

\bibitem{ali}
A. Ali, Nucl. Inst. and Methods, {\bf A 351} (1994) 1; 
G. Ricciardi, preprint DSF-T-95/39, hep-ph/9510447, 
Proc. of the EPS Conference on High Energy Physics, Bruxelles, July 1995.

\bibitem{paver0}
C.A. Dominguez, N. Paver and Riazuddin, Phys. Lett. 
{\bf B214} (1988) 459.

\bibitem{paver}P. Colangelo, C.A. Dominguez, G, Nardulli and
N. Paver, Phys. Lett. {\bf B317} (1993) 183.

\bibitem{ball2}
P. Ball, preprint TUM-T31-43-93 (1993).

\bibitem{braunbkg}A. Ali, V. M. Braun 
and H. Simma, Z. Phys. {\bf C63} (1994) 437.

\bibitem{narison}S. Narison, Phys. Lett. {\bf B327} (1994) 354.

\bibitem{latticeqcd}K. C. Bowler et al., UKQCD Coll., Phys. Rev. Lett.
{\bf 72} (1994) 1398 and Phys. Rev. 
{\bf D51} (1955) 4955; C. Bernard, P. Hsieh and A. Soni, Phys. Rev.
Lett. {\bf 72} (1994) 1402;
A. Abada et al., APE Coll., preprint CERN-TH/95-59.

\bibitem{Casalbuoni1}
R. Casalbuoni, A. Deandrea, N. Di Bartolomeo, R. Gatto and G. Nardulli, 
Phys. Lett. {\bf B312} (1993) 315.

\bibitem{burdman2}
G.Burdman and J.F.Donoghue, Phys. Lett. {\bf B270} (1991) 55.

\bibitem{milana} J. Milana, Phys.Rev.  {\bf D53} (1996) 1403.

\bibitem{a} A. Khodjamiryan, G. Stoll, D. Wyler,
Phys. Lett. {\bf B358} (1995) 129.

\bibitem{k} A. Ali, V. M. Braun, Phys. Lett. {\bf B359} (1995) 223. 

\bibitem{colriaz}
P.Colangelo, G.Nardulli, N.Paver and Riazuddin, Z. Phys. {\bf C45} (1990)
575.

\bibitem{soares} J. M. Soares, Phys.Rev. {\bf D53} 241.

\bibitem{golowich1}E. Golowich and S. Pakvasa, Phys. Lett. {\bf B205} 
(1988) 393.

\bibitem{deshpande}
N.G.Deshpande, J.Trampetic and K.Panose, Phys. Lett. {\bf B214} (1988) 467.

\bibitem{golowich2}E. Golowich and S. Pakvasa, Phys. Rev. {\bf D51}
(1995) 1215.

\bibitem{at} D. Atwood, B. Blok, A. Soni, SLAC-PUB-6635, hep-ph/9408373.

\bibitem{cng} H.Y.Cheng, Phys. Rev. {\bf D51} (1995) 6228.

\bibitem{inami}
T.Inami and C.S.Lim, Progress Theor. Phys. {\bf 65} (1981) 297.


\bibitem{nn}
N. G. Deshpande, Bombay HEP Workshop, 1989, pgs. 538-558,
talk given at the workshop on High Energy Phenomenology (TFR, Bombay, 1989).

\bibitem{scrimieri}P. Colangelo, F. De Fazio, P. Santorelli and E.
Scrimieri, Phys.Rev. {\bf D53} (1996) 3672.

\bibitem{riaz}
C.A.Dominguez, N.Paver and Riazuddin, Z. Phys. {\bf C48} (1990) 55.

\bibitem{appo}
T.Appelquist and H.D.Politzer, Phys. Rev. Lett. {\bf 34} (1975) 43;
E.Eichten, K.Gottfried, T.Kinoshita, J.Kogut, K.D.Lane and T.M.Yan, Phys. Rev.
Lett. {\bf 34} (1975) 369;
T.Appelquist, A.De R\'ujula, H.D.Politzer and S.D.Glashow, Phys. Rev. Lett. 
{\bf 
34} (1975) 365;
R.Barbieri, R.K\"ogerler, Z.Kunszt and R.Gatto, Nucl. Phys. {\bf B105} (1976) 
125;
E.Eichten and K.Gottfried, Phys. Lett. {\bf B66} (1977) 286;
A.B.Henriques, B.H.Kellet and R.G.Moorhouse, Phys. Lett. {\bf B64} (1976)85;
C.Quigg and J.Rosner, Phys. Rep. {\bf 56} (1979) 167;
A.Martin, Phys. Lett. {\bf B93} (1980) 338.

\bibitem{BTye}
W.Buchmueller and S.-H.H.Tye, Phys.Rev. {\bf D24} (1981) 132.

\bibitem{Quigg}
C.Quigg and J.L.Rosner, Phys. Lett. {\bf B71} (1977) 153.

\bibitem{Grant}
A.K.Grant, J.L.Rosner and E.Rynes, Phys. Rev. {\bf D47} (1993) 1981.

\bibitem{BW}
L.S.Brown and W.I.Weisberger, Phys. Rev. {\bf D20} (1979) 3239. 

\bibitem{bbl}
G.T.Bodwin,E.Braaten and G.P.Lepage, Phys. Rev. {\bf D51} (1995) 1125. 

\bibitem{nrqcd}
W.E.Caswell and G.P.Lepage, Phys. Lett. {\bf B167} (1986) 437. 

\bibitem{hqqet1}
B.Grinstein, W.Kilian, T.Mannel and M.Wise, Nucl. Phys. {\bf B363} (1991) 19; 
W.Kilian, P.Manakos and T.Mannel, Phys. Rev. {\bf D48} (1993) 1321; W.Kilian,
T.Mannel and T.Ohl, Phys. Lett. {\bf B304} (1993) 311.

\bibitem{hqqet2} 
T.Mannel and G.A.Schuler, Phys. Lett. {\bf B349} (1995) 181. 

\bibitem{hqqet3}
T.Mannel and G.A.Schuler, Z. Phys. {\bf C67} (1995) 159-180.

\bibitem{noi0}
R.Casalbuoni, A.Deandrea, N.Di Bartolomeo, F.Feruglio, R.Gatto and G.Nardulli,
Phys.Lett. {\bf B302} (1993) 95.

\bibitem{potbc1}
E.Eichten and F.Feinberg, Phys. Rev. {\bf D23} (1981) 2724;
S.Godfrey and N.Isgur, Phys. Rev. {\bf D32} (1985) 189.

\bibitem{potbc2}
S.S.Gershtein, V.V.Kiselev, A.K.Likhoded, S.R.Slabospitskii and A.V.Tkabladze,
Sov. J. Nucl. Phys. {\bf 48} (1988) 327;
S.S. Gershtein, V.V.Kiselev, A.K.Likhoded, A.V.Tkabladze, Phys. Rev. {\bf D51} 
(1995) 3613; see also their review article in Usp. Fiz. Nauk. {\bf 165} (1995) 
1.

\bibitem{delubc}
M.Lusignoli, M.Masetti and S. Petrarca, Z. Phys. {\bf C51} (1991) 549.

\bibitem{potbc3}
W.Kwong and J.L.Rosner, Phys. Rev. {\bf D44} (1991) 212.

\bibitem{potbc4}
E.J.Eichten and C.Quigg, Phys. Rev. {\bf D49} (1994) 5845.

\bibitem{lattbc}
C.T.H.Davies, A.J.Lidsey, K.Hornbostel, G.P.Lepage, J.Shigemitsu and J. Sloan,
hep-lat/9510052, Jul 1995. 4pp. Contributed to International Symposium on 
Lattice Field Theory, Melbourne, Australia, 11-15 Jul 1995.

\bibitem{sumrulbc}
S.Narison, Phys. Lett. {\bf B210} (1988) 238;
V.V. Kiselev and A.V. Tkabladze, Phys. Rev. {\bf D48} (1993) 5208;
P.Colangelo, G.Nardulli and N.Paver, Z. Phys. {\bf C57} (1993) 43;
E.Bagan, H.G.Dosch, P.Gosdzinsksy, S.Narison and J.M.Richard, Z. Phys. {\bf 
C64} (1994) 57. 

\bibitem{probc}
L.Clavelli, Phys. Rev. {\bf D26}, 1610 (1982);
C.R.Ji and F.Amiri, Phys. Rev. {\bf D35}, 3318 (1987), Phys. Lett {\bf B195}, 
593 (1987);
M.Lusignoli, M.Masetti, and S.Petrarca, Phys. Lett. {\bf B266}, 142 (1991);
C.H.Chang and Y.Q.Chen, Phys. Lett. {\bf B284}, 127 (1992); Phys. Rev. {\bf 
D46}, 3845 (1992);
K.Kolodziej, A.Leike and R.Ruckl, Phys. Lett. {\bf B355} (1995) 337;
M.Masetti and F.Sartogo, Phys. Lett. {\bf B357} (1995) 659.

\bibitem{fragbc}
K.Cheung, Phys. Rev. Lett. {\bf 71} (1993) 3413;
E.Braaten, K. Cheung and T.C.Yuan, Phys. Rev. {\bf D48} (1993) 5049;
K.Cheung and T.C.Yuan, Phys. Lett. {\bf B325} (1994) 481. 

\bibitem{hqetbc}
E.Jenkins, M.Luke, A.V.Manohar and M.J.Savage, Nucl. Phys. {\bf 
B390} (1993) 463.

\bibitem{Yann}
T.M.Yan, Phys.Rev. {\bf D22} (1980) 1653.

\bibitem{noi1}
R.Casalbuoni, A.Deandrea, N.Di Bartolomeo, F.Feruglio, R.Gatto and G.Nardulli,
Phys.Lett. {\bf B309} (1993) 163.

\bibitem{dobook}
J.Donoghue, E.Golowich, B.Holstein, ``Dynamics of the Standard Model'',
Cambridge University Press, Cambridge 1992.

\bibitem{barga}
R.Barbieri, R.Gatto and E.Remiddi, Phys. Lett. {\bf B61} (1976) 465;
R.Barbieri, M.Caffo, R.Gatto and E.Remiddi, Phys. Lett. {\bf B95} (1980) 93.

\bibitem{bbl1p1}
G.T.Bodwin, E.Braaten and G.P.Lepage, Phys. Rev. {\bf D46} (1992) R1914. 

\bibitem{proda}
C.Albajar et al., Phys. Lett. {\bf B256} (1991) 112;
F.Abe et al., Phys. Rev. Lett. {\bf 69} (1992) 3704; Phys. Rev. Lett. {\bf 71}
(1993) 2537.

\bibitem{promo}
H.Fritzsch, Phys. Lett. {\bf B67} (1977) 217; B.Guberina, J.H.K\"uhn,
R.D.Peccei and R.R\"uckl, Nucl. Phys. {\bf B174} (1980) 317; for a review and 
a complete list of references see G.A.Schuler, hep-ph/9403387 submitted to 
Phys. Rep.

\bibitem{proth}
E.Braaten, K.Cheung and T.C.Yuan, Phys. Rev. {\bf D48} (1993) 4230;
E.Braaten and T.C.Yuan, Phys. Rev. Lett. {\bf 71} (1993) 1673; E.Braaten and 
T.C.Yuan, Phys. Rev. {\bf D50} (1994) 3176; M.Cacciari and M.Greco, Phys. Rev. 
Lett. {\bf 73} (1994) 1586; E.Braaten, M.A.Doncheski, S.Fleming and M.Mangano, 
Phys. Lett. {B333} (1994)548; D.P.Roy and K.Sridhar, Phys. Lett. {\bf B339} 
(1994) 141; see also M.Mangano, hep-ph/9507353, proceedings of the Xth Topical 
Workshop on Proton-Antiproton Collider Physics.

\bibitem{radpo}
P.Moxhay and J.Rosner, Phys. Rev. {\bf D28} (1983) 1132;
S.Godfrey, Phys. Rev. {\bf D31} (1985) 2375;
S.Godfrey and N.Isgur, Phys. Rev. {\bf D32} (1985) 189;
S.N.Gupta, S.F.Radford and W.W.Repko, Phys. Rev. {\bf D34} (1986) 201;
F.Daghighian and D.Silverman, Phys. Rev. {\bf D36} (1987) 3401;
L.P.Fulcher, Phys. Rev. {\bf D39} (1989)295.

\bibitem{chwi}
P.Cho and M.B.Wise, Phys. Lett. {\bf B346} (1995) 129.

\bibitem{kuatua}
Y.P.Kuang, S.F.Tuan and T.M.Yan, Phys. Rev. {\bf D37} (1988) 1210.

\bibitem{E760}
E760 Collaboration, Phys. Rev. Lett. {\bf 69} (1992) 2337.

\bibitem{1p1}
K.T.Chao, Y.B.Ding and D.H.Qin, Phys. Lett. {\bf B301} (1993) 282.

\bibitem{BC}
L.S.Brown and R.N.Cahn, Phys. Rev. Lett. {\bf 35} (1975) 1.

\bibitem{Abrams}
G.S. Abrams et al., Phys. Rev. Lett. {\bf 34} (1975) 1181.

\bibitem{Argus}
Argus Collaboration, H. Albrecht et al., Z. Phys. {\bf C35} (1987) 283.
                
\bibitem{Bow}
T.Bowcock et al., Phys. Rev. Lett. {\bf 58} (1987) 307.

\bibitem{Ioffe}
B.L.Ioffe and M.A.Shifman, Phys.Lett. {\bf B95} (1980) 99.

\bibitem{Don}
J.F.Donoghue and S.F.Tuan, Phys.Lett. {\bf B164} (1985) 401.

\bibitem{Malt}
K.Maltman, Phys.Rev. {\bf D44} (1991) 751.

\bibitem{DW}
J.F.Donoghue and D.Wyler, Phys.Rev. {\bf D45} (1992) 892.

\end{thebibliography}
\end{document}